\documentclass[12pt]{article}

\usepackage{amstext,amsmath,amssymb,amsfonts,bbm}
\usepackage[latin1]{inputenc}
\usepackage{epsfig}
\usepackage{dsfont}
\usepackage{hyperref}
\usepackage{amsthm}
\usepackage{subfigure}
\usepackage{color}
\usepackage{multirow}
\usepackage{psfrag}
\usepackage{graphicx}
\usepackage{extarrows}

\theoremstyle{plain}

\newtheorem{lemma}{Lemma}

\newtheorem{theorem}{Theorem}
\newtheorem{corollary}{ Corollary }
\newtheorem{definition}{Definition}
\newtheorem{proposition}{Proposition}

\theoremstyle{definition}

\newcommand{\bea}{\begin{eqnarray}}
\newcommand{\eea}{\end{eqnarray}}

\newcommand{\Tr}{{\rm Tr}}
\newcommand{\tr}{{\rm tr}}

\newcommand{\cV}{{\cal V }}
\newcommand{\cE}{{\cal E }}

\newcommand{\cF}{{\cal F }}

\newcommand{\cS}{{\cal S }}

\newcommand{\cT}{ {\cal T} }

\newcommand{\cB}{ {\cal B} }

\newcommand{\be}{\begin{equation}}
\newcommand{\ee}{\end{equation}}

\newcommand{\ba}{\begin{array}}
\newcommand{\ea}{\end{array}}

\makeatletter

\begin{document}

\title{The $1/N$ Expansion of Tensor Models Beyond Perturbation Theory}

\author{Razvan Gurau\footnote{rgurau@cpht.polytechnique.fr; 
CPHT - UMR 7644, CNRS, \'Ecole Polytechnique, 91128 Palaiseau cedex, France  
and Perimeter Institute for Theoretical Physics, 31 Caroline St. N, ON, N2L 2Y5, Waterloo, Canada. }}

\maketitle

\abstract{We analyze in full mathematical rigor the most general quartically perturbed invariant probability measure for a random tensor.
Using a version of the Loop Vertex Expansion (which we call the mixed expansion) we show that the cumulants write as explicit series 
in $1/N$ plus bounded rest terms. The mixed expansion recasts the problem of determining the subleading corrections in $1/N$ into a 
simple combinatorial problem of counting trees decorated by a finite number of loop edges.

As an aside, we use the mixed expansion to show that the (divergent) perturbative expansion of the tensor models
is Borel summable and to prove that the cumulants respect an uniform scaling bound. In particular the quartically perturbed
measures fall, in the $N\to \infty$ limit, in the universality class of Gaussian tensor models.}

\section{Introduction}

Tensor models \cite{review} generalize matrix models \cite{Mehta,matrix} to higher dimensions and provide the analytical tool for the 
study of random geometries in three and more dimensions. Matrix models are probability distributions for random matrices 
and their moments can be evaluated via a perturbative expansion in ribbon Feynman graphs representing surfaces \cite{matrix}. 
A crucial tool in matrix models is the $1/N$ expansion discovered by 't Hooft \cite{'tHooft:1973jz}. The
perturbative series of matrix models can be reorganized as a series in $1/N$ (where $N$ is the size of the matrix)
indexed by the genus. At leading order only planar graphs \cite{Brezin:1977sv} contribute and a matrix model undergoes a phase 
transition to a theory of random continuum surfaces when the coupling constant 
is tuned to some critical value \cite{Kazakov:1985ds,mm}. 

However many questions concerning the $1/N$ expansion of matrix models remain unanswered. 
First and foremost the $1/N$ expansion seems inherently a perturbative tool: in order to perform it one
first performs the expansion in the coupling. The perturbative
series is not summable hence it is not clear if the subsequent steps are mathematically meaningful.
Furthermore, matrix models are defined for a fixed sign of the coupling constant (say positive) for which the 
perturbation is stable. However the phase transition to
continuum surfaces takes place when tuning the coupling constant to a {\it negative} critical value. This tuning is meaningful
after restricting to the leading order planar series (which is absolutely convergent for both signs of the 
coupling constant), but what, if any, is the meaning of this tuning to criticality beyond perturbation theory? Can one actually reach this 
phase transition by an analytic continuation starting from a matrix model? These questions have not yet been satisfactory answered.

Matrix models have been generalized in higher dimensions to tensor models and 
group field theories \cite{oldgft0.5,oldgft1,oldgft2,oldgft3,oldgft4,oldgft5,oldgft6,oldgft10}.
They encode models of random geometries relevant for quantum gravity. However, progress has been slow (beyond 
model building) in higher dimensions mainly due to the lack of a $1/N$ expansion for tensor models. 

This has changed with the advent of the $1/N$ expansion, initially for the colored \cite{color,lost} models, and subsequently
for all invariant tensor models \cite{uncoloring}. Indeed the perturbative series of tensor models supports a $1/N$ 
expansion \cite{expansion1,expansion2,expansion3,expansion4,expansioin5} indexed by the {\it degree}, 
a positive integer which plays in higher dimensions the role of the genus\footnote{Unlike the genus, the degree is {\it not} a topological 
invariant.}. The leading order {\it melonic} \cite{critical,melbp} graphs triangulate the $D$-dimensional sphere 
in any dimension \cite{expansion1,expansion2,expansion3}
and, like their two dimensional counterparts, tensor models undergo a phase transition to a theory of
continuous random spaces when tuning to criticality.
These results led to significant progress in the understanding of random geometries in higher dimensions
\cite{sefu2,Ryan:2011qm,Carrozza:2011jn,Carrozza:2012kt,IsingD,EDT,doubletens,sdequations1,sdequations2,sdequations3,Bonzom:2012sz,Bonzom:2012qx} 
and the related critical phenomena. Tensor models have been generalized to renormalizable (and generically asymptotically 
free) tensor field theories, \cite{BenGeloun:2011rc,BenGeloun:2012pu,
BenGeloun:2012yk,jbgelon,Samary:2013xla,Geloun:2012bz,Carrozza:2012uv,tt2} leading to the formulation of the ``tensor track'' approach 
to quantum gravity \cite{Rivasseau:2011hm,Rivasseau:2012yp}.

Tensor models have been shown to exhibit a powerful universality property \cite{universality}: all invariant models respecting a uniform bound 
on the cumulants become Gaussian in the $N\to \infty$ limit, but the covariance of the limiting Gaussian is a non trivial function of the parameters 
of the model.

The same questions concerning the status of the $1/N$ for matrix models can also be formulated for tensor models. 
Can one give a meaning to the $1/N$ expansion beyond perturbation theory? Can one reach the phase transition point by analytic continuation?
This paper answers the first question and establishes the $1/N$ expansion in the constructive field theory sense
\cite{GlimmJaffe}. In order to achieve this result we consider the most general tensor model with a relevant quartic 
interaction\footnote{Other quartic interactions can be added but are suppressed in $1/N$.}.
We generalize  to tensors
the Loop Vertex Expansion (LVE) \cite{LVE1,LVE2} initially introduced for 
matrices\footnote{The LVE has already been used in the literature  \cite{Magnen:2009at} in a related context.} and analyze it in detail. The LVE 
expresses any cumulant as an absolutely convergent series 
indexed by trees. In this series the coupling constant $\lambda$ appears in two 
places: first as an overall factor for each tree and second in the contribution of each tree. This allows for instance to 
prove that, like in matrix models, the perturbative series of tensor models is Borel summable in $\lambda$ uniformly in $N$. 

However, in the case of tensors the LVE is much more powerful. Although rather involved at first sight, this formulation has two important advantages.
First, in the contribution of each tree, the coupling constant always appears rescaled as $\lambda /N^{D-1}$ where $D$ is the rank of the 
tensor. Performing a Taylor expansion in $\lambda/N^{D-1} $ one proves that the rest term 
is {\it suppressed in powers of $|\lambda|/N^{D-2}$}.
We call this expansion {\it the mixed expansion}. The mixed expansion provides the $1/N$ expansion of 
the model beyond perturbation theory: although shifted from $N^{D-1}$ to $N^{D-2}$, the scaling with 
$N$ suppresses the rest term as long as $D\ge 3$. This does not hold for the case of matrices, $D=2$.
Second, at each order in $1/N$, one must sum the series indexed by trees and 
this can be done explicitly. Somewhat surprisingly, the mixed expansion turns out to be the appropriate computational 
tool  for the study of the $1/N$ series. It reorganizes the corrections in $1/N$ in terms of trees with a finite 
number of loop edges, and counting such trees is a straightforward combinatorial problem. 

The critical behavior of each order in $1/N$ is governed order by order by the divergence of the series indexed 
by trees. In particular this proves that all terms will diverge for the same critical constant.
The mixed expansion (more precisely the $1/N$ expansion derived from it) is an  
explicitly perturbation in $1/N$ around the leading order theory of trees with no loop edges. The leading order (melonic)
theory acts effectively as a new ``vacuum'' around which the subleading terms in $1/N$ act as perturbations.

Beyond providing a tool to analyze order by order the subleading behavior in $1/N$ of the tensor models and an avenue towards 
establishing their double scaling limits, the present study is a needed step in order to analytically continue 
the cumulants to the critical constant of the phase transition to the continuum theory. 

The situation is more subtle for matrix models. The rest terms in the mixed expansion are not suppressed in powers of $1/N$. In order 
to obtain a non perturbative definition of the $1/N$ series for matrices one needs to refine further the mixed expansion introduced in the 
present paper. We believe that an in depth study of the corrections in $1/N$ in the case of tensors will provide the guide 
towards obtaining the non perturbative $1/N$ expansion for matrices.

We emphasize that the techniques of this paper rely heavily on the LVE formalism: 
some familiarity with \cite{LVE1,LVE2} would greatly benefit the reader. 

This paper is organized as follows. Section \ref{sec:prereq} is an introductory section. In it we present in some details the framework 
of tensor models, we briefly recall the notion of Borel summability and we introduce at length the various notions 
used in the sequel.
In section \ref{sec:quart} we introduce the most general model with a relevant quartic perturbation, 
and we state and comment our main theorems. 
Section \ref{sec:proofs} contains the proofs of our results.

\section{Prerequisites}\label{sec:prereq}

In this section we present the general setting of random tensor models and introduce the various  notions and notations we will use in
the sequel.

\subsection{Generalities on Tensor Models}

We start by a brief overview of the general framework of invariant tensor models presented in detail in 
\cite{uncoloring,universality}. This allows us to introduce some relevant notions and notations and to state the universality
theorem for random tensors. Later on we will prove that the quartically perturbed measure we deal with in this paper obeys
indeed the universality theorem.

We consider rank $D$ covariant tensors ${\mathbb T}_{n^1\dots n^D}$, with $n^1, n^2,\dots n^D \in \{ 1,\dots N\}$.
having {\bf no symmetry} under permutation of their indices. The tensors transform under the
{\it external} tensor product of $D$ fundamental representations of the unitary group $U(N)$
(that is the unitary group acts independently on each index). The complex conjugate tensor, $\bar {\mathbb T}_{  n^1 \dots n^D }$
is a rank $D$ contravariant tensor. A tensor and its complex conjugate can be seen as 
collections of $N^D$ complex numbers supplemented by the requirement of covariance under base change. 
\bea
&& {\mathbb T}_{a^1\dots a^D} = \sum_{n^1\dots n^D}U_{a^1n^1}\dots V_{a^Dn^D} {\mathbb T}_{n^1\dots n^D}  \; ,\crcr
&&  \bar {\mathbb T}_{ \bar a^1\dots  \bar a^D} = \sum_{ \bar n^1\dots \bar n^D}
\bar U_{\bar a^1 \bar n^1 }\dots \bar V_{  \bar a^D \bar n^D} \bar {\mathbb T}_{ \bar n^1\dots \bar n^D}  \; ,
\eea
where the indices of the complex conjugated tensor are denoted conventionally with a bar.
We emphasize that the unitary operators $U,\dots V$ are all {\it independent}\footnote{One can consider 
more generally tensors transforming under the external tensor product of fundamental representations of 
unitary groups of different sizes $U(N_1)\boxtimes U(N_2)\boxtimes \dots \boxtimes U(N_D)$ with $N_i \neq N_j$.}.
We denote $\vec n$ the $D$-tuple of integers $(n^1, \dots n^D)$, and we take $D\ge 3$.

Any invariant polynomial in the tensor entries can be expressed in terms of the {\bf trace invariants}
built by contracting in all possible ways pairs of covariant and contravariant indices in a product 
of tensor entries (see \cite{collins,ColSni} for a direct proof relying on averaging
over the unitary group). The trace invariants are one to one with closed $D$-colored graphs.

\begin{definition}\label{def:colorgraph}
A {\bf bipartite closed $D$-colored graph} is a graph $\cB = \bigl(\cV(\cB) ,\cE(\cB) \bigr)$ with vertex set $\cV(\cB)$
and edge set $\cE(\cB)$
such that:
\begin{itemize}
\item  $\cV(\cB)$ is bipartite, i.e. the vertex set writes as $\cV(\cB)  = {\cal A}(\cB) \cup \bar {\cal A}(\cB) $, such that  
 $\forall l\in\cE(\cB)$, then $ l = (v,\bar v )$ with $v\in {\cal A}(\cB)$ and $\bar v\in\bar {\cal A}(\cB)$. Their cardinalities
satisfy $|\cV(\cB)| = 2| {\cal A}(\cB) | = 2|\bar {\cal A}(\cB) |$. We call $v\in {\cal A}(\cB) $ the white vertices and $ \bar v\in\bar {\cal A}(\cB)$
the black vertices of $\cB$.
\item  The edge set is partitioned into $D$ subsets $\cE (\cB)= \bigcup_{i  =1}^{D} \cE^i(\cB)$, where $\cE^i(\cB)=\{l^i=(v,\bar v)\}$ is the subset
of edges with color $i$.
\item  All vertices are $D$-valent with all edges
incident  to a given vertex having distinct colors.
\end{itemize}
\end{definition}

\begin{figure}[htb]
\begin{center}
\psfrag{T}{$ \mathbb T$}
\psfrag{D}{$ \bar{ \mathbb T} $}
\includegraphics[width=8cm]{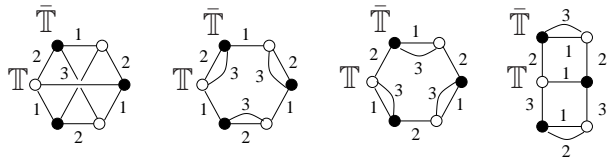}
\caption{$D$-colored graphs representing trace invariants.}
\label{fig:tensobs}
\end{center}
\end{figure}

The graph associated to an invariant (see Figure \ref{fig:tensobs} for some examples) is obtained as follows.
We represent every ${\mathbb T}_{ n^1\dots n^D}$ (respectively $ \bar {\mathbb T}_{ \bar n^1\dots \bar n^D}$) 
by a white vertex $v$ (respectively a black vertex $\bar v$). The position of an index becomes a {\it color}:
$n^1$ has color $1$, $n^2$ has color $2$ and so on. We represent by an edge the contraction of an index 
$n^i$ on ${\mathbb T}_{ n^1\dots n^D}$ with an index $\bar n^i$
of $ \bar{\mathbb T}_{ \bar n^1\dots \bar n^D}$. The edges $l^i = (v,\bar v) \in \cE^i(\cB)$ inherit the color $i$ of the indices 
and always connect a white and a black vertex.

The trace invariant associated to $\cB$ is 
\bea
\Tr_{\cB}({\mathbb T},\bar {\mathbb T} ) = \sum_{n,\bar n}  \delta^{\cB}_{n\bar n}  \; \prod_{v,\bar v \in \cV(\cB) }
 {\mathbb T}_{\vec n_v} \bar {\mathbb T}_{\vec {\bar n }_{\bar v} }
\; , \qquad \delta^{\cB}_{n\bar n} = \prod_{i=1}^D \prod_{l^i = (v,\bar v)\in \cE^i(\cB)} \delta_{  n^i_v \bar n^i_{\bar v} } \; .
\eea

The trace invariant associated to $\cB$ factors over its connected components. We call the invariant {\it connected} if $\cB$ is
connected. We denote $k(\cB)$ the number of 
white (or black) vertices of the graph $\cB$ and $C(\cB)$ the number of connected components of $\cB$, which we label $\cB_{\rho}$.
Colored graphs are dual to $D$-dimensional abstract simplicial pseudo-manifolds \cite{color,lost,pezzana,FG}.

A random tensor is a collection of $N^{D}$ complex random variables whose joint distribution is encoded in the cumulants (connected moments)
of $k$ entries $\mathbb{T}$ and $\bar k$ entries $\bar {\mathbb T}$, 
$ \kappa_{2k } [ {\mathbb T}_{ { \vec n}_{ 1 } } ,  \bar {\mathbb T}_{ { \vec {\bar n} }_{ \bar  1  } } \dots  {\mathbb T}_{ {\vec {\bar n} }_{  k   } },
 \bar {\mathbb T}_{ {\vec {\bar n} }_{{\bar  k}  } } ] $. We consider only even distributions, that is the cumulants are nontrivial only if $k=\bar k$.

\begin{definition}
 The probability distribution $\mu_N$ of the $N^D$ complex random variables ${\mathbb T}_{\vec n}$ is called
 {\bf trace invariant} if its {\bf cumulants} are linear combinations of trace invariant operators,
\bea\label{eq:cumul}
 \kappa_{2k } [ {\mathbb T}_{ { \vec n}_{ 1 } } ,  \bar {\mathbb T}_{ { \vec {\bar n} }_{ \bar  1  } } \dots  {\mathbb T}_{ {\vec {\bar n} }_{  k   } },
 \bar {\mathbb T}_{ {\vec {\bar n} }_{{\bar  k}  } } ]
  = \sum_{ \cB, \; k(\cB)=k}  {\mathfrak K} (  \cB  ,\mu_N )  \prod_{\rho=1}^{ C(\cB )  } \delta^{\cB_{\rho} }_{n\bar n}  \; ,
\eea
where the sum runs over {\bf all} the $D$-colored graphs $\cB $ with $2k$ vertices. 
\end{definition}

There exists a unique $D$-colored graph with $2$ vertices (it has $D$ edges connecting all the two vertices),
called the $D$-dipole and denoted $\cB^{(2)}$. 
We are interested in the large $N$ behavior of a trace invariant probability measure.
In order for such a limit to exist, the cumulants must scale with $N$. We denote
the rescaled cumulants
\bea\label{eq:cumul1.5}
  \frac{ {\mathfrak K} (  \cB ,\mu_N )  }{   N^{ - 2 (D-1) k (\cB)+ D- C(\cB) }}  \equiv K(\cB,N) \; ,
\eea
and we call $ K(\cB^{(2)},N) $ the covariance of the distribution.

\begin{definition}\label{def:unifbound}
 We say that the trace invariant probability distribution is {\bf properly uniformly bounded} at large $N$ if
\bea\label{eq:cumul2}
&&    \lim_{N\to \infty} K(\cB^{(2)}, N ) = K(\cB^{(2)}) <\infty \; ,\crcr
&&    K(\cB,N) \le K(\cB) \;,\quad  \forall \cB \neq \cB^{(2)} \text{ and } N \text{ large enough }\;.
\eea
\end{definition}

The scaling in equations \eqref{eq:cumul1.5} and \eqref{eq:cumul2} is {\it the only} scaling leading to a large $N$ limit.
This is a nontrivial statement and the reader should consult \cite{universality} for detailed explanations on this point.
The simplest example of a probability distribution for random tensors is the normalized Gaussian distribution of covariance $\sigma^2$ 
\bea \label{eq:gaussian1}
    e^{-N^{D-1}
 \frac{1}{\sigma^2}\sum_{\vec n, {\vec {\bar n} }   } {\mathbb T}_{\vec n } \delta_{ \vec n {\vec {\bar n} }   } \bar {\mathbb T}_{  {\vec {\bar n} } } }
\prod_{ \vec n } \Bigl(  \frac{N^{D-1}}{\sigma^2} \frac{d{\mathbb T}_{\vec n} d\bar {\mathbb T}_{ {\vec n } } } { 2 \pi \imath } \Bigr) \; ,
\eea
characterized by the expectations of the connected trace invariants
\bea\label{eq:moments}
\Big\langle \Tr_{\cB}({\mathbb T},\bar {\mathbb T}) \Big\rangle_{\sigma^2}
 =  \int \Big(   \prod_{ \vec n } \frac{N^{D-1}}{\sigma^2}  \frac{d{\mathbb T}_{\vec n} d\bar {\mathbb T}_{ {\vec n } } } { 2 \pi \imath} \Big{)}
  \;   e^{-N^{D-1}
\frac{1}{\sigma^2}\sum_{\vec n  {\vec {\bar n} }  } {\mathbb T}_{\vec n } \delta_{ \vec n { \vec {\bar n} }   } \bar {\mathbb T}_{ {\vec {\bar n} }  }
 } \; \Tr_{\cB}({\mathbb T},\bar {\mathbb T})  \; .
\eea
The moments of the Gaussian distribution are rather non trivial and have been studied in \cite{universality}.
For any connected graph $\cB$ with $2k(\cB)$ vertices there exist two non-negative integers, $\Omega(\cB)$ and $R(\cB)$ such that
\bea\label{eq:moments1}
 \lim_{N\to \infty} \ N^{ -1 + \Omega(\cB) }  \Big\langle \Tr_{\cB}({\mathbb T},\bar {\mathbb T}) \Big\rangle_{\sigma^2}
 =  \sigma^{2k(\cB)}  R(\cB) \; .
\eea
The normalization in eq.\eqref{eq:gaussian1} is the {\it only
normalization} which ensures that the convergence order is positive and, more importantly, {\it for all}
$\cB$, there exists an {\it infinite} family of invariants (graphs $\cB'$)
such that $\Omega(\cB) =\Omega(\cB')$. Again, this is a nontrivial statement and the reader
should consult \cite{universality} for more details.

\begin{definition}
A random tensor ${\mathbb T}$ distributed with the probability measure $\mu_N$ {\bf converges in distribution} to the distributional
limit of a Gaussian tensor model of covariance $\sigma^2$
if, for any connected trace invariant $\cB$,
\bea
 \lim_{N\to \infty}  N^{ -1 + \Omega(\cB) } \mu_N \Bigl[ \Tr_{\cB} (\mathbb{T},\bar {\mathbb T} )\Bigr]
= \sigma^{2k(\cB)}  R(\cB)   \; .
\eea

\end{definition}

The universality theorem for tensor models is \cite{universality} 
\begin{theorem}[Universality]\label{thm:mare}
Let $N^D$ random variables ${\mathbb T}_{\vec n }$ whose joint distribution is trace invariant and properly uniformly bounded
of covariance $K(\cB^{(2)},N)$.
Then, in the large $N$ limit, the tensor $ \mathbb{T}_{\vec n}$
converges in distribution to a Gaussian tensor of covariance $K(\cB^{(2)}) = \lim_{N\to \infty} K (\cB^{(2)}, N) $.
\end{theorem}

\subsection{Borel Summability}

The perturbative expansion of tensor models (and of quantum field theories in general)
is not summable. The root of the problem is that one usually performs an expansion in some coupling constant $\lambda$ around $\lambda=0$. However the interaction is stable for $\lambda>0$ but unstable for $\lambda<0$. The partition function and the cumulants are analytic 
in some domain in the complex plane outside the negative real axis. Hence $\lambda=0$ belongs to the boundary of the analyticity
domain of the cumulants. A Taylor expansion around a point belonging to the boundary of analyticity domain of some function is 
not absolutely convergent. However, in some cases, such Taylor expansions turn out to be Borel summable.

\begin{theorem}[Nevanlinna-Sokal, \cite{NevSok}]
 A function $f(\lambda,N)$ with $\lambda\in \mathbb{C}$ and $N \in \mathbb{R}_+$ is said to be Borel summable in $\lambda$ uniformly in $N$ if
 \begin{itemize}
  \item $f(\lambda,N)$ is analytic in a disk $\Re{\lambda^{-1}}>R^{-1}$ with $R\in \mathbb{R}_+$ independent of $N$.
  \item $f(\lambda,N)$ admits a Taylor expansion at the origin
        \bea
        f(\lambda,N) = \sum_{ k =0}^{r-1} f_{N,k} \lambda^k + R_{N,r}(\lambda) \; , \qquad |R_{N,r}(\lambda)| \le K \sigma^r r! |\lambda|^r \;,
        \eea
       for some constants $K$ and $\sigma$ independent of $N$.
 \end{itemize}

If $f(\lambda,N)$ is Borel summable in $\lambda$ uniformly in $N$ then $B(t,N) =\sum_{k=0}^{\infty} \frac{1}{k!} f_{N,k} t^k $
is an analytic function for $|t|<\sigma^{-1}$ which admits an analytic continuation in the strip
 $\{ z | \; | \Im z | < \sigma^{-1} \} $ such that $|B(t,N)| \le B e^{t/R}$ for some constant $B$ independent of $N$ and $f(\lambda,N)$ is
represented by the absolutely convergent integral
 \bea
   f(\lambda,N ) = \frac{1}{\lambda} \int_0^{\infty} dt \; B(t,N) e^{-\frac{t}{\lambda}} \; .
 \eea
\end{theorem}

That is the Taylor expansion of $f(\lambda,N)$ at the origin is Borel summable, and $f(\lambda,N)$ is its Borel sum.
The important thing about Borel summability is that it provides a uniqueness criterion: if a divergent series 
is the Taylor expansion of a Borel summable function $f(\lambda,N)$ at $\lambda=0$, then $f(\lambda,N)$ 
is the {\it unique} Borel summable function whose Taylor series is the original series.

\begin{figure}[htb]
   \begin{center}
 \includegraphics[width=4cm]{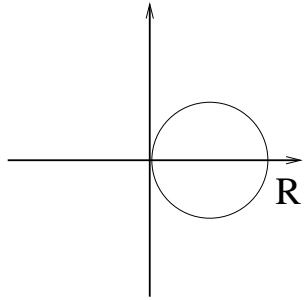}
 \caption{A Borel disk.}
  \label{fig:borel}
   \end{center}
\end{figure}

The set $\{\lambda | \Re{\lambda^{-1}}>R^{-1}, R\in \mathbb{R}_+ \} $ is a disk (which we call a Borel disk) in the complex
plane centered at $\frac{R}{2}$ and of radius $\frac{R}{2}$ (hence tangent to the imaginary axis, see Figure \ref{fig:borel}) as, denoting $\lambda = \frac{R}{2}+ae^{\imath \gamma}$,
\bea
\Re{\lambda^{-1}}>R^{-1} \Leftrightarrow \frac{R^2}{4} > a \; . 
\eea

\subsection{Decorated Trees}\label{sec:trees}

All our results rely on expansions indexed by various kinds of trees which we describe at length in this section.
We will denote $\vec \sigma$ a $D$-tuple of permutations over $k$ elements, $\vec \sigma = (\sigma_1,\dots \sigma_D)$.

\bigskip

\begin{figure}[htb]
   \begin{center}
 \includegraphics[width=8cm]{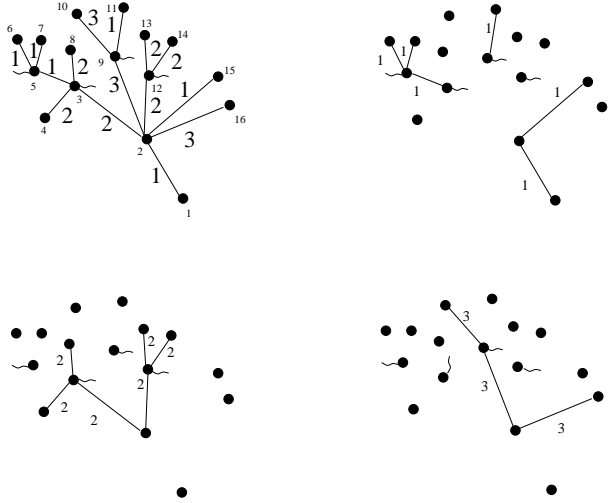}
 \caption{A ciliated plane tree with colored edges.}
  \label{fig:planetree}
   \end{center}
\end{figure}

\noindent{\bf Unrooted plane trees with colored, oriented edges and marked vertices $\cT_{n,\iota}^{\circlearrowright}  $.} An unrooted plane tree is a tree with
a cyclic ordering (say clockwise) of the edges at every vertex. We denote the total number of vertices of the tree by $n$ and label them $1,2,\dots n$.
The edges $(i,j)$ of the tree are oriented either from $i$ to $j$ or from $j$ to $i$ and have a color $c\in\{1,2\dots D\}$. Plane trees with marked
vertices $\iota=\{i_1,\dots i_k\}$ are obtained by selecting a preferred starting point of the cyclic ordering at the vertices $i_1, \dots i_k$.
The starting point is represented as a mark (or cilium) on the vertex\footnote{Such trees are sometimes called ciliated plane trees.}.
We denote such a tree $\cT_{n,\iota}^{\circlearrowright} $ with $\iota=\{ i_1,\dots i_k\}$ and  we denote the abstract tree associated to 
$\cT_{n,\iota}^{\circlearrowright} $ by $T_n$. Note that several plane trees are associated to the same abstract tree, and a vertex can have at most one cilium.
An example is presented in Figure \ref{fig:planetree}.

All the notions we introduce in the sequel refer to a fixed plane tree $\cT_{n,\iota}^{\circlearrowright} $. To simplify notations we consider 
this dependence implicit.

Every plane tree has an associated contour walk (Harris walk or Dyck path) consisting in the ordered list of 
vertices encountered when walking clockwise around the tree starting from some position. We identify a tree with its contour walk.
Whenever the walk encounters a cilium, the walk steps twice (first a step to 
the cilium and then a step from the cilium). We signal the presence of a cilium by a semicolon (rather than a comma)
separating the two entries in the walk. The contour walk of the tree in Figure \ref{fig:planetree} is therefore
\bea
&& ( 1, 2, 3,4,3,5 ; 5,6,5,7,5,3,8,3 ; 3,2,9, 10 ,9, 11 ,9 ; 9 ,2 , 12  , 13  , 12 , 14  , 12 ; \crcr
&& \qquad 12 ,2, 15  , 2  , 16  , 2 ) \; .
\nonumber
\eea
The contour walk of $ \cT_{n,\iota}^{\circlearrowright} $
has $2n-2+k$ {\it steps}, that is entries, denoted $q = 1 ,2 ,\dots 2n-2+k$. The walk in the previous example has $34$ steps. 
We denote the vertex corresponding to the step $q$ by $i(q)$. The contour walk is cyclic.

For every tree $\cT_{n,\iota}^{\circlearrowright}  $ we consider its connected subgraphs having all the vertices of
$\cT_{n,\iota}^{\circlearrowright}  $ but only the edges of color $c$ of $ \cT_{n,\iota}^{\circlearrowright}$ (see Figure \ref{fig:planetree}).
Each such subgraph is comprised of several plane trees which we denote $f^c$.
We call them {\it faces of color $c$} of $  \cT_{n,\iota}^{\circlearrowright} $. 
The walk of $ \cT_{n,\iota}^{\circlearrowright}   $ induces a contour walk 
for each of its faces $f^c $ obtained by selecting only the (ordered set of) steps $q(f^c )$ at which 
the contour walk of $ \cT_{n,\iota}^{\circlearrowright}   $ encounters a vertex of $ f^c  $. 
For instance, the face of color $1$ with vertices $1,2,15$ has an induced contour walk $f^1 =(1,2,2,2,2,15,2,2)$ consisting 
in the steps $q(f^1 )=\{ 1,2,16,23,30,31,32,34 \}$. Note that the walk of $  \cT_{n,\iota}^{\circlearrowright}  $ 
can leave and re-intersect several times $f^c $ (in the previous example it leaves $f^1 $ after the step $2$ and re-intersects 
it at the step $16$). We partition the faces into two subsets:
\begin{itemize}
 \item The faces $f^c$ having no cilium. We call them the {\bf internal faces of color $c$} of $\cT_{n,\iota}^{\circlearrowright}  $.
  For the tree in figure \ref{fig:planetree}, $f^1 =(1,2,2,2,2,15,2,2)$ or $f^1 =(4)$ are examples of internal faces of color $1$.
 \item The faces $f^c$ having at least a cilium. We call them the {\it external faces of color $c$} of $\cT_{n,\iota}^{\circlearrowright}  $.
  Say the vertices $i_{k_1}\dots i_{k_d}$ are the ciliated
  vertices of $f^c$ properly ordered,
  that is $f^c = ( \dots  i_{k_1} ; i_{k_1} \dots i_{k_2} ;i_{k_2} \dots i_{k_d} ; i_{k_d} \dots   )$. We can further subdivide $f^c$ into the walks
  $ f^{c; i_{k_1}\to i_{k_2}}$, $f^{c; i_{k_2}\to i_{k_3}} $ up to $f^{c ; i_{k_{d } \to i_{k_1} } }$\footnote{Where we use the cyclicity of 
  the walk to rearrange the last list in the appropriate order.}. We call these walks the
  {\bf external strands} of $\cT_{n,\iota}^{\circlearrowright}$. For the example of \ref{fig:planetree}, the external face of color $1$
  $f^1 =(3,3, 5 ; 5,6,5,7,5,3 ,3;3) $ subdivides into the external strands of color $1$ $f^{1;3\to 5} =(3,3,3,  5 )$
  and $f^{1; 5\to 3}=(5, 6,5,7,5,3,3) $. The strands are comprised of the (ordered) entries separated by semicolons in the walks of the faces.
  Every strand $f^{c;i\to i'} $ has an induced contour walk, made of the steps $q(f^{c;i\to i'})$.
\end{itemize}

Any vertex belongs to exactly $D$ faces, either internal or external, one for each color.
The cilia are identified as the encounters of semicolons (separating the same label)
in the contour walk of the tree. We denote $q_1 \dots q_k$ the positions of the cilia in the contour walk of the tree $ \cT_{n,\iota}^{\circlearrowright}   $ 
(that is $i(q_l)$ precedes a semicolon and $i(q_l)=i(q_l+1)$). Thus $q_1$ is the step at which the 
first cilium is encountered, $q_2$ the step at which the second cilium is encountered an so on.
All the faces $f^c $ to which the ciliated vertices $i(q_l)$ belong  are external.

The external faces and strands of color $c$ can be encoded by a permutation  $\xi_c$ over $1,\dots k$.
The encoding goes as follows. Consider the ciliated vertex $ i( q_1) $.
It belongs to the external face $f^c $ of color $c$.
Starting from $i( q_1)$, the first ciliated vertex (which is of course unique),
we encounter in the contour walk of $f^c  $ is of the form 
$ i( q_{l}) $ for some $l$ (as it is one of the ciliated vertices of $ \cT_{n,\iota}^{\circlearrowright}   $)  . We set $\xi_c(1)=l$. We repeat the procedure for $q_2$ and
so on and obtain a permutation $\xi_c$ over $1,\dots k$. For the example in figure \ref{fig:planetree} we have
$q_1 =6 ,q_2=14,q_3=21,q_4=28$ (and $i(q_1) = 5, i(q_2)=3, i(q_3)=9, i(q_4)=12$). The permutations $\xi_c$ write 
in cycle decomposition as $\xi_1 = (1,2)(3)(4)$, $\xi_2 =(1)(2,4)(3)$ and $\xi_3=(1)(2)(3)(4)$.
The external faces of color $c$ correspond to the cycles of $\xi_c$
\bea
&&  f^{c}  = \Bigl(  i( q_l)
\dots i ( q_{\xi_c(l)} ) ;  i( q_{\xi_c(l)} )
\dots i( q_{\xi_c^2(l)}) ;  \crcr
&& \qquad \qquad \qquad i( q_{\xi_c^2(l)} )
 \dots i( q_{\xi_c^{d-1}(l)}) ;  i( q_{\xi_c^{d-1} (l)} )
\dots   i( q_l)   \Bigr) \; , \qquad \xi_c^{d } (l) =l \; ,
\eea
and the strands write as
\bea
&&  f^{c; i( q_l) \to   i( q_{\xi_c(l)} ) }
=  \Bigl(  i( q_l)
\dots i( q_{\xi_c(l)} )  \Bigr) \; ,  \crcr
&& f^{c; i( q_{\xi_c(l)}) \to   i( q_{\xi^2_c(l)} ) }
=  \Bigl(  i( q_{\xi_c(l)})
\dots i( q_{\xi^2_c(l)} )  \Bigr) \; , \dots
  \eea
Consider for instance the example of \ref{fig:planetree} and $\xi_1 = (1,2)(3)(4)$.
The external faces and strands of color $1$ are 
\bea
&& f^1 = (3,3, 5 ; 5,6,5,7,5,3 ,3;3) \Rightarrow  f^{1; 3\to 5} = (  3 ,3,3,5  ) \;,\;  f^{1;5\to 3} = ( 5,6,5,7,5,3,3 ) \crcr
&& f^1 = (9,9,11,9;9) \Rightarrow  f^{1;9\to 9} = (  9,9,9,11,9) \crcr
&& f^1 = (12,12,12;12 ) \Rightarrow f^{1;12\to 12} =(12,12,12,12) \; .
\eea 

We denote the set of all internal faces of $\cT_{n,\iota}^{\circlearrowright}$ by $\cF^{ \rm{ int} }( \cT_{n,\iota}^{\circlearrowright} ) $,
the set of all external strands of $\cT_{n,\iota}^{\circlearrowright} $ by ${\cal S}^{\rm ext}(\cT_{n,\iota}^{\circlearrowright})$, and the set of all faces of 
$\cT_{n,\iota}^{\circlearrowright}$ by $\cF ( \cT_{n,\iota}^{\circlearrowright} ) $. 

A tree can be built by adding one by one edges connecting univalent vertices. At every step of this procedure the total number of faces of a tree increases by 
$D-1$ (every vertex connected by an edge of color $c$ will bring a new face for every color $c'\neq c$), and taking into account that the tree with a unique vertex has $D$ 
faces, we obtain
\bea\label{eq:facestree}
 | \cF ( \cT_{n,\iota}^{\circlearrowright} )| = D + (n-1)(D-1) = | \cF^{ \rm{ int} }( \cT_{n,\iota}^{\circlearrowright} ) | + \sum_{c=1}^D C(\xi_c) \; ,
\eea 
where $C(\xi_c)$ denotes the number of cycles of the permutation $\xi_c$.

\bigskip

\begin{figure}[htb]
   \begin{center}
 \includegraphics[width=8cm]{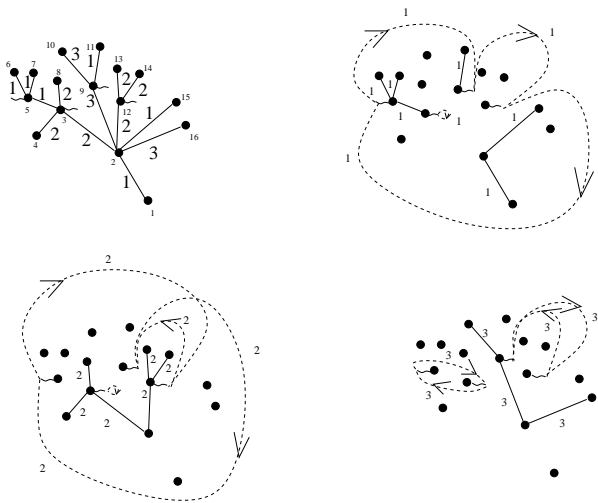}
 \caption{A ciliated plane tree with external edges.}
  \label{fig:planetree1}
   \end{center}
\end{figure}

\noindent{\bf Plane trees with external edges.} A notion closely related to ciliated plane trees is the one of trees with external edges.
For every $c$, we add to the tree dashed oriented edges of color $c$ connecting pairs of cilia such that every cilium 
has exactly an incoming and an outgoing dashed edge (we allow an edge to be incoming and outgoing on the same cilium). We call the dashed edges {\bf external edges}.
We presented in figure \ref{fig:planetree1} and example of tree decorated by external edges (we represented the external edges of color $c$ decorating the 
faces of color $c$).

The external edges of color $c$ can be encoded by a permutation $\tau_c$ over $k$ elements. The outgoing external edge of color $c$ on the cilium 
on the vertex $i(q_1)$ is an incoming external edge on some other cilium, say the one on the vertex $i(q_l)$. We set $\tau_c(1) = l$, and repeat the 
procedure for all the cilia.

The example in figure \ref{fig:planetree} corresponds to the permutations with cycle 
decomposition $\tau_1=(2)(1,3,4)$, $\tau_2=(2)(1,4,3)$ and $\tau_3=(1,2)(3,4)$
(recall that $q_1 =6 ,q_2=14,q_3=21,q_4=28$ and $i(q_1) = 5, i(q_2)=3, i(q_3)=9, i(q_4)=12$).

We denote the tree decorated by external edges $\cT_{n,\iota,\vec \tau}^{\circlearrowright}  $, with $\vec \tau = (\tau_1,\dots \tau_D)$.
The external edges recombine the external strands of $\cT_{n,\iota}^{\circlearrowright}$ into {\bf external faces} 
of $\cT_{n,\iota,\vec \tau }^{\circlearrowright}  $. They are defined as follows.

Starting from the external faces $f^{c}$ of $\cT_{n,\iota}^{\circlearrowright} $,
\bea
 && f^{c} = \Bigl(  i( q_l)
\dots i( q_{\xi_c(l)} ) ; i( q_{\xi_c(l)} )
\dots i( q_{\xi_c^2(l)}); \crcr
&& \qquad \qquad \qquad i( q_{\xi_c^2(l)} ) \dots
 i( q_{\xi_c^{d-1}(l)}); i( q_{\xi_c^{d-1} (l)} )
\dots   i (q_l)   \Bigr) \; , \qquad  \xi_c^{d } (l) =l \; ,
\eea
with strands
\bea
  f^{c; i( q_l) \to   i( q_{\xi_c(l)} ) } 
=  \Bigl(  i( q_l)
\dots i( q_{\xi_c(l)} )  \Bigr) \; ,
  \eea
we build the external faces of $\cT_{n,\iota,\vec \tau }^{\circlearrowright}  $ indexed by the cycles of $\tau_c\xi_c$
\bea
&& f^{c}(  \cT_{n,\iota,\vec \tau }^{\circlearrowright}   ) =
\Bigl( i (q_l)
\dots i( q_{\xi_c(l)} ) ; i( q_{ \tau_c\xi_c(l)} )
\dots  i( q_{ \xi_c \tau_c\xi_c(l)} ); \crcr
&& \qquad \qquad \qquad \qquad  i( q_{ \tau_c \xi_c \tau_c\xi_c(l)} ) \dots i( q_{  \xi_c( \tau_c \xi_c)^{d-1}(l) } ) \Bigr) \; , \qquad ( \tau_c\xi_c )^{d}(l) =l \; .
\eea
Consider for instance the example of \ref{fig:planetree1}. Recall that we have $\xi_1 = (1,2)(3)(4)$, $\tau_1 = (2)(1,3,4)$ and 
\bea
&& f^{1; 3\to 5} = (  3 ,3,3,5  ) \;, \;  f^{1;5\to 3} = ( 5,6,5,7,5,3,3 ) \; , \crcr
 && f^{1;9\to 9} = (  9,9,9,11,9) \;,\;  f^{1;12\to 12} =(12,12,12,12) \; .
\eea 
We have $\tau_1 \xi_1 = (1,2,3,4)$ and the associated external face of $\cT_{n,\iota,\vec \tau }^{\circlearrowright}  $ is
\bea
f^c( \cT_{n,\iota,\vec \tau }^{\circlearrowright}  ) = (5,6,5,7,5,3,3 ;  3 ,3,3,5 ;  9,9,9,11,9 ; 12,12,12,12 )
\eea 
The internal faces of $ \cT_{n,\iota, \vec \tau }$ are the internal faces of $\cT_{n,\iota } $ and
we denote $ {\cal F}( \cT_{n,\iota, \vec \tau }) $ the set of all the faces, internal and external, of
$ \cT_{n,\iota, \vec \tau } $. We have
\bea\label{eq:facestreeext}
&& |\cF (  \cT_{n,\iota, \vec \tau}^{\circlearrowright}  )| = |\cF^{\text{int}} (\cT_{n,\iota }^{\circlearrowright} )| + \sum_c C(\tau_c \xi_c) \crcr
&& = D + (n-1)(D-1) -  \sum_{c=1}^D C(\xi_c) +  \sum_c C(\tau_c \xi_c) \; .
\eea

\bigskip

\begin{figure}[htb]
   \begin{center}
 \includegraphics[width=8cm]{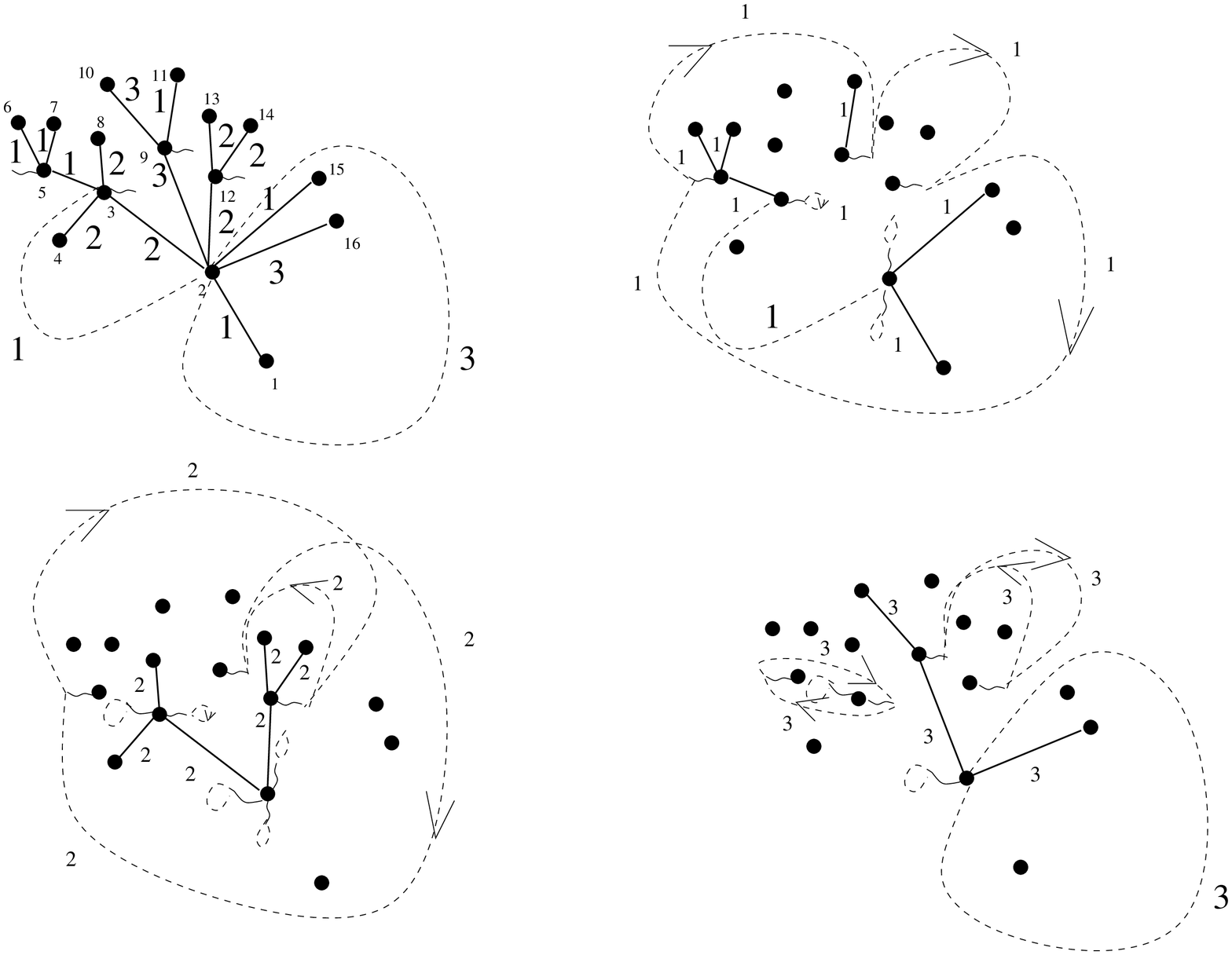}
 \caption{A ciliated plane tree with external edges and loop edges.}
  \label{fig:planetree2}
   \end{center}
\end{figure}

\noindent{\bf Plane trees with external edges and loop edges}. The last type of decorated trees we will use is trees with loop edges.
We add to the plane tree with external edges $\cT_{n,\iota,\vec \tau}^{\circlearrowright} $ 
$2s$ new cilia located on the vertices $j_1$, $j_{1'}$, $j_2, j_{2'}$ up to $j_s,j_{s'}$. The new cilia are allowed to be located anywhere on the tree,
including on one of the vertices in $\iota$. We assign to these new cilia a color such that $j_1$ and $j_{1'}$ 
have both color $c_1$, $j_2$ and $j_{2'}$ have both color $c_2$ an do on. We connect the cilia on $j_1$ and $j_{1'}$ by a dashed edge of color $c_1$
and so on. We call these new edges {\bf loop edges}. The loop edges are {\bf not} oriented.
We denote the graph thus obtained $ \cT_{n,\iota,\vec \tau,{\cal L} }^{\circlearrowright} $ with 
${\cal L}=\{(j_1,j_{1'}), (j_2,j_{2'}),\dots (j_s,j_{s'})\}$. We present in figure \ref{fig:planetree2}, in the upper left corner, a plane tree 
with two loop edges: one of color $c_1=3$ and one of color $c_2=1$ with $j_1 = 2, j_{1'}=2, j_2=2, j_{2'}=3$. The loop edges have no orientation. 

The presence of the loop edge has several consequences. First, the walk around the tree  is modified
by the presence of the new cilia. For the example in figure 
 \ref{fig:planetree2}, in the presence of the loop edge, the walk writes 
 \bea
&&  ( 1, 2\fbox{;}2 \fbox{;} 2, 3,4,3\fbox{;}3,5 ; 5,6,5,7,5,3,8,3 ; 3,2,9, 10 ,9, 11 ,9 ;\crcr
  && \qquad  9 ,2 , 12  , 13  , 12 , 14  , 12 ; 12 ,2 \fbox{;}2, 15  , 2  , 16  , 2 ) \; .
 \eea 
where we boxed the semicolons representing the new cilia (a new cilium transforms $i,$ into $i;i$ in the walk).
Furthermore the presence of a loop edge modifies the faces. Consider the cilia on $j_p$ and $j_{p'}$. They can either
\begin{itemize}
 \item belong to two distinct faces of color $c_p$, 
        \bea 
        f^{c_p}_1=(\dots j_p;j_p \dots )\qquad \text{and} \qquad f^{c_p}_2=(\dots j_{p'};j_{p'}\dots) \;.
        \eea 
      Then the two faces are merged into the face $f^{c_p}=(\dots j_p ; j_{p'}\dots j_{p'} ; j_p \dots )$.
 \item belong to a unique face of color $c_p$. Then $f^{c_p}= (\dots j_p;j_p \dots j_{p'} ; j_{p'} \dots)$ splits
      into two faces $(j_p \dots j_{p'})$ and $(j_{p'}\dots j_p)$.
\end{itemize}

More precisely, the tree decorated by loop edges $ \cT_{n,\iota,\vec \tau,{\cal L} }^{\circlearrowright}  $ is the tree with external edges 
$   \cT_{n,\iota',\vec \tau'}^{\circlearrowright}$ obtained by adding the $2s$ new cilia on $  j_1,j_{1'},j_2,j_{2'},\dots  j_s, j_{s'} $ 
(hence $\iota' = \iota \cup \{  j_1,j_1',j_2,j_2',\dots  j_s, j'_s\}$) and connecting by external edges 
\begin{itemize}
\item $j_p$ with itself (and $j_{p'}$ with itself) by a pair of external edges of opposite orientations for every color $c \neq c_p$.
\item $j_p$ with $j_{p'}$ by two external edges (one for each orientation) of color $ c_p $. 
\end{itemize}

The permutation $\vec \xi'$ and $\vec \tau'$ associated to $\cT_{n,\iota',\vec \tau'}^{\circlearrowright} $ can easily be 
identified: for the example of figure \ref{fig:planetree2} we have
\bea
&&  q_1 = 2 \;, \;q_2 =3 \;, \; q_3 =7 \;, \; q_4 = 9 \;, \; q_5 = 17 \;, \; q_6 = 24 \;, \;q_7 = 31 \;, \; q_8 =33 \crcr
&& \xi'_1 = (1,2,8) (3,4,5) (6) (7) \crcr
&& \xi'_2 = (1,2,3,5,7,8) (4) (6) \crcr
&& \xi'_3 = (1,2,6,8) (3,5) (4) (7) \crcr
&& \tau'_1 = (1)(2,3)(4,6,7) (5) (8) \crcr
&& \tau'_2 = (1)(2)(3)(4,7,6) (5)(8)\crcr
&& \tau'_3 = (1,8) (2)(3)(4,5) (6,7) \; .
\eea 

However, writing them requires a bit of care as the presence of the new cilia shifts the steps in the contour walk.
It is best to present $  \cT_{n,\iota',\vec \tau'}^{\circlearrowright} $ in two stages.
The cilia of $\cT_{n,\iota' }^{\circlearrowright} $ are either cilia of $ \cT_{n,\iota}^{\circlearrowright} $ or
they are among the $2s$ new cilia. The step $q_l$ 
corresponding to the $l$'th cilium encountered in the walk around $ \cT_{n,\iota }^{\circlearrowright} $ becomes the step $q'_{m(l)}$
corresponding to the $m(l)$'th cilium in the walk around $\cT_{n,\iota' }^{\circlearrowright} $. The other steps, denoted
$q'_{t(p)}$ and $ q'_{t(p')}$ in the walk around $\cT_{n,\iota' }^{\circlearrowright} $ corresponding to the new cilia $j_p$ and $j_{p'}$. 
It follows that the steps $q'_r$ corresponding to cilia in the walk around $ \cT_{n,\iota' }^{\circlearrowright}  $ are partitioned into
three categories: $r=m(l)$ for some $l$, $r=t(p)$ for some $p$ or $r=t(p')$ for some $p'$. 

For the example of Figure \ref{fig:planetree2} we have 
\bea
 && m(1)= 4 \;, \; m(2) = 5\;, \; m(3) = 6 \;, \; m(4) = 7 \; , \crcr
 && t(1) = 1 \; , \; t(1') = 8 \; , \;  t(2) = 2 \;, \; t(2')= 3 \; .
\eea 

{\it Step 1}. The presence of new cilia modifies the cycle decomposition of $\xi'_c$. A cycle of $\xi_c$ of the form $(\dots l \xi_c(l) \dots )$
becomes $(\dots m(l) \dots t(p) \dots t(q') \dots \xi_c(m(l)) \dots )$, where $t(p)$ and $t(q')$ denote the (ordered list of) 
new cilia inserted on the face corresponding to the cycle of $\xi_c$ (for example the cycle $(3,4,5)$ in $\xi'_1$ which comes from the cycle $(1,2)$ in $\xi_1$). 
Note that if $j_p$ and $j_{q'}$ are inserted on some internal face of $\cT_{n,\iota}^{\circlearrowright}   $
then they will form a cycle $\dots t(p) \dots t(q') \dots$ in the permutation $\xi_c'$ having no corespondent in $\xi_c$ (for example the cycle $(1,2,8)$ in 
$\xi'_1$). We reconnect the new cilia with trivial external edges
\bea
 \tilde \tau_c(r) = \begin{cases}
               m(\tau_c(l))  \quad &  \text{ if } r=m(l)  \\
               t(p)  \quad & \text{ if } r=t(p) \\
               t(p')  \quad &  \text{ if } r=t(p') 
               \end{cases} \; .
\eea 
We thus obtain an intermediate tree with external edges $ \cT_{n,\iota',\vec {\tilde \tau} }^{\circlearrowright}  $. 
The permutations $\tilde \tau_c$ acquire extra cycles of length 1 with respect to $\tau_c$. As a function of the cycle structure of $\xi_c'$, one can at 
most convert internal faces of color $c$ into external faces of colors $c$, (if $\xi'_c$ has more cycles than $\xi_c$) but the total 
number of faces does not change
\bea
 | {\cal F}(\cT_{n,\iota',\vec {\tilde \tau} }^{\circlearrowright} )   | = |  {\cal F}( \cT_{n,\iota ,\vec {  \tau} }^{\circlearrowright}  ) | \; .
\eea 

{\it Step 2.} We convert the permutation $\tilde \tau_c$ into the permutations $\tau'_c$ defined as 
\bea
 \tau_c' (r) = \begin{cases}
               m(\tau_c(l))  \quad &  \text{ if } r=m(l) \text{ for some } l \text{ (} \tau'_1(4) = 6 \text{ )}\\
               t(p')  \quad & \text{ if } r=t(p) \text{ for some } p \text{ and } c=c_p \text{ (} \tau'_1(2) = 3 \text{ )} \\
               t(p)   \quad & \text{ if } r=t(p') \text{ for some } p'\text{ and } c=c_p \text{ (} \tau'_1(3) = 2 \text{ )} \\
               t(p)   \quad & \text{ if } r=t(p) \text{ for some } p \text{ and } c\neq c_p \text{ (} \tau'_2(2) = 2 \text{ )}\\
               t(p')  \quad &  \text{ if } r=t(p') \text{ for some } p' \text{ and } c\neq c_p \text{ (} \tau'_2(3) = 3 \text{ )}
               \end{cases} \; .
\eea 

We now obtain the tree with external edges $  \cT_{n,\iota',\vec \tau'}^{\circlearrowright}  $ corresponding to the tree with loop edges
$  \cT_{n,\iota,\vec \tau,{\cal L}}^{\circlearrowright}  $. For every loop edge of color $c$, two cycles of length one in 
$\tilde \tau_c$ are merged into a cycle of length two of $\tau_c'$.

On the other hand, for any permutation $\xi$, if $\tau'$ is obtained from $\tilde \tau$ by merging two cycles of length 1 into a cycle of length two,
\bea
 C( \tau' \xi' ) \le C(\tilde \tau \xi) + 1 \; ,
\eea 
hence 
\bea\label{eq:facestreeextloop}
  && \sum_c C(\tau'_c \xi'_c) \le \sum_{c} C( \tilde \tau_c \xi'_c) + s \Rightarrow 
  |{\cal F}( \cT_{n,\iota, \vec \tau,{\cal L} }^{\circlearrowright} )| \le |{\cal F}( \cT_{n,\iota, \vec \tau }^{\circlearrowright}  ) | + s \crcr
  && \Rightarrow  |{\cal F}( \cT_{n,\iota, \vec \tau,{\cal L} }^{\circlearrowright} )| \le 
  D + (n-1)(D-1) -  \sum_{c=1}^D C(\xi_c) +  \sum_c C(\tau_c \xi_c)  +s
  \; ,
\eea 
where ${\cal F}( \cT_{n,\iota, \vec \tau,{\cal L} }^{\circlearrowright}  )$ is the set of all the faces of $  \cT_{n,\iota, \vec \tau,{\cal L} }$.

We will associate to each step $q$ in the contour walk of a tree (or of a tree with external edges, or of a tree with external and loop 
edges) a positive real parameter $\alpha_q$. 

\subsection{Interpolated Gaussian Measure}

Consider an abstract tree $T_n$ with $n$ vertices labeled $1, 2,\dots n$.
To every vertex $1,2,\dots n$ we associate $D$ matrices (one for each color) of size 
$N\times N$. We denote the matrices associated to the vertex $i$ by $\sigma^{(i)_1},\sigma^{(i)_2}, \dots \sigma^{(i)_D} $.
We associate to every edge of the tree $ (i,j)\in T_n $ a real variable $u^{ij}$. To every couple of vertices $k$ and $l$ we associate the function
\bea
    w^{kk}(T_n,u) = 1 \qquad w^{kl}(T_n,u) = \inf_{(i,j)\in { P }_{k\to l}(T_n) } u^{ij} ,
\eea
with $ {  P }_{k\to l}(T_n)  $ the unique path in the tree $T_n$ joining the vertices $k$ and $l$. We denote 
$\mu_{w^{ij}(T_n,u)  1^{\otimes D} } (\sigma  )   $
the normalized Gaussian measure of covariance
\bea
 && \int d \mu_{w^{ij}( T_n,u)  1^{\otimes D} } (\sigma ) \; \; \sigma^{(k)_c}_{ab}  (\sigma^{(l)_{c'} \dagger})_{b'a'} \crcr
 && =
 \int d \mu_{w^{ij}( T_n,u)  1^{\otimes D} } (\sigma ) \; \; \sigma^{(k)_c}_{ab} \bar \sigma^{(l)_{c'}}_{a'b'} =
   w^{kl}(   T_n,u   ) \;  \delta_{aa'} \delta_{bb'} \delta^{cc'}\; .
\eea
The existence and uniqueness of this measure follows form the positivity of the real symmetric matrix $w^{kk}(T_n,u)$
(see eg \cite{BKAR}).

\subsection{The Graph of D Permutations Over k Elements }

\begin{figure}[htb]
   \begin{center}
     \psfrag{1}{$\bar 1$}
   \psfrag{b1}{$ 1$}
   \psfrag{2}{$\bar 2 $}
   \psfrag{b2}{$ 2$}
   \psfrag{c}{$c$}
   \psfrag{bs1}{$\overline{\sigma_c(1)}$}
   \psfrag{s1}{$\sigma_c(1)$}
 \includegraphics[width=8cm]{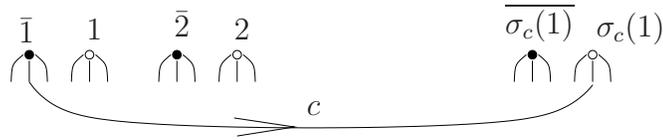}
 \caption{Graph associated to a $D$-tuple of permutations.}
  \label{fig:external}
   \end{center}
\end{figure}

Any $D$-tuple $\vec \sigma$ can be represented as a $D$ colored graph $\cB_{ \vec \sigma }$ with labelled vertices. We draw $k$ black vertices labeled
$\bar 1, \dots  \bar k$ and $k$ white vertices labeled $ 1, \dots  k$ and we connect the vertex $\bar l $ to the vertex $\sigma_c(l)$ by 
an edge of color $c$ oriented from $\bar l$ to $\sigma_c(l)$, see Figure \ref{fig:external}.
Conversely, to every colored graph $\cB$ with $2k$ vertices labelled $1,\dots k,\bar 1 ,\dots \bar k$ we associate a unique $D$-tuple of permutations
$\vec\sigma(\cB)$ encoding the connectivity of its edges.

\subsection{The Weingarten Function}

The Weingarten function introduced in \cite{collins,ColSni} arises naturally when one considers integrals over the unitary group,
namely 
\bea\label{eq:weing}
&& \int_{U(N)} [dU] \prod_{j=1}^k U_{n_j p_j} U^{\dagger}_{p'_jn'_j} = \crcr
&& \qquad =\sum_{\sigma,\tau} \delta_{n_1 n'_{\sigma(1)}}
  \dots \delta_{n_k n'_{\sigma(k)}} \delta_{p_1 p'_{\tau(1)}} \dots \delta_{p_k p'_{\tau(k)}} {\rm Wg}(N,\sigma\tau^{-1}) \; ,
\eea
where the sum runs over all the permutations $\sigma$ and $\tau$ of $k$ elements and the Weingarten function is
 \bea
{\rm Wg}(N,\tau) = \frac{1}{k!^2} \sum_{ \pi } \frac{ \chi^{\pi}(1)^2 \chi^{\pi} (\tau) }{s_{\pi,N}(1)} \; ,
\eea 
where the sum runs over the partitions $\pi$ of $N$, $\chi^{\pi}$ is the character of the symmetric group corresponding to $\pi$ and $  s_{\pi,N}(x) $ 
is the Schur function of the unitary group (hence $s_{\pi,N}(1)  $ is the dimension of the irreducible representation of $U(N)$ associated with $\pi$).
We will in particular use the following properties of the Weingarten function \cite{collins,ColSni}
\bea \label{eq:weingbound}
&& {\rm Wg}(N, (1)  ) = \frac{1}{N} \; ,\crcr
&& \lim_{N\to \infty} N^{2k-C(  \sigma ) } {\rm Wg}(N,\sigma  ) =
   \prod_{s=1}^{ C(\sigma) } (-1)^{| C_s(\sigma) |-1 }  \frac{1}{ |C_s(\sigma) |   } \binom{ 2 |C_s(\sigma) | -2  }{ |C_s(\sigma) | -1 } \; ,
   \eea
where $C(\sigma)$ denotes the number of cycles of the permutation $\sigma$ and $|C_s(\sigma)|$ denotes the length of the $s$'th cycle.
It follows that for $N$ large enough we have 
\bea
| {\rm Wg}(N,\sigma  ) | < \frac{1}{  N^{2k-C(  \sigma ) }   } 2^{2k} \; ,
\eea 
and we will always assume in this paper that $N$ is large enough such that this bound is respected.

\section{The Quartically Perturbed Gaussian Measure}\label{sec:quart}

\begin{figure}[htb]
   \begin{center}
 \includegraphics[width=2.5cm]{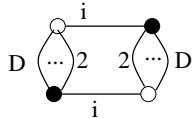}
 \caption{The graphs of the quartic perturbation terms.}
  \label{fig:quart}
   \end{center}
\end{figure}

Our starting point is the quartically perturbed Gaussian tensor measure 
\bea\label{eq:measure}
&& d\mu^{(4)} =  \frac{1}{Z(\lambda,N)}  \Big{(}\prod_{ \vec n } N^{D-1} \frac{d \mathbb{T}_{\vec n} d\bar {\mathbb{T} }_{ {\vec n } } } { 2 \pi \imath } \Big{)} \;
   e^{-N^{D-1}  S^{(4)}( \mathbb{T} ,\bar { \mathbb{T} } ) } \; , \\
&&  S^{(4)}( \mathbb{T},\bar { \mathbb{T} } ) =   \sum_{\vec n } \mathbb{T}_{\vec n } \delta_{ \vec n  \bar {\vec n}   } \bar { \mathbb{T} }_{ \bar {\vec n} }
    + \lambda \sum_{i=1}^D \sum_{n  \bar n  }
    \mathbb{T}_{\vec n }  \bar { \mathbb{T}}_{ \vec {\bar m} }
\mathbb{T}_{ \vec m }  \bar { \mathbb{T}}_{ \vec {\bar n} } \;  \delta_{n^i\bar m^i} \delta_{m^i \bar n^i}
\prod_{j\neq i}  \delta_{n^j \bar n^j} \delta_{m^j \bar m^j} \; , \nonumber
\eea 
with $Z(\lambda,N) $ some normalization constant. 
The quartic perturbation corresponds to a sum of  invariants whose graphs are represented in figure \ref{fig:quart}.
 
One can in principle consider more general quartic perturbations (in the associated graphs the vertices share $q$ 
and $D-q$ lines respectively with $q>1$) however such perturbations are suppressed in powers of $1/N$. 
The generating function of the moments of $\mu^{(4)} $ is
\bea
  Z(J,\bar J ;  \lambda,N) = \int  \Big{(}\prod_{  \vec n  } N^{D-1}\frac{d  {\mathbb T}_{\vec n}
d \bar  {\mathbb T}_{\vec n }  } { 2 \pi \imath } \Big{)} \;
  e^{-   N^{D-1}    S^{(4)}( \mathbb{T} ,\bar { \mathbb{T} } ) +  \sum_{\vec {\bar  n} } \bar {\mathbb T}_{ \vec {\bar n} } J_{ \vec {\bar n} }
  + \sum_{\vec {   n} }  {\mathbb T} _{  n  } \bar J_{ n   }    } \; ,
\eea
and the generating function of the cumulants is $ W(J,\bar J ;   \lambda ,N) = \ln Z(J ,\bar J; \lambda  ,N) $,
\bea\label{eq:cumudefconst}
 \kappa \bigl({\mathbb T}_{\vec n_1 }, \bar {\mathbb T}_{ \vec{\bar n  }_1} ,
  \dots {\mathbb T}_{ \vec n_k    } , \bar {\mathbb T}_{ \vec{ \bar n }_k } \bigr)
= \frac{\partial^{(2k)}} { \partial \bar J_{ \vec n_1  } \partial J_{  \vec{\bar n  }_1   }
    \dots  \partial \bar J_{    \vec n_k   }  \partial J_{  \vec{ \bar n }_k   }
   }
 W( J ,\bar J ; \lambda,N  ) \Big{\vert}_{J =\bar J =0} \; .
\eea
 
Our first result concerns the constructive expansion of $ W( J ,\bar J ; \lambda,N  ) $. 
\begin{theorem}[Constructive Expansion 1]\label{thm:constructive} The generating function of the cumulants of $\mu^{(4)}$ in eq.\eqref{eq:measure} 
writes as a sum over plane trees $ \cT_{n,\iota}^{\circlearrowright}   $ 
\bea \label{eq:WLVE}  
&&  W( J, \bar J; \lambda,N  ) =
   \sum_{n\ge 1} \frac{1}{n!} (-\lambda)^{n-1}
  \sum_{k=0}^n \frac{1}{k!} \frac{1}{  N^{(k+n-1)(D-1)} }
\sum_{ \stackrel{ i_{ 1} , i_{ 2} \dots , i_{ k} =1 }{i_d\neq i_{d'}} }^n  \sum_{  \cT^{\circlearrowright}_{n,\iota} } 
{\mathfrak T}(\cT^{\circlearrowright}_{n,\iota} ) \, ,
\eea
where the contribution of a tree is 
\bea 
&&  {\mathfrak T}(\cT^{\circlearrowright}_{n,\iota} ) =  \int_{0}^1 \Bigl( \prod_{ (i,j) \in T_n } du^{ij} \Bigr) 
 \int d\mu_{w^{ij}( T_n,u)  1^{ \otimes D}  } (\sigma )  \int \Big( \prod_{q=1}^{2n-2+k} d\alpha_q\Big)
 \; e^{-\sum_{q=1}^{2n-2+k} \alpha_q } \crcr
&&\;\;\times \prod_{f^c \in \cF^{ \rm{ int} }( \cT_{n,\iota}^{\circlearrowright} ) }
\Tr \Big[ \prod^{\rightarrow}_{ q \in q(f^c)} e^{ -\alpha_q \sqrt{\frac{\lambda}{N^{D-1}}} (\sigma^{i(q)_c} - \sigma^{i(q)_c\dagger}) } \Big]
\sum_{ \{ p^c_l,n^c_l \} } \prod_{l=1}^k  \bar J_{p^1_l,\dots p^D_l  }  J_{n^1_l,\dots n^D_l}
\crcr
&& \;\;\times \prod_{ f^{c;i(q_l) \to i( q_{ \xi_c(l)}  ) } \in \cS^{ \rm{ ext} }( \cT_{n,\iota}^{\circlearrowright} )  } 
\Bigl[ \prod^{\rightarrow}_{ q \in q( f^{c;i(q_l) \to i(q_{ \xi_c(l)} )  } ) }
e^{ -\alpha_q \sqrt{\frac{\lambda}{N^{D-1}}} (\sigma^{i(q)_c} - \sigma^{i(q)_c\dagger}) } 
\Bigr]_{p^c_l  n^c_{\xi_c(l)}}  
 \; , \nonumber
\eea 
where we used the notations of section \ref{sec:trees} and $ \rightarrow $ means that the products 
are ordered.
\end{theorem}

This expansion is the generalization to tensor models of the constructive Loop Vertex Expansion (LVE) introduced 
in \cite{LVE1} for matrix models. Note that our expansion looks somewhat different from the LVE of \cite{LVE1},
notably the step parameters $\alpha_p$ have no equivalent in the initial formulation. The usual LVE is 
recovered from \ref{thm:constructive} by restricting to trees having no ciliated vertices. In this case the 
integrals over $\alpha_p$ can be computed explicitly and one recovers the formulation in terms of resolvents of \cite{LVE1}.

As a consequence of the (LVE) we can derive an expansion for the cumulants of our measure.
\begin{theorem}[Constructive Expansion 2]\label{thm:constructive2} 
The cumulants of the measure $\mu^{(4)}$ in eq.\eqref{eq:measure} are  trace invariants
\bea
  \kappa \bigl({\mathbb T}_{\vec p_1 }, \bar {\mathbb T}_{ \vec { n  }_1 } ,
  \dots {\mathbb T}_{ \vec p_k    } , \bar {\mathbb T}_{ \vec{ n }_k } \bigr)
  = \sum_{ \cB, \; k(\cB)=k}  {\mathfrak K} (  \cB  ,\mu^{(4)}_N )  \prod_{\rho=1}^{ C(\cB )  } \delta^{\cB_{\rho} }_{n\bar n}  \; ,
\eea
where the sum runs over $D$-colored graphs with $2k$ vertices labelled $1,\dots k,\bar 1,\dots \bar k$ and
admit an expansion as a sum over plane trees with external edges $  \cT_{n,\iota,\vec \tau }$
\bea\label{eq:KLVE}
&& {\mathfrak K} (  \cB  ,\mu^{(4)}_N )  =   \sum_{n\ge k} \frac{1}{n!} (-\lambda)^{n-1}
   \frac{1}{  N^{(k+n-1)(D-1)} }
\sum_{ \stackrel{ i_{ 1} , i_{ 2} \dots , i_{ k} =1 }{i_d\neq i_{d'}} }^n \sum_{  \cT^{\circlearrowright}_{n,\iota} }  \sum_{ \vec \tau }  
{\mathfrak T}^{ {\mathfrak E}}(\cT^{\circlearrowright}_{n,\iota,\vec \tau}  ) \; ,
\eea
where the contribution of a tree with external edges is
\bea 
&& {\mathfrak T}^{\mathfrak{E}}(\cT^{\circlearrowright}_{n,\iota,\vec \tau}  ) =   k!  \Bigl( \prod_{c=1}^D {\rm Wg}\bigl(N,\tau_c \sigma_c^{-1}(\cB) \bigr) \Bigr) 
\int_{0}^1 \Bigl( \prod_{ (i,j) \in T_n } du^{ij} \Bigr)  
 \int d\mu_{w^{ij}( T_n,u)  1^{ \otimes D}  } (\sigma ) \crcr
 &&   \times \int \Big( \prod_{q=1}^{2n-2+k} d\alpha_q\Big)
  e^{-\sum_{q=1}^{2n-2+k} \alpha_q } \;     
\prod_{f^c \in \cF ( \cT_{n,\iota,\vec \tau}^{\circlearrowright} ) }
\Tr \Big[ \prod^{\rightarrow}_{ q \in q(f^c)} e^{ -\alpha_q \sqrt{\frac{\lambda}{N^{D-1}}} (\sigma^{i(q)_c} - \sigma^{i(q)_c\dagger}) } \Big] 
 \; . \nonumber
\eea
\end{theorem}

The main advantage of the constructive expansion of the cumulants is that, unlike the perturbative expansion in $\lambda$, it leads to 
a series which is absolutely convergent uniformly in $N$.
\begin{theorem}[Absolute Convergence]\label{thm:absconv}
The series in eq. \eqref{eq:KLVE} is absolutely convergent for $\lambda \in \mathbb{R}$, $\lambda \in[0, 2^{-3}  D^{-1})$. Moreover, 
the cumulants are bounded by 
\bea\label{eq:scalingbound}
 |{\mathfrak K} (  \cB  ,\mu^{(4)}_N )| \le   N^{D - 2 k(D-1)-C(\cB)} |\lambda|^{k-1} K(\cB)\;,
\eea 
for some constant $K(\cB)$ independent of $N$ (and independent of $\lambda$ for $|\lambda|$ small enough).
\end{theorem}
   The scaling bound of equation \eqref{eq:scalingbound} coincides with the proper uniform bound of definition \ref{def:unifbound}.
   However we can not yet conclude that $\mu^{(4)}$ is properly uniformly bounded. Indeed, in order to conclude this, one still needs to prove 
   that the second cumulant converges when $N\to \infty$. We will show this later in this paper. 
   
   The domain of convergence $0\le \lambda <2^{-3}  D^{-1}$ is optimal. We will see below that the leading order
   in $1/N$, the ``melonic'' sector yields a series whose radius of convergence is exactly $2^{-3}  D^{-1} $. It is 
   not surprising that the full non perturbative expression diverges when one reaches the radius of convergence of the leading order.
 
   The series\eqref{eq:KLVE} computes the cumulants for real, positive (and small) coupling constant. 
   The cumulants can be analytically continued to some domain in the complex plane.
   
\begin{corollary}\label{col:continuation}
The cumulants $ {\mathfrak K} (  \cB  ,\mu^{(4)}_N )$ can be analytically continued for $\lambda = |\lambda|e^{\imath \varphi}$ with 
$\varphi \in (-\pi,\pi)$ and $|\lambda|< \bigl(\cos\frac{\varphi}{2}\bigr)^2 2^{-3} D^{-1}$. In this domain they are
represented by the absolutely convergent series
\bea
&& {\mathfrak K} (  \cB  ,\mu^{(4)}_N )  =   \sum_{n\ge k} \frac{1}{n!} (-\lambda)^{n-1}
   \frac{1}{  N^{(k+n-1)(D-1)} }
\sum_{ \stackrel{ i_{ 1} , i_{ 2} \dots , i_{ k} =1 }{i_d\neq i_{d'}} }^n \sum_{  \cT^{\circlearrowright}_{n,\iota} }\sum_{ \vec \tau } 
{\mathfrak T}^{\mathfrak{E}}( \cT^{\circlearrowright}_{n,\iota,\vec \tau} ) \; , \\
&&
{\mathfrak T}^{\mathfrak{E}}( \cT^{\circlearrowright}_{n,\iota,\vec \tau} )  =  k!  \Bigl( \prod_{c=1}^D {\rm Wg}\bigl(N,\tau_c \sigma_c^{-1}(\cB) \bigr) \Bigr) 
\int_{0}^1 \Bigl( \prod_{ (i,j) \in T_n } du^{ij} \Bigr) 
 \int d\mu_{w^{ij}( T_n,u)  1^{ \otimes D}  } (\sigma ) \crcr
&& \qquad \qquad \times
 \int \Big( \prod_{q=1}^{2n-2+k} e^{-\imath \frac{\varphi}{2}}d\alpha_q\Big)
 e^{-\sum_{q=1}^{2n-2+k} e^{-\imath \frac{\varphi}{2}}  \alpha_q } \crcr
&& \qquad \qquad  \times \prod_{f^c \in \cF ( \cT_{n,\iota,\vec \tau}^{\circlearrowright} ) }
\Tr \Big[ \prod^{\rightarrow}_{ q \in q(f^c)} e^{ -\alpha_q \sqrt{\frac{|\lambda|}{N^{D-1}}} (\sigma^{i(q)_c} - \sigma^{i(q)_c\dagger}) } \Big] 
 \; . \nonumber
\eea 
\end{corollary}

\begin{figure}[htb]
   \begin{center}
   \psfrag{a}{$2^{-3}  D^{-1}$}
 \includegraphics[width=4cm]{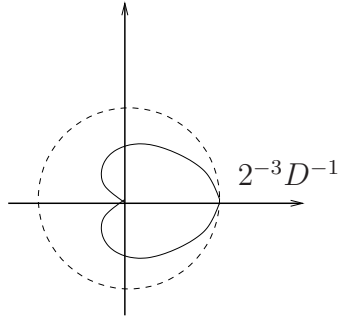}
 \caption{Domain of convergence.}
  \label{fig:domain}
   \end{center}
\end{figure}

We have thus a well defined expression for the cumulants in a heart-shaped domain (see Figure \ref{fig:domain} below where we represented 
by a dashed circle the circle of radius $2^{-3}D^{-1}$). As expected, $\lambda=0$ is a point belonging to the boundary
of this analyticity domain.

The interplay between the non perturbative LVE expansions presented so far and the $1/N$ expansion is captured by 
the following theorem.  Consider the rescaled cumulants, which according to eq.\eqref{eq:KLVE} write as
\bea\label{eq:cumulant1}
&& K(\cB, N) =  N^{ - D +2 k(D-1) + C(\cB)} {\mathfrak K} (  \cB  ,\mu^{(4)}_N ) \crcr
&& =   \sum_{n\ge k} \frac{1}{n!} (-\lambda)^{n-1} \sum_{ \stackrel{ i_{ 1} , i_{ 2} \dots , i_{ k} =1 }{i_d\neq i_{d'}} }^n
\sum_{  \cT^{\circlearrowright}_{n,\iota} } \sum_{\vec \tau}  T^{E}( \cT^{\circlearrowright}_{n,\iota,\vec\tau}  ) \; ,
\eea
where the rescaled contribution of each tree with external edges $\cT^{\circlearrowright}_{n,\iota,\vec \tau}  $
is
\bea
&&   T^{E}( \cT^{\circlearrowright}_{n,\iota,\vec\tau}  )  =   N^{ - D +2 k(D-1) + C(\cB) - (k+n-1)(D-1)  } \; k!
 \Bigl( \prod_{c=1}^D {\rm Wg}\bigl(N,\tau_c \sigma_c^{-1}(\cB) \bigr) \Bigr) \crcr
&& \qquad \qquad \times \int_{0}^1 \Bigl( \prod_{ (i,j) \in T_n } du^{ij} \Bigr)   \int d\mu_{w^{ij}( T_n,u)  1^{ \otimes D}  } (\sigma )  \crcr
&& \qquad \qquad  \times \int \Big( \prod_{q=1}^{2n-2+k} e^{-\imath \frac{\varphi}{2}}d\alpha_q\Big)
 \; e^{-\sum_{q=1}^{2n-2+k} e^{-\imath \frac{\varphi}{2}}  \alpha_q } \crcr
&& \qquad \qquad  \times \prod_{f^c \in \cF ( \cT_{n,\iota,\vec \tau}^{\circlearrowright} ) }
\Tr \Big[ \prod^{\rightarrow}_{ q \in q(f^c)} e^{ -\alpha_q \sqrt{\frac{|\lambda|}{N^{D-1}}} (\sigma^{i(q)_c} - \sigma^{i(q)_c\dagger}) } \Big] \; .
\eea 

\begin{theorem}[The Mixed Expansion]\label{thm:mixedexp}
 The contribution of a tree with external edges $ \cT^{\circlearrowright}_{n,\iota} $ admits an expansion in terms of trees 
 with external edges and loop edges $  \cT_{n,\iota,\vec \tau,{\cal L}}^{\circlearrowright} $  
 \bea
 &&  T^{E}( \cT^{\circlearrowright}_{n,\iota,\vec\tau}  )  
   = \sum_{q=0}^{s-1} T^{(q)}   (  \cT^{\circlearrowright}_{n,\iota,\vec\tau}   )
   + R^{(s)}  (  \cT^{\circlearrowright}_{n,\iota,\vec\tau}    )\;, \crcr
  &&  T^{(q)}  (  \cT^{\circlearrowright}_{n,\iota,\vec\tau}   ) = 
  \sum_{ {\cal L}, |{\cal L}| = q } \; \sum_{c_1\dots c_q }  T^{EL, (q) }( \cT^{\circlearrowright}_{n,\iota,\vec\tau,{\cal L}}  )  
  \eea  
 where ${\cal L}$ runs over all possible ways to decorate $ \cT_{n,\iota,\vec \tau}^{\circlearrowright}  $ with $q$ loop edges 
$(j_1,j_{1'})$, $(j_2,j_{2'})$ up to $(j_q,j_{q'})$ and $c_1,\dots c_q$ run over the possible colorings of the loop edges and
the contribution of a tree with external edges and loop edges is
\bea
 && T^{EL, (q) }( \cT^{\circlearrowright}_{n,\iota,\vec\tau,{\cal L}}  )  =
  \frac{1}{q!}  \Bigl( - \frac{ \lambda }{ N^{D-1} } \Bigr)^q  k!
  \Bigl( \prod_{c=1}^D {\rm Wg}\bigl(N,\tau_c \sigma_c^{-1}(\cB) \bigr) \Bigr)  \crcr
 &&  \qquad \qquad \times \int_{0}^1 \Bigl( \prod_{ (i,j) \in T_n } du^{ij} \Bigr) \prod_{p=1}^s w^{j_pj_{p'}}(T_n,u)
 \crcr
 && \qquad \qquad \times N^{ - D +2 k(D-1) + C(\cB) - (k+n-1)(D-1) +   | \cF ( \cT_{n,\iota,\vec \tau,{\cal L}}^{\circlearrowright} ) |    }  \; ,
\eea 
while the rest term is 
\bea 
&& R^{(s)} (  \cT^{\circlearrowright}_{n,\iota,\vec\tau}   ) = \int_{0}^1 dt (1-t)^{s-1} \Bigg[ 
  N^{ - D +2 k(D-1) + C(\cB) - (k+n-1)(D-1)  } \crcr
  && \qquad\qquad \times \;  \frac{1}{(s-1)!}  \Bigl( - \frac{ \lambda }{ N^{D-1} } \Bigr)^s  k!
 \Bigl( \prod_{c=1}^D {\rm Wg}\bigl(N,\tau_c \sigma_c^{-1}(\cB) \bigr) \Bigr)  
 \sum_{ {\cal L}, | {\cal L} | = s } \;\;\sum_{c_1\dots c_s }     \crcr
&& \qquad\qquad\times  
 \int_{0}^1 \Bigl( \prod_{ (i,j) \in T_n } du^{ij} \Bigr)  \prod_{p=1}^s w^{j_pj_p'}(T_n,u)  \int d\mu_{w^{ij}( T_n,u)  1^{ \otimes D}  } (\sigma )  \crcr
&& \qquad \qquad \times \int_0^{\infty} \Big( \prod_{q=1}^{2n-2+k+2s} e^{-\imath \frac{\varphi}{2}}d\alpha_q\Big)  \; e^{-\sum_{q=1}^{2n-2+k+2s} e^{-\imath \frac{\varphi}{2}}  \alpha_q }  
\crcr && 
\qquad\qquad \times
  \prod_{f^c \in \cF ( \cT_{n,\iota,\vec \tau,{\cal L}}^{\circlearrowright} ) }
\Tr \Big[ \prod^{\rightarrow}_{ q \in q(f^c)} e^{ -\alpha_q \sqrt{t}\sqrt{\frac{|\lambda|}{N^{D-1}}} (\sigma^{i(q)_c} - \sigma^{i(q)_c\dagger}) } \Big] 
 \Bigg] \; .
\eea 
Furthermore the terms in the mixed expansion admit the bounds
\bea
&& \big{|} T^{(q)}( \cT_{n,\iota,\vec \tau }^{\circlearrowright}) \big{|} \le \frac{|\lambda|^q}{N^{q(D-2)}} ( k! 2^{2Dk} D^q )
 \frac{ (2n+2q+k-3)! }{q! (2n+k-3)!} \\
&&  |R^{(s)} ( \cT_{n,\iota,\vec \tau }^{\circlearrowright}  ) |\le \frac{1}{\bigl( \cos \frac{\varphi}{2} \bigr)^{2n+2s+k-2}  }
 \frac{|\lambda|^s}{N^{s(D-2)}} ( k! 2^{2Dk} D^s) 
 \frac{ (2n+2s+k-3)! }{(s-1)! (2n+k-3)!} \; . \nonumber 
\eea 
\end{theorem}
We call this expansion the mixed expansion because it is at the same time an expansion in $\lambda$ {\it and} an expansion in $1/N$. 
More precisely, being an expansion in $\frac{\lambda}{N^{D-2}}$, one can use it to establish the 
Borel summability of the cumulants or, alternatively, one can use it to establish the $\frac{1}{N}$ expansion 
of the cumulants at all orders. 

\begin{theorem}[Borel Summability]\label{thm:borelsumability}
The rescaled cumulants 
\bea
K(\cB, N) \equiv  N^{ - D +2 k(D-1) + C(\cB)} {\mathfrak K} (  \cB  ,\mu^{(4)}_N ) \; ,
\eea
are Borel summable in $\lambda$ uniformly in $N$. 
\end{theorem}

A crucial point is that, as we are interested in the large $N$ regime, both the convergence of the constructive expansion 
in its analyticity domain and the Borel summability around $\lambda=0$ are uniform in $N$. 

\begin{theorem}[The $1/N$ expansion of the cumulants]\label{thm:1/N} 
Using the mixed expansion, the rescaled cumulants of $\mu^{(4)} $ write as 
\bea
 K(\cB, N) && =\sum_{n\ge k} \frac{1}{n!} (-\lambda)^{n-1} \sum_{ \stackrel{ i_{ 1} , i_{ 2} \dots , i_{ k} =1 }{i_d\neq i_{d'}} }^n
\sum_{  \cT^{\circlearrowright}_{n,\iota} } \sum_{\vec \tau}  \sum_{q=0}^{s-1} 
\sum_{ {\cal L}, |{\cal L}| = q } \; \sum_{c_1\dots c_q }  T^{EL, (q) }( \cT^{\circlearrowright}_{n,\iota,\vec\tau,{\cal L}}  ) \crcr 
&& + R^{(s)}_{N}(\cB,\lambda) \; ,
\eea 
and for $|\lambda|< 5^{-2} 2^{-1} D^{-1} \bigl( \cos \frac{\varphi}{2}\bigr)^2$ the rest term admits the bound
\bea\label{eq:rest1/N}
  |  R^{(s)}_{N}(\cB,\lambda) | \le  K  \sigma^s  s!  \frac{1}{N^{s(D-2)}}  \frac{|\lambda|^{s+k-1}}{\bigl( \cos \frac{\varphi}{2} \bigr)^{2s+3k-1}  }\; .
\eea 
for some constants $K$ and $\sigma$.
\end{theorem}

One can now use the $1/N$ expansion as follows: the terms up to order $N^{-s(D-2)}$ are indexed by trees with 
at most $s$ loop edges. Such corrections can be evaluated order by order. In particular all corrections at fixed order in $1/N$ will 
reach criticality when the sum over $n$ becomes critical, i.e. all terms in the $1/N$ expansion 
will diverge for {\it the same} critical constant. 

The factorial bound in eq.\eqref{eq:rest1/N} suggests that the cumulants are Borel summable in $1/N$. This is most likely the case, 
however the attentive reader will notice that we did not yet establish analyticity of the cumulants in $1/N$. This is difficult,
because, besides the explicit occurrences of $N$, we also must take into account that $N$ is the size of the matrices $\sigma^{(i)}_c$.
In order to establish analyticity in $1/N$ one needs to find a better representation of the cumulants in which $N$ appears exclusively as a 
parameter. We postpone this to future work.

Before we conclude this paper we present as an example the leading order behaviour of the two point cumulant. A number of simplifications 
arise in this case : $k=1$,  ${\rm Wg}\bigl(N,(1) \bigr) = \frac{1}{N}$\; , $C(\cB_{\vec{(1)} }) = 1$. Using the $1/N$ expansion 
up to order $s=1$ we get 
\bea
 && K(\cB^{(2)}, N)  =\sum_{n\ge k} \frac{1}{n!} (-\lambda)^{n-1} \sum_{ i_{ 1} =1   }^n
\sum_{  \cT^{\circlearrowright}_{n,\{i_1\} } }    
   T^{EL, (0) }( \cT^{\circlearrowright}_{n, \{i_1\}  ,\vec{(1) } , \emptyset }  ) + R^{(1)}_{N}(\cB^{(2)},\lambda) \; , \crcr
&&  T^{EL, (0) }( \cT^{\circlearrowright}_{n, \{i_1\}  ,\vec{(1) } , \emptyset }  ) =
\frac{1}{N^D}
N^{ - D +2  (D-1) + 1- (1+n-1)(D-1) +   | \cF ( \cT_{n,\iota,\vec{(1)},\emptyset}^{\circlearrowright} ) |    }  \; ,\crcr
&&  R^{(1)}_{N}(\cB^{(2)},\lambda) \le \frac{1}{N^{D-2}} K  \frac{|\lambda| }{\bigl( \cos \frac{\varphi}{2} \bigr)^{4}  } \; .
\eea 
The leading order behavior can be resummed, as 
\bea 
 | \cF ( \cT_{n,\iota,\vec{(1)},\emptyset}^{\circlearrowright} ) |  = D + (n-1)(D-1) \Rightarrow 
 T^{EL, (0) }( \cT^{\circlearrowright}_{n, \{i_1\}  ,\vec{(1) } , \emptyset }  ) =1 \; ,
\eea
and the sum over trees can be computed explicitly (we do this for arbitrary trees in equation \eqref{eq:counttrees} below),
\bea
 K(\cB^{(2)}, N)  =\sum_{n\ge k}  (-2D\lambda)^{n-1} \frac{(2n-2)!}{(n-1)! n!} 
 + \frac{1}{N^{D-2}} K  \frac{|\lambda| }{\bigl( \cos \frac{\varphi}{2} \bigr)^{4}  } \; .
\eea 
\begin{corollary}[The large $N$ limit] For $ \frac{|\lambda| }{\bigl( \cos \frac{\varphi}{2} \bigr)^{2}  } $ small enough
\bea
\lim_{N\to \infty} K(\cB^{(2)}, N) = \frac{-1+ \sqrt{1 + 8D\lambda}}{ 4D\lambda }   \; .
\eea  
\end{corollary}

In particular this coupled with the uniform bound in theorem \ref{thm:absconv} proves that the measure $\mu^{(4)}$ is properly
uniformly bounded, hence according to theorem \ref{thm:mare}, becomes Gaussian in the large $N$ limit.

\section*{Acknowledgments}

The author thanks Vincent Rivasseau for numerous discussions and for a detailed read of this manuscript.

\section{Proofs}\label{sec:proofs}

Before going to the core of the proofs of the various theorems in the text we establish a number of results we will use recurrently
in the sequel.

\subsection{Technical prerequisites}
 
\noindent{\bf The main inequality on permutations.} Recall that $C(\sigma)$ denotes the number of cycles of the 
permutation $\sigma$ and $C(\cB_{ \vec \sigma})$ denotes the number of 
connected components of the graphs associated to the $D$-tuple of permutations $\vec \sigma$. 

\begin{lemma}\label{lem:faces}
Let $\vec \xi $ and $\vec \sigma$ and $\vec \tau$ be three $D$-tuples of permutations over $k$ elements 
such that $\vec \xi$ are permutations encoding the external faces of a tree. We have the bound 
\bea
 \sum_c C(\tau_c \sigma_c^{-1}) -\sum_c C(\xi_c) + \sum_c C(\tau_c \xi_c ) \le (D+1) k - C(\cB_{ \vec \sigma}) \; .
\eea  
\end{lemma}

This bound is a trivial consequence of the following two propositions.

\begin{proposition}
   Let $\xi$ and $\sigma$ be any two fixed permutations over $k$ elements, and let $\tau$ be any permutation over $k$ elements. Then
\bea
      C(\tau \sigma^{-1}) + C(\tau\xi)   \le  C( \xi^{-1} \sigma^{-1} ) + k \; .
\eea 
\end{proposition}

\noindent{\bf Proof:} The bound is saturated if $\tau\xi $ is the identity permutation.

\begin{figure}[htb]
   \begin{center}
     \psfrag{1}{$\overline{  \xi (p) }$}
   \psfrag{bs1}{$\overline{ \tau (\xi(p))  }$}
   \psfrag{s1}{$  \tau( \xi( p) ) $}
   \psfrag{t-1}{$ \tau^{-1} (p) $}
   \psfrag{bt1}{$ p  $}
 \includegraphics[width=8cm]{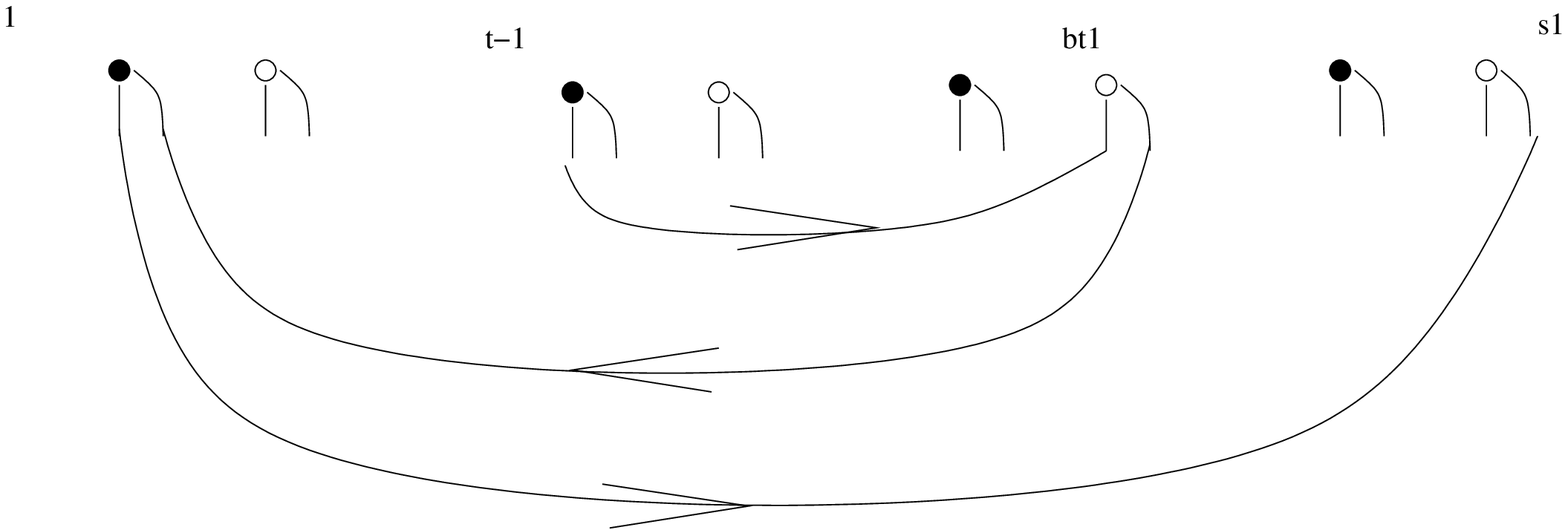}
 \caption{Graph associated to $\xi^{-1},\tau $.}
  \label{fig:external1}
   \end{center}
\end{figure}

Consider then the sum $ C(\tau \sigma^{-1}) + C(\tau\xi)  $ and suppose that $\tau\xi$ is not the identity permutation. 
The cycle structure of the permutation $\tau \xi$ can be easily read of by drawing the graph of the two permutations
$\xi^{-1},\tau$ (where we reverse the orientation of the edges representing the permutation $\xi^{-1}$), see Figure \ref{fig:external1}: 
a moment of reflection reveals that the number of cycles of $\tau\xi$ is the number 
of connected components of the graph as the  $\tau\xi$ jumps from a white vertex to the next white vertex following the 
arrows.

\begin{figure}[htb]
   \begin{center}
     \psfrag{1}{$\overline{  \xi (p) }$}
   \psfrag{bs1}{$\overline{ \tau (\xi(p))  }$}
   \psfrag{s1}{$  \tau( \xi( p) ) $}
   \psfrag{t-1}{$ \tau^{-1} (p) $}
   \psfrag{bt1}{$ p  $}
 \includegraphics[width=8cm]{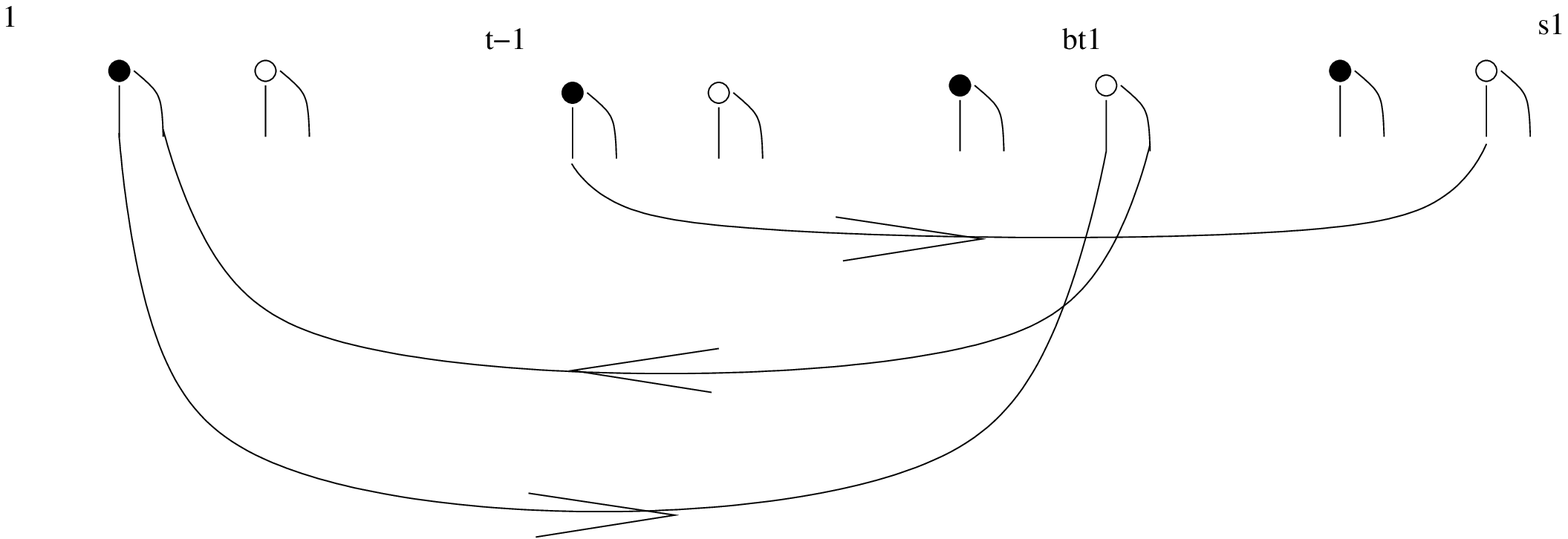}
 \caption{Graph associated to $\xi^{-1},\tau' $.}
  \label{fig:external2}
   \end{center}
\end{figure}

If $\tau\xi$ is not the identity, then there exists $p$ such that $\xi(p)\neq \tau^{-1}(p)$ (and $ \tau(\xi(p)) \neq p$). We compare
$   C(\tau \sigma^{-1}) + C(\tau\xi)  $ with $ C(\tau' \sigma^{-1}) + C(\tau'\xi) $, with $\tau'$ defined as
\bea
 \tau'(q) = \begin{cases}
              \tau(q) \quad & \forall q\neq \tau^{-1}(p), \xi(p) \\ 
              \tau'\bigl(\tau^{-1}(p) \bigr) = \tau\bigl( \xi(p) \bigr) \\ 
              \tau'\bigl( \xi(p) \bigr)  = p 
            \end{cases} \; .
\eea 
The graph of the permutations $\xi^{-1},\tau'$ is represented in Figure \ref{fig:external2}.
By substituting $\tau$ with $\tau'$ we created a new connected component, $  C(\tau'\xi) = C(\tau\xi) + 1 $. 
Representing now the graph of the permutations $\sigma, \tau$ and $\sigma, \tau'$, we see that replacing $\tau$ by $\tau'$
amounts to permuting the endpoints of the edges representing the $\tau$ permutation originating in $\tau^{-1}(p)$ and $\xi(p)$, hence
the number of cycles of $\tau\sigma^{-1}$ can not decrease by more than $1$. Thus 
\bea
    C(\tau \sigma^{-1}) + C(\tau\xi)  \le C(\tau' \sigma^{-1}) + C(\tau'\xi) \; .
\eea 
Iterating, we find that the maximum is achieved for $\tau$ such that $\tau\bigl(\xi(p) \bigr) = p$ for all $p$.

\qed

\begin{figure}[htb]
   \begin{center}
     \psfrag{l}{$l$}
   \psfrag{bl}{$\bar l$}
 \includegraphics[width=6cm]{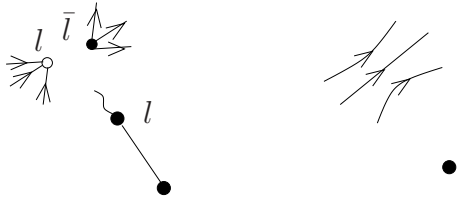}
 \caption{Ciliated vertex and the associated vertices of $\cB_{\vec \sigma}$.}
  \label{fig:lemlem}
   \end{center}
\end{figure}

\begin{proposition}  Let $\vec \xi $ and $\vec \sigma$ be two $D$-tuples of permutations (with $D\ge 2$) such that $\vec \xi$ are permutations 
encoding the external faces of a tree. Then 
\bea
\sum_c C( \sigma_c \xi_c  )   - \sum_c C(\xi_c) + C(\cB_{\sigma}) \le k .
\eea 
\end{proposition}
\noindent{\bf Proof:} For every cilium $l$ we draw a black  and white vertex 
and represent the graph $\cB_{\vec \sigma}$ associated to the permutation $\vec \sigma$, see Figure \ref{fig:lemlem}.

Consider the tree with associated permutations $\vec \xi$. As the univalent vertices with no cilia of the tree have 
no bearing over the permutations $\vec \xi$ we can eliminate them. Consider a ciliated univalent vertex $l$ in 
the tree, and say that the line touching it has color $c$. It follows that we are in one of the two cases
\bea
 \begin{cases}
  \text{case I } & \xi_c(l) \neq l\; , \qquad \xi_{c_1}(l) = l \quad \forall c_1\neq c , \\
  \text{case II } & \xi_c(l)= l \; ,\qquad \xi_{c_1}(l) = l \quad \forall c_1\neq c .
 \end{cases}
\eea 
The permutations $\vec \sigma$ fall in one of the following three categories
\bea
 \begin{cases}
    \text{case 1} & \sigma_c(l) \neq l \;, \qquad \forall c\; , \\
    \text{case 2} & \sigma_{c_i}(l) = l \;, i=1\dots q <D \; , \;\; \sigma_{c_j}(l) \neq l \qquad \{ c_i\} \cup \{c_j\} = \{1,\dots D\}\; , \\
    \text{case 3} & \sigma_c(l) = l \;, \qquad \forall c \; .
 \end{cases} 
\eea 
We build the graph obtained by eliminating the ciliated vertex $l$ in the tree (by cutting the tree line touching it) and 
deleting the vertices $l$ and $\bar l$ in $\cB_{\vec \sigma}$ and reconnecting the $\sigma$ lines respecting the colors (see Figure \ref{fig:lemlem} 
on the right).
The new graph is characterized by permutations $\vec \xi'$ and $\vec \sigma'$ over $\{ 1,\dots k\} \setminus \{l\}$ 
\bea
\begin{cases}
 \text{case I} & \xi'_c(p) = \xi_c(p) \; , \forall p\neq \xi_c^{-1}(l)\; , \qquad \xi'_c\bigl(  \xi^{-1}_c(l) \bigr) = \xi_c(l) \\
              &  \xi'_{c_1}(p) = \xi_{c_1}(p) \; , \forall p\neq l \\
 \text{case II} & \xi_{c}(p) = \xi_{c}(p) \; , \forall p\neq l \\
              & \xi'_{c_1}(p) = \xi_{c_1}(p) \; , \forall p\neq l \\
\end{cases} \; , \crcr
\begin{cases}
 \text{case 1} & \sigma'_c(p) = \sigma_c(p) \; , \forall p \neq \sigma^{-1}_c(l) \; , \qquad   \sigma'_c\bigl(\sigma^{-1}_c(l) \bigr) = \sigma_c(l) \\
 \text{case 2} & \sigma'_{c_j} (p) = \sigma_{c_j} (p) \; , \forall p \neq \sigma^{-1}_{c_j} (l) \; , \qquad   \sigma'_{c_j} \bigl(\sigma^{-1}_{c_j} (l) \bigr) 
                     = \sigma_{c_j}(l)  \\
               & \sigma'_{c_i}(p) = \sigma_{c_i}(p) \; , \forall p\neq l \\
 \text{case 3} &   \sigma'_{c }(p) = \sigma_{c }(p) \; , \forall p\neq l
\end{cases} \; .
\eea 
Consider now the graph of two permutations $\xi_c^{-1},\sigma_c$. The permutations $(\xi_c')^{-1},\sigma_c'$ correspond to the graph whose
vertices $l$ and $\bar l$ have been deleted and whose edges have been reconnected coherently. The number of cycles
changes as
\begin{itemize}
 \item if $\sigma_c(l)\neq l$, and $\xi_c(l) \neq l$ then $C(\sigma_c \xi_c  )\le C(\sigma_c' \xi_c'  ) +1 $,
 \item if $\sigma_c(l)\neq l$, and $\xi_c(l) = l$ then $C(\sigma_c \xi_c  ) =  C(\sigma_c' \xi_c')    $,
 \item if $\sigma_c(l)= l$, and $\xi_c(l) \neq l$ then $C(\sigma_c \xi_c  ) =  C(\sigma_c' \xi_c')     $,
 \item if $ \sigma_c(l) =l$ and $\xi_c(l) = l  $ then $C(\sigma_c \xi_c  ) =  C(\sigma_c' \xi_c')   +1  $.
\end{itemize}
Finally, the number of connected components of $\cB_{\vec \sigma}$ changes as
\begin{itemize}
 \item case 1, $\cB_{\vec \sigma}\le \cB_{ \vec \sigma'} +1$,
 \item case 2, $ \cB_{\vec \sigma}=\cB_{ \vec \sigma'} $,
 \item case 3, $\cB_{\vec \sigma}= \cB_{ \vec \sigma'} +1$.
\end{itemize}

As a function of the case we are into, we then have
 \bea
 I1 :&&  \sum_c C( \sigma_c \xi_c ) \le \sum_c C(\sigma_c' \xi_c'  ) +1 \; , \crcr
   &&     \quad  \sum_c C(\xi_c) = \sum_c C(\xi'_c) +D-1 \; , \quad  C(\cB_{\vec \sigma}) \le  C(\cB_{\vec \sigma'}) +1 \;, \crcr
 II1: && \sum_c C(\sigma_c \xi_c ) = \sum_c C(\sigma_c' \xi_c'  )  \; , \crcr
    &&  \quad  \sum_c C(\xi_c) = \sum_c C(\xi'_c) +D \; , \quad  C(\cB_{\vec \sigma}) \le  C(\cB_{\vec \sigma'}) +1 \;, \crcr
I2: && \sum_c C( \sigma_c \xi_c ) \le \sum_c C(\sigma_c' \xi_c'  ) +q+ 1 \; , \crcr
    && \quad  \sum_c C(\xi_c) = \sum_c C(\xi'_c) +D-1 \; , \quad  C(\cB_{\vec \sigma}) =  C(\cB_{\vec \sigma'})   \;, \crcr
II2: && \sum_c C( \sigma_c \xi_c ) = \sum_c C(\sigma_c' \xi_c'  ) +q \; , \crcr
    && \quad  \sum_c C(\xi_c) = \sum_c C(\xi'_c) +D  \; , \quad  C(\cB_{\vec \sigma}) =  C(\cB_{\vec \sigma'})   \;, \crcr
I3: && \sum_c C(\sigma_c \xi_c) = \sum_c C(\sigma_c' \xi_c'  )) +D -1 \; , \crcr
    && \quad  \sum_c C(\xi_c) = \sum_c C(\xi'_c) +D-1 \; , \quad  C(\cB_{\vec \sigma}) =  C(\cB_{\vec \sigma'}) +1 \;, \crcr
II3:&& \sum_c C( \sigma_c \xi_c ) = \sum_c C(\sigma_c' \xi_c'  ) + D  \; , \crcr
    &&  \quad  \sum_c C(\xi_c) = \sum_c C(\xi'_c) +D \; , \quad  C(\cB_{\vec \sigma}) =  C(\cB_{\vec \sigma'}) +1 \;. 
   \eea 
In all cases 
\bea
 \sum_c C( \xi_c \sigma_c^{-1} )   - \sum_c C(\xi_c) + C(\cB_{\sigma}) \le 
 \sum_c C( \xi'_c (\sigma'_c) ^{-1} )   - \sum_c C(\xi'_c) + C(\cB_{\vec\sigma'}) +1 \;,
\eea 
and the bound can be attained only for $I3$, $II3$, $I2$ with $q=D-1$, or $I1$ if $D=2$. We iterate the procedure,
taking into account that after eliminating the univalent vertex $l$ of the tree one might need to eliminate some new univalent 
vertices with no cilia (as the latter have no bearing on the permutations $\vec \xi'$). Iterating up to permutations of 
$1$ element $\xi_c = \sigma_c = (1)\;  \forall c$, we get 
\bea
 \sum_c C( \xi_c \sigma_c^{-1} )   - \sum_c C(\xi_c) + C(\cB_{\sigma}) \le  (k-1) + 1 \; .
\eea 
\qed

\bigskip

\noindent{\bf Evaluating derivatives.} We will repeatedly use in the sequel the following result.

\begin{lemma}\label{lem:dervi} For any $N\times N$ matrices $F $ and $G $ and any function $H(\alpha_1,\alpha_2)$ we have
\bea
&& \int_{0}^{\infty} d\alpha_1 d\alpha_2 \; H(\alpha_1,\alpha_2) \sum_{p_1n_1p_2n_2} F_{n_1p_1} G_{n_2p_2}\crcr
 && \qquad  \times \sum_{ab} \frac{\partial}{\partial \sigma_{  a b }^{(i)_c} } 
 \Bigl[e^{ -\alpha_1  \sqrt{\frac{ \lambda }{N^{D-1}} } (\sigma^{(i)_c}- \sigma^{(i)_c\dagger})   }  \Bigr]_{ p_1n_1} \crcr
&& \qquad \times \frac{\partial}{\partial \sigma_{  b a }^{(j)_c  \dagger }  } 
 \Bigl[e^{ -\alpha_2  \sqrt{\frac{ \lambda }{N^{D-1}} } (\sigma^{(j)_c}- \sigma^{(j)_c\dagger})   }  \Bigr]_{ p_2n_2 } \crcr
 && = - \frac{\lambda}{N^{D-1}} \int_{0}^{\infty} d\beta_1 d\gamma_1 d\beta_2 d\gamma_2 \; H( \beta_1 + \gamma_1,\beta_2+\gamma_2 ) \crcr
&& \qquad  \times \Tr\Bigg[ F e^{-\beta_1\sqrt{\frac{ \lambda }{N^{D-1}} }(\sigma^{(i)_c}- \sigma^{(i)_c\dagger})} 
  e^{-\gamma_2\sqrt{\frac{ \lambda }{N^{D-1}} }(\sigma^{(j)_c}- \sigma^{(j)_c\dagger})} \crcr
  && \qquad \qquad \qquad  G 
   e^{-\beta_2\sqrt{\frac{ \lambda }{N^{D-1}} }(\sigma^{(j)_c}- \sigma^{(j)_c\dagger})} 
   e^{-\gamma_1\sqrt{\frac{ \lambda }{N^{D-1}} }(\sigma^{(i)_c}- \sigma^{(i)_c\dagger})} 
  \Bigg]  \; .
\eea 
\end{lemma}

\noindent{\bf Proof:} Expanding the exponentials and evaluating the derivatives the integral writes 
\bea
&& =\int_{0}^{\infty} d\alpha_1 d\alpha_2  \; H(\alpha_1,\alpha_2) \sum_{p_1n_1p_2n_2} F_{n_1p_1} G_{n_2p_2}\\
 && \qquad \sum_{ab}
  \Bigg[\sum_{u_1,v_1\ge 0} \frac{(-\alpha_1)^{u_1+v_1+1}}{(u_1+v_1+1)!} \sqrt{\frac{ \lambda }{N^{D-1}} }^{u_1+v_1+1} \crcr
&& \qquad \qquad  \Bigl[  (\sigma^{(i)_c}- \sigma^{(i)_c\dagger})^{u_1}  \Bigr]_{ p_1a }  \Bigl[ (\sigma^{(i)_c}- \sigma^{(i)_c\dagger})^{ v_1}  \Bigr]_{ bn_1 }  \Bigg]
  \crcr && \qquad  
    \Bigg[\sum_{u_2,v_2\ge 0} \frac{(-\alpha_2)^{u_2+v_2+1}}{(u_2+v_2+1)!} (-1) \sqrt{\frac{ \lambda }{N^{D-1}} }^{u_2+v_2+1} \crcr
&& \qquad \qquad  
   \Bigl[  (\sigma^{(j)_c}- \sigma^{(j)_c\dagger})^{u_2}  \Bigr]_{ p_2b }  
  \Bigl[ (\sigma^{(j)_c}- \sigma^{(j)_c\dagger})^{v_2}  \Bigr]_{ an_2 }  \Bigg] \; . \nonumber
 \eea 
Using $\frac{1}{(u_1+v_1+1)!} = \int_{0}^1 dx_1   \frac{(1-x_1)^{u_1}}{u_1!} \frac{ x_1^{v_1}}{v_1!} $ this rewrites as
\bea
 && - \frac{\lambda}{N^{D-1}}
 \int_{0}^{\infty} d\alpha_1 d\alpha_2  \;  H(\alpha_1,\alpha_2)  \int_{0}^1 \alpha_1 dx_1 \alpha_2 dx_2 
 \sum_{p_1n_1p_2n_2} F _{n_1p_1} G _{n_2p_2}\crcr
 && \qquad \sum_{u_1,v_1,u_2,v_2\ge 0} \frac{ [ -\alpha_1 (1-x_1)]^{u_1}}{u_1!} \frac{ [ -\alpha_1 x_1 ]^{v_1}}{v_1!}
  \frac{ [ -\alpha_2 (1-x_2)]^{u_2}}{u_2!} \frac{ [ -\alpha_2 x_2 ]^{v_2}}{v_2!} \crcr
  && \qquad \sqrt{\frac{ \lambda }{N^{D-1}} }^{u_1+v_1+u_2+v_2} 
  \Bigl[ (\sigma^{(i)_c}- \sigma^{(i)_c\dagger})^{u_1} (\sigma^{(j)_c}- \sigma^{(j)_c\dagger})^{v_2}  \Bigr]_{p_1n_2} \crcr
&&  \qquad 
\Bigl[  (\sigma^{(j)_c}- \sigma^{(j)_c\dagger})^{u_2}   (\sigma^{(i)_c}- \sigma^{(i)_c\dagger})^{ v_1}     \Bigr]_{p_2n_1} \; ,
\eea 
and changing variables to $\beta_1 = \alpha_1 (1-x_1) , \gamma_1 = \alpha_1 x_1$ and similarly for $2$ and summing
over $u_1,v_1,u_2,v_2$ the lemma follows
\qed

\bigskip

\noindent{\bf Combinatorial countings.} We count the number of plane trees with $n$ vertices, $k$ ciliated vertices $i_1,\dots i_k$
and colored oriented edges.
Every combinatorial tree $T_n$ with degrees of the vertices $d_1,\dots d_n$,
has $ (2D)^{n-1} d_{i_1}! \dots d_{i_k}! \prod_{i\neq i_k} (d_{i}-1)!$ associated plane trees with colored oriented edges and marked vertices
$\cT_{n,\iota}^{\circlearrowright} $, corresponding to the two possible orientations of the edges, the $D$ possible colorings 
of every edge and the permutations of all but one of the halfedges touching each vertex (plus a choice $d_{i_r}$ of where to place
the cilium on the marked vertices). The number
of combinatorial trees with assigned degrees $d_1, \dots d_n$ is $ \frac{(n-2)!}{(d_1-1)! \dots (d_n-1)!}$ and we get
\bea\label{eq:counttrees}
&& \Bigl( \sum_{ \cT^{\circlearrowright}_{n,\iota}  } 1\Bigr)_{n,\iota \text{ fixed }} = \crcr
&& = (2D)^{n-1} \sum_{ \stackrel{d_1,\dots d_n =1}{ \sum d_i = 2n-2}}^n \frac{(n-2)!}{(d_1-1)! \dots (d_n-1)!  }
  d_{i_1}! \dots d_{i_k}! \prod_{i\neq i_k} (d_{i_k}-1)! \crcr
&&=  (2D)^{n-1}  (n-2)!  \sum_{ \stackrel{d_1,\dots d_n =1}{ \sum d_i = 2n-2}}^n d_{i_1} \dots d_{i_k}
  = (2D)^{n-1} \frac{(2n+k-3)!}{(n+k-1)!} \; ,
\eea
as the sums over $d_i$ yield the coefficient of the term of degree $x^{2n-2}$ in the expansion of
\bea
 \Bigr[ x \bigl(\frac{1}{1-x} \bigr)' \Bigl]^k  \frac{x^{n-k}}{(1-x)^{n-k}} = \frac{x^n}{(1-x)^{n+k}} = x^n \sum_{p} \binom{n+k+p-1}{p} x^p \; .
\eea

For every tree $\cT_{n,\iota}^{\circlearrowright} $ one has $ k!^D  $ trees with external edges $ \cT_{n,\iota,\vec \tau}^{\circlearrowright}$,
\bea
 \Bigl( \sum_{\vec \tau} 1\Bigr)\Bigl{|}_{ \cT_{n,\iota}^{\circlearrowright} \text{ fixed } } = k!^D \; .
\eea 
The number of plane trees decorated by loop edges is counted as follows. The walk around $\cT_{n,\iota}^{\circlearrowright} $ has $2n-2+k$ steps.
The insertion of $s$ loop edges consists in the choice of $2s$ positions to insert new cilia. One can insert a new cilium at every step of the walk. 
However, with every insertion of a cilium the walk acquires a new step, hence one has
\bea
[2(n-1)+k] [2(n-1)+k+1] \dots [2(n-1)+k+2s-1] = \frac{(2n+2s+k-3)!}{(2n+k-3)!}  \; ,
\eea
ways to connect the $2s$ loop edges on $\cT_{n,\iota}^{\circlearrowright} $. Furthermore for each choice, one has
$D^s$ possible colorings of the loop edges, hence 
\bea\label{eq:countloops}
 \Bigl( \sum_{ {\cal L} , |{\cal L} | =s} \sum_{c_1,\dots c_s} 1 \Bigr) 
 \Bigl{|}_{  \cT_{n,\iota,\vec \tau} \text{ fixed }  }  = D^s \frac{(2n+2s+k-3)!}{(2n+k-3)!} \; .
\eea 

Finally,
\bea\label{eq:n}
 \frac{1}{n!} \sum_{ \stackrel{ i_{ 1} , i_{ 2} \dots , i_{ k} =1 }{i_d\neq i_{d'}} }^n 1 = \frac{1}{(n-k)!} \; .
\eea 

\bigskip

\noindent{\bf Computing the logarithm of a Gaussian integral.} The logarithm of a Gaussian integral can be computed using
the universal Brydges-Kennedy-Abdesselam-Rivasseau forest formula \cite{BKAR} and a replica trick.

\begin{lemma}\label{lem:log} Let $X$ be a complex vector of components $X_1,\dots X_N$  and let $I$ be the Gaussian integral of covariance $C$
defined as
\bea
 I =  \int d\mu_C (X) \; e^{V(\bar X,X)} \;, \qquad \int  d\mu_C (X) \; X_a \bar X_{\bar b} = C_{a \bar b} \; .
\eea 
Then
 \bea 
 && \ln I =
 \sum_{n\ge 0} \frac{1}{n!} \sum_{ T_n }
 \int_{0}^1 \Bigl( \prod_{ (i,j) \in T_n } du^{ij} \Bigr) \;
 \int d\mu_{w^{ij}(T_n,u) C }  (X^{(i)} ) \crcr
&& \qquad \times \Bigg[  \prod_{ (i,j)\in T_n } \Bigg( \sum_{a\bar b}
\frac{\partial}{\partial X_a^{(i)} } C_{ a \bar b}
\frac{\partial}{\partial \bar X_{\bar b}^{(j)}  } + \sum_{a\bar b}
\frac{\partial}{\partial X_a^{(j)} } C_{ a \bar b}
\frac{\partial}{\partial \bar X_{\bar b}^{(i)}  }
    \Bigg) \Bigg] \crcr
&& \qquad \times \prod_{i=1}^n V(\bar X^{(i)},X^{(i)} ) \; ,
\eea 
where $T_n$ runs over all the combinatorial trees with $n$ vertices labelled $1$, $2$ up to $n$, $(i,j)$ denotes the 
tree edge connecting the vertices $i$ and $j$, the parameters $w^{ij}(T_n,u)$ are defined as
\bea
 w^{ii}(T_n,u) = 1 \; , \qquad  w^{ij}(T_n,u) = \inf_{(k,l)\in {  P }_{i\to j}(T_n) } u^{kl} , 
\eea 
where $ { P }_{i\to j}(T_n) $ denotes the unique path in the tree $T_n$ joining $i$ and $j$ and 
the interpolated Gaussian measure $ d\mu_{w^{ij}(T_n,u) C }  (X^{(i)} ) $ is  
\bea
\int  d\mu_{w^{ij}(T_n,u) C }  (X^{(i)} ) \; X^{(i)}_a \bar X^{(j)}_{\bar b} = w^{ij}(T_n,u) C_{a\bar b} \; .
\eea 
\end{lemma}

\noindent{\bf Proof:} We Taylor expand in $V(X,\bar X)$ to get
\bea
  I =  \int d\mu_C (X) \; e^{V(\bar X,X)} = \int d\mu_C (X)  \; \sum_{n\ge 0} \frac{1}{n!} \; V(\bar X,X)^n \; .
\eea
The term of degree $n$ can be rewritten as a Gaussian integral over $n$ replicas $X^{(1)}$, $X^{(2)} $ up to $X^{(n)}$
with degenerate covariance between the replicas $ C^{(i,j)}_{a \bar b} = C_{a \bar b} $, 
\bea
 I =  \sum_{n\ge 0}  \frac{1}{n!} \int d\mu_{C^{(i,j)}_{a\bar b} }(X^{(i)} )
 \;\; \prod_{i=1}^n V(\bar X^{(i)},X^{(i)} ) \; .
\eea
Each term in this expansion is a function of parameters $x^{ij}=x^{ji}$, evaluated for $x^{ij}=1$, corresponding 
to a Gaussian measure with covariance
\bea
C^{(i,i)}_{a\bar b} = C_{a\bar b}\;, \qquad C^{(i,j)}_{a \bar b} = x^{ij} C_{a\bar b} \;,\; i\neq j \; .
\eea 
 
Consider $n$ vertices labeled $1,2 \dots n$ and a function $f$ depending on $\frac{n(n-1)}{2}$ edge variables $x^{ij}$ with $i\neq j$. 
The universal Brydges-Kennedy-Abdesselam-Rivasseau forest formula \cite{BKAR} states that 
\bea
&& f(1,\dots 1) = \sum_{F_n} \int_{0}^1 \Bigl( \prod_{ (i,j) \in F_n } du^{ij} \Bigr) \;
 \Bigg( \frac{\partial^{ |\cE(F_n)| } f }{     \prod_{ (i,j) \in F_n } \partial  x^{ij}}  \Bigg)\Bigg{|}_{ x^{kl} = w^{kl} (F_n,u)} \; ,\crcr
&& w^{kl}\bigl(F_n,u \bigr)  =  \inf_{(i,j)  \in {\cal P}_{k\to l} (\cF_n ) }  u^{ij} \; ,
\eea
where $F_n$ runs over all the forests built over the $n$ sites, $|\cE(F_n)|$ denotes the number of edges in the forest,
${\cal P}_{k\to l}(F_n)$ is the unique path in $F_n$ joining the vertices $k$ and $l$, and the infimum is set to zero
if $k$ and $l$ do not belong to the same tree in the forest. In order to apply the BKAR formula to the term of degree $n$ 
we evaluate  
\bea
&& \frac{\partial^{ |\cE(F_n)| }   }{     \prod_{ (i,j) \in \cF_n } \partial  x^{ij}} \Bigl[
 \int d\mu_{x^{ij}C_{a\bar b} }(X^{(i)})
 \;\; \prod_{i=1}^n V(\bar X^{(i)},X^{(i)} ) \Bigr] \crcr
&&= \frac{\partial^{ |\cE (F_n) | }   }{     \prod_{ (i,j) \in \cF_n } \partial  x^{ij}} \Bigl[
e^{    \sum_{a,b,i}  \frac{\partial}{\partial X_a^{(i)} } C_{ a \bar b}
\frac{\partial}{\partial \bar X_{\bar b}^{(i)} } +
 \sum_{a,b,i\neq j} x^{ij} \frac{\partial}{\partial X_a^{(i)} } C_{ a \bar b}
\frac{\partial}{\partial \bar X_{\bar b}^{(j)} }    } \crcr
&& \qquad \qquad \qquad \qquad \qquad \prod_{i=1}^n  V(\bar X^{(i)},X^{(i)} )  \Bigr{]} \Bigg{|}_{X^{(i)}=\bar X^{(i)} =0} \crcr
&& =
 e^{   \sum_{a,b,i}  \frac{\partial}{\partial X_a^{(i)} } C_{ a \bar b}
\frac{\partial}{\partial \bar X_{\bar b}^{(i)} } +
 \sum_{a,b,i\neq j} x^{ij} \frac{\partial}{\partial X_a^{(i)} } C_{ a \bar b}
\frac{\partial}{\partial \bar X_{\bar b}^{(j)} }   }  \crcr
&& \qquad \times \Bigg[  \prod_{ (i,j)\in F_n }  \Bigg(  \sum_{a\bar b}
\frac{\partial}{\partial X_a^{(i)} } C_{ a \bar b}
\frac{\partial}{\partial \bar X_{\bar b}^{(j)}  } + \sum_{a\bar b}   \frac{\partial}{\partial X_a^{(j)} } C_{ a \bar b}
\frac{\partial}{\partial \bar X_{\bar b}^{(i)}  }
  \Bigg) \Bigg] \crcr
&& \qquad \qquad \qquad \qquad \qquad \prod_{i=1}^n V(\bar X^{(i)},X^{(i)} )  \Bigg{|}_{X^{(i)},\bar X^{(i)} =0} \crcr
&& =  \int d\mu_{x^{ij}C_{a\bar b} }(X^{(i)} )  \Bigg[ \prod_{ (i,j)\in \cF_n }   \Bigl(  \sum_{a\bar b}
\frac{\partial}{\partial X_a^{(i)} } C_{ a \bar b}
\frac{\partial}{\partial \bar X_{\bar b}^{(j)}  } +  \sum_{a\bar b} \frac{\partial}{\partial X_a^{(j)} } C_{ a \bar b}
\frac{\partial}{\partial \bar X_{\bar b}^{(i)}  }
 \Bigr) \Bigg] \crcr
&& \qquad \qquad \qquad \qquad \qquad \prod_{i=1}^n V(\bar X^{(i)},X^{(i)} ) \; ,
\eea
where we have taken into account that $x^{ij} = x^{ji}$. Thus 
\bea
&& I = \sum_{n\ge 0} \frac{1}{n!} \sum_{ F_n }
 \int_{0}^1 \Bigl( \prod_{ (i,j) \in  F_n } du^{ij} \Bigr) \;
 \int d\mu_{w^{ij}( F_n,u) C_{a\bar b} }  (X^{(i)} ) \crcr
&& \qquad \times \Bigg[  \prod_{ (i,j)\in F_n } \Bigg( \sum_{a\bar b}
\frac{\partial}{\partial X_a^{(i)} } C_{ a \bar b}
\frac{\partial}{\partial \bar X_{\bar b}^{(j)}  } + \sum_{a\bar b}
\frac{\partial}{\partial X_a^{(j)} } C_{ a \bar b}
\frac{\partial}{\partial \bar X_{\bar b}^{(i)}  }
    \Bigg) \Bigg] \crcr
&& \qquad \qquad \qquad \qquad \qquad  \prod_{i=1}^n V(\bar X^{(i)},X^{(i)} ) \; .
\eea
The lemma follows by noticing that the Gaussian integral factors over the trees in the forest and 
recalling that the logarithm of a function which is a sum over forests of contributions factored over the 
trees is the sum over trees of the tree contribution.

\qed
 
The most important feature of the BKAR
formula is that the matrix $ w^{ij}(T_n,u) $ is {\it positive} \cite{BKAR}. The 
Gaussian measure is thus well defined and the expectation of any function of 
$X,\bar X$ is bounded by its supremum.

\subsection{Proofs of the Theorems}

In the reminder of this section we present the proofs of the theorems enunciated in the text.

\subsubsection{Proof of the first constructive expansion theorem \ref{thm:constructive}}

The Loop Vertex Expansion of $W(J,\bar J; \lambda,N)$ in eq.\eqref{eq:WLVE}
is obtained by combining the Hubbard Stratonovich intermediate field representation and 
the BKAR formula. We will sometimes drop the bar over the indices of $\bar {\mathbb{ T} } $.

\bigskip

\noindent{\it Step 1: Hubbard Stratonovich intermediate field representation}.
For any complex numbers $Z_1$ and $Z_2$, $e^{-Z_1Z_2}$ can be represented as a Gaussian
integral
\bea
 && \int \frac{d\bar z dz}{2\imath \pi} e^{-z\bar z - z Z_1 + \bar z Z_2}
  \xlongequal[\bar z = x-\imath y]{z=x+ \imath y} 
  \int \frac{dxdy}{\pi} e^{-x^2-y^2 - x(Z_1-Z_2) -\imath y (Z_1+Z_2)} \crcr
 && \qquad = e^{\frac{(Z_1-Z_2)^2}{4} - \frac{(Z_1+Z_2)^2}{4}} = e^{-Z_1Z_2} \; .
\eea
It follows that a term in the quartic perturbation of the measure in eq.\eqref{eq:measure}
can be represented using $N\times N$ integration variables $\sigma^c_{ab}$ as
\bea
 && e^{  -  N^{D-1}   \lambda    \sum
    \mathbb{T}_{\vec n }  \bar { \mathbb{T}}_{ \vec {\bar m} }
\mathbb{T}_{ \vec m }  \bar { \mathbb{T}}_{ \vec {\bar n} } \;  \delta_{n^c\bar m^c} \delta_{m^c \bar n^c}
\prod_{c'\neq c}  \delta_{n^{c'} \bar n^{c'}} \delta_{m^{c'} \bar m^{c'}}  } \crcr
&& =  e^{  -  N^{D-1}   \lambda   \sum  \delta_{n^c\bar m^c} \delta_{m^c \bar n^c}
 \bigl(\sum   \mathbb{T}_{\vec n }   \bar { \mathbb{T}}_{ \vec {\bar n} } \prod_{c'\neq c}  \delta_{n^{c'} \bar n^{c'}} \bigr)
\bigl( \sum   \bar { \mathbb{T}}_{ \vec {\bar m} } \mathbb{T}_{ \vec m } \prod_{c'\neq c}  \delta_{m^{c'} \bar m^{c'}}   \bigr)}
 \crcr
&&  \qquad
  = \int  \Big{(}\prod_{   a b } \frac{d  \sigma^c_{ab }
d \bar { \sigma   }^c_{ ab  } } { 2 \pi \imath }  \Big{)} e^{ - \sum_{ ab} \sigma^c_{a b } \bar \sigma^c_{ ab } } \crcr
&& e^{- \sqrt{  \lambda }  N^{\frac{D-1}{2}}
 \sum  \bigl( \mathbb{T}_{\vec n }   \bar { \mathbb{T}}_{ \vec {\bar n} } \prod_{c'\neq c}  \delta_{n^{c'} \bar n^{c'}} \bigl) \sigma^c_{  n^c \bar n^c }
 +   \sqrt{  \lambda }  N^{\frac{D-1}{2}} \sum \bigl(  \bar { \mathbb{T}}_{ \vec {\bar m} } \mathbb{T}_{ \vec m } \prod_{c'\neq c}  \delta_{m^{c'} \bar m^{c'}} \bigl)
       \bar \sigma^c_{ \bar m^c m^c } }   \;.
\eea
The new integration variables $\sigma^c_{ab }$ form $D$  matrices of size $N\times N$, known as intermediate
fields. Denoting $1$ the identity matrix of size $N \times N $ we write more compactly
\bea
&&  e^{  -  N^{D-1} \lambda \sum_{c=1}^D \sum_{n  \bar n  }
    \mathbb{T}_{\vec n }  \bar { \mathbb{T}}_{ \vec {\bar m} }
\mathbb{T}_{ \vec m }  \bar { \mathbb{T}}_{ \vec {\bar n} } \;  \delta_{n^c\bar m^c} \delta_{m^c \bar n^c}
\prod_{c'\neq c}  \delta_{n^{c'} \bar n^{c'} } \delta_{m^{c'} \bar m^{c'} }   } \\
 && \quad = \int  \Big{(}\prod_{ c ab } \frac{d  \sigma^c_{ ab}
d \bar { \sigma   }^c_{ ab} } { 2 \pi \imath }  \Big{)} \;
  e^{ - \tr ( \sigma^c \sigma^{c\dagger} )- N^{D-1}  \sqrt{\frac{ \lambda }{N^{D-1}} }   \sum
     \mathbb{T}_{\vec n }  \Bigl(
     \sum_c  1^{\otimes c-1}\otimes (\sigma^c- \sigma^{c\dagger})  \otimes 1^{\otimes D-c}  \Bigr)  \bar {\mathbb T}_{ \vec {\bar n} } 
       } \; , \nonumber
\eea
thus $ Z(J,\bar J; \lambda,N)$ becomes
\bea
&&  Z(J,\bar J; \lambda,N) = \int  \Big{(}\prod  N^{D-1}\frac{d  {\mathbb T}_{\vec n}
d \bar {\mathbb T}_{ {\vec n } } } { 2 \pi \imath } \Big{)}   \Big{(}\prod  \frac{d  \sigma^c_{ab}
d \bar { \sigma   }^c_{ab  } } { 2 \pi \imath }  \Big{)} \;  e^{ - \sum_c  \tr ( \sigma^c  \sigma^{c\dagger} ) }   \\
&& \; \times e^{  -  N^{D-1} \sum     \mathbb{T}_{\vec n }
   \Bigl(1^{\otimes D } + \sqrt{\frac{ \lambda }{N^{D-1}} }  \sum_c  1^{\otimes c-1}\otimes (\sigma^c- \sigma^{c\dagger})  \otimes 1^{\otimes D-c}  \Bigr)
   \bar {\mathbb T}_{ \vec {\bar n} }
   +   \sum \bar {\mathbb T}_{ \vec {\bar n}  } J_{ \vec{\bar n} }
   +  \sum  {\mathbb T}_{ \vec { n}  } \bar J_{ \vec{ n}   } } \; . \nonumber
\eea
As $\sigma - \sigma^{\dagger}$ is anti hermitian the operator
$R(\sigma) =   \Bigl[1^{\otimes D } +
\sqrt{\frac{ \lambda }{N^{D-1}} }  \sum_c  1^{\otimes c-1}\otimes (\sigma^c- \sigma^{c\dagger})  \otimes 1^{\otimes D-c} \Bigr]^{-1}$,
which we call the resolvent, is well defined. The intermediate field representation renders the integration
over ${\mathbb T}\bar {\mathbb T}$ Gaussian, thus
\bea
  Z(J,\bar J; \lambda,N) &=& \int  \Big{(}\prod_{c;  ab } \frac{d  \sigma^c_{ab}
d \bar { \sigma   }^c_{ a b } } { 2 \pi \imath }  \Big{)}
e^{ - \sum_c \tr \sigma^c\sigma^{c\dagger}  + \tr \ln \bigl(  R(\sigma) \bigr)
   + \frac{1}{N^{D-1} } \langle \bar J | R(\sigma) | J\rangle  }
\; ,
\eea
where  $\langle \bar J | R(\sigma) | J\rangle   = \sum_{\vec n, \vec m} \bar J_{\vec n} R(\sigma)_{\vec n \vec m} J_{\vec m}$.

\bigskip

\noindent{\it Step 2: Extracting the logarithm.} Using lemma \ref{lem:log}
the generating function of the connected moments  $W( J, \bar J; \lambda,N  )$ is
\bea
&& W( J,\bar J; \lambda,N  ) =  \sum_{n\ge 1} \frac{1  }{n!}\sum_{ T_n}  \int_{0}^1 \Bigl( \prod_{ (i,j) \in  T_n } du^{ij} \Bigr) \;
 \int d\mu_{w^{ij}( T_n,u)  1^{\otimes D} } (\sigma) \\
&& \quad \times \Bigg[ \prod_{ (i,j)\in T_n } \sum_{c=1}^D
\bigg( \sum_{ab}    \frac{\partial}{\partial \sigma_{  a b }^{(i)_c} }
\frac{\partial}{\partial    \sigma^{(j)_c\dagger} _{ b a }  } +
\sum_{ab}    \frac{\partial}{\partial \sigma_{  a b }^{(j)_c} }
\frac{\partial}{\partial    \sigma^{(i)_c\dagger} _{ b a }  }
\bigg)  \Bigg] \crcr
&& \qquad \times \prod_{i=1}^n
  \Bigl{\{ }    \tr \ln \bigl[ R(\sigma^{(i) } ) \bigr]
+   \frac{1}{N^{D-1} } \langle \bar J | R(\sigma^{(i)}) |  J \rangle  \Bigr{\}} \; , \nonumber
\eea
where the interpolated Gaussian measure writes formally as
\bea
&& d\mu_{w^{ij}( T_n,u)  1^{\otimes D} } (\sigma ) = \\
&&  e^{ \sum_{c=1}^D  \Bigg( \sum_i \sum_{ ab} \Bigl[ \frac{\partial}{ \partial \sigma_{ab}^{(i)_c} } \frac{\partial}{ \partial \sigma_{ba}^{(i)_c\dagger} }
   + \frac{\partial}{ \partial \sigma_{ab}^{(i)_c}} \frac{\partial}{ \partial \sigma_{ba}^{(i)_c\dagger} }
 \Bigr]
 + \sum_{i< j}  x^{ij} \sum_{  ab} \Bigl[ \frac{\partial}{ \partial \sigma_{ab}^{(i)_c} } \frac{\partial}{ \partial \sigma_{ba}^{(j)_c\dagger} }
   + \frac{\partial}{ \partial \sigma_{ab}^{(j)_c}} \frac{\partial}{ \partial \sigma_{ba}^{(i)_c\dagger} }
 \Bigr] \Bigg) } \; .\nonumber 
\eea

Expanding the product over $i$, we get
\bea
&& W( J,\bar J; \lambda,N  ) =  \sum_{n\ge 1} \frac{1}{n!} \sum_{ T_n}  \int_{0}^1 \Bigl( \prod_{ (i,j) \in  T_n } du^{ij} \Bigr) 
\int d\mu_{w^{ij}( T_n,u)  1^{ \otimes D} } (\sigma^{(i)_c})
\\
&&  \qquad \times 
  \Bigg[ \prod_{ (i,j)\in  T_n } \sum_{c=1}^D
\bigg( \sum_{ab}    \frac{\partial}{\partial \sigma_{  a b }^{(i)_c} }
\frac{\partial}{\partial    \sigma^{(j)_c\dagger} _{ b a }  } +
\sum_{ab}    \frac{\partial}{\partial \sigma_{  a b }^{(j)_c} }
\frac{\partial}{\partial    \sigma^{(i)_c\dagger} _{ b a }  }
\bigg)  \Bigg]
   \nonumber\\
&& \qquad  \times \sum_{k=0}^n \frac{1}{k!}  \frac{1}{N^{k(D-1)} } \sum_{ \stackrel{ i_{ 1} , i_{ 2} \dots , i_{ k} =1 }{i_d\neq i_{d'}} }^n
    \langle \bar J | R(\sigma^{(i_1)}) | J \rangle    \dots  \langle \bar J | R(\sigma^{(i_k)}) | J \rangle \crcr
&& \qquad \times \prod_{\stackrel{i=1}{ i\neq i_{ 1},\dots i_{ k} } }^n
    \tr \ln \bigl[ R(\sigma^{(i) } ) \bigr]  \; . \nonumber
\eea
We need to evaluate the action of the derivative operators on the product resolvents.
This is done in Lemma \ref{lem:contrib} below. For each combinatorial tree $ T_n $  one obtains
a sum over all the plane trees $ \cT_{n,\iota}^{\circlearrowright}$ with colored oriented edges compatible with it, and
indexing the sum by these plane trees we get
\bea\label{eq:cumustrand}
&&  W( J, \bar J; \lambda,N  ) =
   \sum_{n\ge 1} \frac{1}{n!} (-\lambda)^{n-1}
  \sum_{k=0}^n \frac{1}{k!} \frac{1}{  N^{(k+n-1)(D-1)} }
\sum_{ \stackrel{ i_{ 1} , i_{ 2} \dots , i_{ k} =1 }{i_d\neq i_{d'}} }^n
\sum_{  \cT^{\circlearrowright}_{n,\iota} }  \\
&& \qquad \times
 \int_{0}^1 \Bigl( \prod_{ (i,j) \in T_n } du^{ij} \Bigr) \int d\mu_{w^{ij}( T_n,u)  1^{ \otimes D}  } (\sigma )  \int \Big( \prod_{q=1}^{2n-2+k} d\alpha_q\Big)
 \; e^{-\sum_{q=1}^{2n-2+k} \alpha_q } \crcr
&& \prod_{f^c \in \cF^{ \rm{ int} }( \cT_{n,\iota}^{\circlearrowright} ) }
\Tr \Big[ \prod^{\rightarrow}_{ q \in q(f^c)} e^{ -\alpha_q \sqrt{\frac{\lambda}{N^{D-1}}} (\sigma^{i(q)_c} - \sigma^{i(q)_c\dagger}) } \Big] 
\sum_{pn}  \prod_{l=1}^k  \bar J_{p^1_l,\dots p^D_l  }  J_{n^1_l,\dots n^D_l}
\crcr
&&  \prod_{ f^{c;i(q_l) \to i( q_{ \xi_c(l)}  ) } \in \cS^{ \rm{ ext} }( \cT_{n,\iota}^{\circlearrowright} )  } 
\Bigl[ \prod^{\rightarrow}_{ q \in q( f^{c;i(q_l) \to i(q_{ \xi_c(l)} )  } ) } e^{ -\alpha_q \sqrt{\frac{\lambda}{N^{D-1}}} (\sigma^{i(q)_c} - \sigma^{i(q)_c\dagger}) } 
\Bigr]_{p^c_l  n^c_{\xi_c(l)}} \; , \nonumber
\eea 
which proves the theorem. 

\qed

\begin{lemma}\label{lem:contrib} The contribution of a tree $T_n$ with marked vertices $i_1,\dots i_k$ is
\bea\label{eq:contrib}
&& \Bigg[ \prod_{ (i,j)\in  T_n } \sum_{c=1}^D
\bigg( \sum_{ab}    \frac{\partial}{\partial \sigma_{  a b }^{(i)_c} }
\frac{\partial}{\partial    \sigma^{(j)_c\dagger} _{ b a }  } +
\sum_{ab}    \frac{\partial}{\partial \sigma_{  a b }^{(j)_c} }
\frac{\partial}{\partial    \sigma^{(i)_c\dagger} _{ b a }  }
\bigg)  \Bigg] \\
&& \qquad \qquad \times \; \langle \bar J |  R(\sigma^{(i_1)}) | J \rangle   \dots  \langle \bar J | R(\sigma^{(i_k)}) | J \rangle
\prod_{\stackrel{i=1}{ i\neq i_{ 1},\dots i_{ k} } }^n
    \tr \ln \bigl[ R(\sigma^{(i) } ) \bigr] \crcr
 && = \frac{(-\lambda)^{n-1}}{N^{(n-1)(D-1)}}   \sum_{  \cT^{\circlearrowright}_{n,\iota} }
 \int \Big( \prod_{q=1}^{2n-2+k} d\alpha_q\Big)
 \; e^{-\sum_{q=1}^{2n-2+k} \alpha_q } \crcr
&& \prod_{f^c \in \cF^{ \rm{ int} }( \cT_{n,\iota}^{\circlearrowright} ) }
\Tr \Big[ \prod^{\rightarrow}_{ q \in q(f^c)} e^{ -\alpha_q \sqrt{\frac{\lambda}{N^{D-1}}} (\sigma^{i(q)_c} - \sigma^{i(q)_c\dagger}) } \Big] 
\sum_{pn} \prod_{l=1}^k  \bar J_{p^1_l,\dots p^D_l  }  J_{n^1_l,\dots n^D_l} 
\crcr
&&  \prod_{ f^{c;i(q_l) \to i( q_{ \xi_c(l)}  ) } \in \cS^{ \rm{ ext} }( \cT_{n,\iota}^{\circlearrowright} )  } 
\Bigl[ \prod^{\rightarrow}_{ q \in q( f^{c;i(q_l) \to i(q_{ \xi_c(l)} )  } ) } e^{ -\alpha_q \sqrt{\frac{\lambda}{N^{D-1}}} (\sigma^{i(q)_c} - \sigma^{i(q)_c\dagger}) } 
\Bigr]_{p^c_l  n^c_{\xi_c(l)}} \; , \nonumber
\eea
where the sum runs over all the plane trees with colored oriented edges and ciliated vertices $i_1,\dots i_k$ which reduce
to the combinatorial tree $T_n$.
\end{lemma}
\noindent{\bf Proof:} We orient the edge $(i,j)$ of the tree from $i$ to $j$ for the term $ \partial_{\sigma^{(i)}} \partial_{\sigma^{(j)\dagger}}$.
Taking into account the sum over $c$ we obtain the sum over trees with colored, oriented edges.
In order to compute the contribution of each such tree we need to evaluate the action of the derivative
operators on the product of traces. 

We set $M( \sigma^{(i)} ) = [ R(\sigma^{(i)}) ]^{-1}=    1^{\otimes D } +
\sqrt{\frac{ \lambda }{N^{D-1}} }  \sum_{c=1}^D  1^{\otimes c-1}\otimes (\sigma^{(i)_c}- \sigma^{(i)_c\dagger})  \otimes 1^{\otimes D-c}   $
and we represent the resolvents with the help of a new parameter
\bea 
R(\sigma^{(i)}) = \int_{0}^{\infty} d\alpha  \;\; e^{-\alpha } 
   \bigotimes_{c=1}^D e^{ -\alpha  \sqrt{\frac{ \lambda }{N^{D-1}} } (\sigma^{(i)_c}- \sigma^{(i)_c\dagger})   } \; .
\eea 
Note that, denoting $M_{\hat p \hat q} $ the minor of $M$ with the line $p$ and column $q$ deleted, we have
$
 \frac{\partial}{\partial m_{pq}} \tr \ln M =  \frac{\partial}{\partial m_{pq}} \ln \det M = \frac{1}{\det M} \frac{\partial}{\partial m_{pq}} \det M
 = \frac{1}{\det M}  (-1)^{p+q} M_{\hat p \hat q} = [ M^{-1} ]_{qp} 
$, hence the derivatives of the vertices are 
\bea
&& \frac{\partial}{\partial \sigma_{  a b }^{(i)_c} } \tr \ln \bigl[ R(\sigma^{(i) } ) \bigr] =
 -  \frac{\partial}{\partial \sigma_{  a b }^{(i)_c} } \tr \ln M(\sigma^{(i) } )
 = - \sum_{\vec n \vec p} \frac{\partial   m_{\vec n \vec p}  }{\partial \sigma_{  a b }^{(i)_c} }  
 \; \frac{\partial}{\partial m_{\vec n \vec p}  } \tr \ln M  (\sigma^{(i) } )
   \crcr
&& \qquad   =  - \sqrt{\frac{ \lambda }{N^{D-1}} } \sum_{\vec n \vec p}  \bigl( \delta_{n^c a} \delta_{p^c b}  \prod_{ c' \neq c}\delta_{n^{c'} p^{c'}} \bigr)   \;
 R(\sigma^{(i)})_{\vec p \vec n}  \crcr
&& \qquad =   - \sqrt{\frac{ \lambda }{N^{D-1}} } \int_{0}^{\infty} d\alpha  \;\;  e^{-\alpha} 
\Bigl[e^{ -\alpha  \sqrt{\frac{ \lambda }{N^{D-1}} } (\sigma^{(i)_c}- \sigma^{(i)_c\dagger})   }  \Bigr]_{ ab} \crcr
&& \qquad \qquad \qquad \times
\prod_{c'\neq c}
\Tr \Bigl[e^{ -\alpha  \sqrt{\frac{ \lambda }{N^{D-1}} } (\sigma^{(i)_{c'} }- \sigma^{(i)_{c'}\dagger})   }  \Bigr] \; ,
\crcr
 && \frac{\partial}{\partial \sigma_{  ba }^{(i)_c\dagger } } \tr \ln \bigl[ R(\sigma^{(i) } ) \bigr] = \crcr
&& \qquad  = \sqrt{\frac{ \lambda }{N^{D-1}} } \int_{0}^{\infty} d\alpha  \;\;  e^{-\alpha} 
\Bigl[e^{ -\alpha  \sqrt{\frac{ \lambda }{N^{D-1}} } (\sigma^{(i)_c}- \sigma^{(i)_c\dagger})   }  \Bigr]_{ ba }
\crcr
&& \qquad \qquad \qquad \times \prod_{c'\neq c}  \Tr \Bigl[e^{ -\alpha  \sqrt{\frac{ \lambda }{N^{D-1}} } (\sigma^{(i)_{c'} }- \sigma^{(i)_{c'}\dagger})   }  \Bigr]
  \; .
\eea
It follows that a tree made of one line of color $c$ connecting two non ciliated vertices $i$ and $j$ will yield a contribution 
\bea
&& \frac{(-\lambda) }{N^{D-1}} \int_{0}^{\infty} d\alpha_1 d\alpha_2 \;\; e^{-\alpha_1 -\alpha_2}
\Tr \Bigl[ e^{ -\alpha_1  \sqrt{\frac{ \lambda }{N^{D-1}} } (\sigma^{(i)_{c } }- \sigma^{(i)_{c }\dagger})   }  
e^{ -\alpha_2 \sqrt{\frac{ \lambda }{N^{D-1}} } (\sigma^{(j)_{c } }- \sigma^{(j)_{c }\dagger})   }  
\Bigr] 
\crcr
&& \qquad \times  \prod_{c'\neq c}
\Tr \Bigl[e^{ -\alpha_1  \sqrt{\frac{ \lambda }{N^{D-1}} } (\sigma^{(i)_{c'} }- \sigma^{(i)_{c'}\dagger})   }  \Bigr]
\prod_{c'\neq c}
\Tr \Bigl[e^{ -\alpha_2  \sqrt{\frac{ \lambda }{N^{D-1}} } (\sigma^{(j)_{c'} }- \sigma^{(j)_{c'}\dagger})   }  \Bigr]   \; ,
\eea 
which reproduces the equation \eqref{eq:contrib} for the tree with two vertices labelled $i$ and $j$ connected by one line. 
Note that this tree has $2D-2$ internal faces of colors $c'\neq c$ and one internal face of color $c$. 
On the other hand, a ciliated vertex writes as 
\bea
 &&  \langle \bar J | R(\sigma^{(i_1)}) | J \rangle =
 \int_{0}^{\infty} d\alpha \;\; e^{-\alpha} \sum_{pn} \prod_{c=1}^D 
 \Bigl[  e^{ -\alpha  \sqrt{\frac{ \lambda }{N^{D-1}} } (\sigma^{(i_1)_{c} }- \sigma^{(i_1)_{c}\dagger})   }     \Bigr]_{p^c n^c}
  \;\; \bar J_{ \vec p  }  J_{ \vec n } \; ,
\eea
reproducing eq. \eqref{eq:contrib} for the tree with a unique ciliated vertex $i_1$. 

The proof proceeds by induction on the number of vertices. 
Note that any tree can be obtained by adding one by one its edges. 
At each step two trees are joined by the new edges. As the lemma holds for the initial trees (as they have less vertices),
when evaluating the derivative with respect to $ \sigma^{(i)}$ and $\sigma^{(j)\dagger}$ one obtains a sum over terms, one 
for each occurrence of $\sigma^{(i)}$ and $\sigma^{(j)\dagger} $, i.e. a sum over all possible ways to join the two plane 
trees together into a plane tree with $n$ vertices. Lemma \ref{lem:dervi} shows that the right hand side 
of equation\eqref{eq:contrib} is reproduced.

\qed

\subsubsection{Proof of the second constructive expansion theorem \ref{thm:constructive2}}

We add $D$ fictitious integral over the unitary group $U(N)$, i.e. we write
\bea
 W( J,\bar J; \lambda,N  ) =\int_{U(N)} [dU^1] \dots \int  [dU^D]  \; W( J,\bar J; \lambda,N  )\; ,
\eea
which of course holds as $\int_{U(N)} [dU] =1 $. Now, for all fixed $U^c$,
we perform the change of variables of Jacobian $1$, $ \sigma^{(i)_c} \to U^{c\dagger} \sigma^{(i)_c} U^c $.
in eq.\eqref{eq:cumustrand}.
The Gaussian measure is invariant under this change of variables, hence
\bea
 &&  W( J, \bar J; \lambda,N  ) =
   \sum_{n\ge 1} \frac{1}{n!} (-\lambda)^{n-1}
  \sum_{k=0}^n \frac{1}{k!} \frac{1}{  N^{(k+n-1)(D-1)} }
\sum_{ \stackrel{ i_{ 1} , i_{ 2} \dots , i_{ k} =1 }{i_d\neq i_{d'}} }^n
\sum_{  \cT^{\circlearrowright}_{n,\iota} } \\
&&   \times \int_{0}^1 \Bigl( \prod_{ (i,j) \in T_n } du^{ij} \Bigr) 
 \int d\mu_{w^{ij}( T_n,u)  1^{ \otimes D}  } (\sigma )  \int \Big( \prod_{q=1}^{2n-2+k} d\alpha_q\Big)
 \; e^{-\sum_{q=1}^{2n-2+k} \alpha_q } \crcr
&& \times \prod_{f^c \in \cF^{ \rm{ int} }( \cT_{n,\iota}^{\circlearrowright} ) }
\Tr \Big[ \prod^{\rightarrow}_{ q \in q(f^c)} e^{ -\alpha_q \sqrt{\frac{\lambda}{N^{D-1}}} (\sigma^{i(q)_c} - \sigma^{i(q)_c\dagger}) } \Big] \crcr
&& \times \sum_{pn,uv} 
\prod_{l=1}^k \Bigl(  \bar J_{p^1_l,\dots p^D_l  }  J_{n^1_l,\dots n^D_l} \prod_{c=1}^D 
U^{c\dagger}_{  p^c_l   u^c(q_l)       }
 U^c_{ v^c ( q_{ \xi_c(l)}  )   n^c_{\xi_c(l)} } 
\Bigr)
\crcr
&& \times \prod_{ f^{c;i(q_l) \to i( q_{ \xi_c(l)}  ) } \in \cS^{ \rm{ ext} }( \cT_{n,\iota}^{\circlearrowright} )  } 
\Bigl[ \prod^{\rightarrow}_{ q \in q( f^{c;i(q_l) \to i(q_{ \xi_c(l)} )  } ) }
e^{ -\alpha_q \sqrt{\frac{\lambda}{N^{D-1}}} (\sigma^{i(q)_c} - \sigma^{i(q)_c\dagger}) } 
\Bigr]_{u^c(q_l) v^c ( q_{ \xi_c(l)}  )  } \; , \nonumber 
\eea 
where the indices $u^c(q_l)$ and $ v^c ( q_{ \xi_c(l)}  )   $ are summed.
The integral over the unitary group of a product of matrix elements is, 
according to eq. \eqref{eq:weing} (see \cite{collins,ColSni} 
for details)
\bea
&&   \int_{U(N)} [ dU^c ] \; \prod_{l=1}^k  \; 
 U^c_{ v^c ( q_{ \xi_c(l)}  )   n^c_{\xi_c(l)} }  U^{c\dagger}_{  p^c_l   u^c(q_l)       } \crcr
 &&\qquad = \sum_{\sigma_c,\tau_c}  {\rm Wg}(N,\tau_c \sigma_c^{-1} )
   \prod_{l=1}^k \delta_{  v^c ( q_{ \xi_c(l)}  )  u^c(q_{\tau_c(l)} )             } \delta_{ n^c_{\xi_c(l) }    p^c_{\sigma_c(l)}   } \crcr
 &&\qquad =  \sum_{\sigma_c,\tau_c}  {\rm Wg}(N,\tau_c \sigma_c^{-1} )
   \prod_{l=1}^k \delta_{  v^c ( q_{  l  }  )  u^c(q_{\tau_c(l)} )             } \delta_{ n^c_{ l  }    p^c_{\sigma_c(l)}   }  \; ,
\eea
where $\sigma_c$ and $\tau_c$ run over all the permutations of $k$ elements 
(and in the second line we shifted both $\sigma_c$ and $\tau_c$ by the permutation $\xi_c^{-1}$).
We obtain
\bea
 &&  W( J, \bar J; \lambda,N  ) =
   \sum_{n\ge 1} \frac{1}{n!} (-\lambda)^{n-1}
  \sum_{k=0}^n \frac{1}{k!} \frac{1}{  N^{(k+n-1)(D-1)} }
\sum_{ \stackrel{ i_{ 1} , i_{ 2} \dots , i_{ k} =1 }{i_d\neq i_{d'}} }^n
\sum_{  \cT^{\circlearrowright}_{n,\iota} }\\
&& \times
 \int_{0}^1 \Bigl( \prod_{ (i,j) \in T_n } du^{ij} \Bigr) 
 \int d\mu_{w^{ij}( T_n,u)  1^{ \otimes D}  } (\sigma )  \int \Big( \prod_{q=1}^{2n-2+k} d\alpha_q\Big)
 \; e^{-\sum_{q=1}^{2n-2+k} \alpha_q }  \crcr
 &&  \times     \sum_{\vec \sigma, \vec \tau }  \Bigl( \prod_{c=1}^D {\rm Wg}(N,\tau_c \sigma_c^{-1} ) \Bigr) \crcr
&&  \times \prod_{f^c \in \cF^{ \rm{ int} }( \cT_{n,\iota}^{\circlearrowright} ) }
\Tr \Big[ \prod^{\rightarrow}_{ q \in q(f^c)} e^{ -\alpha_q \sqrt{\frac{\lambda}{N^{D-1}}} (\sigma^{i(q)_c} - \sigma^{i(q)_c\dagger}) } \Big] \crcr
&&  \times \sum_{pn,uv}  \Bigl( \prod_{l=1}^k   \prod_{c=1}^D  \delta_{  v^c ( q_{  l  }  )  u^c(q_{\tau_c(l)} )             } \Bigr) \; 
\prod_{l=1}^k \Bigl(  \bar J_{p^1_l,\dots p^D_l  }  J_{n^1_l,\dots n^D_l} \prod_{c=1}^D 
\delta_{ n^c_{ l  }    p^c_{\sigma_c(l)}   }
\Bigr)
\crcr
&& \times \prod_{ f^{c;i(q_l) \to i( q_{ \xi_c(l)}  ) } \in \cS^{ \rm{ ext} }( \cT_{n,\iota}^{\circlearrowright} )  } 
\Bigl[ \prod^{\rightarrow}_{ q \in q( f^{c;i(q_l) \to i(q_{ \xi_c(l)} )  } ) } e^{ -\alpha_q \sqrt{\frac{\lambda}{N^{D-1}}} (\sigma^{i(q)_c} - \sigma^{i(q)_c\dagger}) } 
\Bigr]_{u^c(q_l) v^c ( q_{ \xi_c(l)}  )  }   \; . \nonumber
\eea 
The external sources group into trace invariants: each permutation $\vec \sigma$ is one to one with a $D$-colored graph
$\cB_{\vec \sigma}$ with $2k$ labelled vertices.
The external strands recompose along the faces of the plane tree with external edges
$  \cT_{n,\iota,\vec \tau }^{\circlearrowright}   $ to yield
\bea
 &&  W( J, \bar J; \lambda,N  ) =
   \sum_{n\ge 1} \frac{1}{n!} (-\lambda)^{n-1}
  \sum_{k=0}^n \frac{1}{k!} \frac{1}{  N^{(k+n-1)(D-1)} }
\sum_{ \stackrel{ i_{ 1} , i_{ 2} \dots , i_{ k} =1 }{i_d\neq i_{d'}} }^n
\sum_{  \cT^{\circlearrowright}_{n,\iota} }\crcr
&&  \times \sum_{\vec \sigma, \vec \tau }  \Bigl( \prod_{c=1}^D {\rm Wg}(N,\tau_c \sigma_c^{-1} ) \Bigr) \;\; \Tr_{\cB_{\vec \sigma}} ( J,\bar J ) \crcr 
&& \times
 \int_{0}^1 \Bigl( \prod_{ (i,j) \in T_n } du^{ij} \Bigr) 
 \int d\mu_{w^{ij}( T_n,u)  1^{ \otimes D}  } (\sigma )  \int \Big( \prod_{q=1}^{2n-2+k} d\alpha_q\Big)
 \; e^{-\sum_{q=1}^{2n-2+k} \alpha_q } \;    \crcr
&& \times
\prod_{f^c \in \cF ( \cT_{n,\iota,\vec \tau}^{\circlearrowright} ) }
\Tr \Big[ \prod^{\rightarrow}_{ q \in q(f^c)} e^{ -\alpha_q \sqrt{\frac{\lambda}{N^{D-1}}} (\sigma^{i(q)_c} - \sigma^{i(q)_c\dagger}) } \Big] 
 \; .
\eea 
It follows that the cumulants of the measure $\mu^{(4)}$ are sums over graphs $\cB $
\bea
  \kappa \bigl({\mathbb T}_{\vec p_1 }, \bar {\mathbb T}_{ \vec { n  }_1 } ,
  \dots {\mathbb T}_{ \vec p_k    } , \bar {\mathbb T}_{ \vec{ n }_k } \bigr)
&&= \frac{\partial^{(2k)}} { \partial \bar J_{ \vec p_1  } \partial J_{  \vec{  n  }_1   }
    \dots  \partial \bar J_{    \vec p_k   }  \partial J_{  \vec{  n }_k   }
   }
 W( J ,\bar J ; \lambda,N  ) \Big{\vert}_{J =\bar J =0} \crcr
&&   = \sum_{ \cB, \; k(\cB)=k}  {\mathfrak K} (  \cB  ,\mu^{(4)}_N )  \prod_{\rho=1}^{ C(\cB )  } \delta^{\cB_{\rho} }_{n\bar n}  \; ,
\eea
where, denoting the (unique) permutation $\vec \sigma$ associated to the $D$-colored graph $\cB$ with labelled vertices 
by $\vec \sigma(\cB)$, we have
\bea
 && {\mathfrak K} (  \cB  ,\mu^{(4)}_N ) =   \sum_{n\ge k} \frac{1}{n!} (-\lambda)^{n-1}
   \frac{1}{  N^{(k+n-1)(D-1)} }
\sum_{ \stackrel{ i_{ 1} , i_{ 2} \dots , i_{ k} =1 }{i_d\neq i_{d'}} }^n
\sum_{  \cT^{\circlearrowright}_{n,\iota} } \sum_{ \vec \tau }   \\
&&  \times  k!  \Bigl( \prod_{c=1}^D {\rm Wg}(N,\tau_c \sigma_c^{-1}(\cB) ) \Bigr)  \int_{0}^1 \Bigl( \prod_{ (i,j) \in T_n } du^{ij} \Bigr)
\int d\mu_{w^{ij}( T_n,u)  1^{ \otimes D}  } (\sigma ) 
\crcr
&&  \times
  \int \Big( \prod_{q=1}^{2n-2+k} d\alpha_q\Big)
 \; e^{-\sum_{q=1}^{2n-2+k} \alpha_q }
\prod_{f^c \in \cF ( \cT_{n,\iota,\vec \tau}^{\circlearrowright} ) }
\Tr \Big[ \prod^{\rightarrow}_{ q \in q(f^c)} e^{ -\alpha_q \sqrt{\frac{\lambda}{N^{D-1}}} (\sigma^{i(q)_c} - \sigma^{i(q)_c\dagger}) } \Big] 
 \; , \nonumber
\eea
and Theorem \ref{thm:constructive2} holds.

\qed
 
\subsubsection{Proof of the absolute convergence theorem \ref{thm:absconv}}

The operator $ \sigma^{i(q)_c} - \sigma^{i(q)_c\dagger} $ is an anti-Hermitian operator
hence, denoting $|| \cdot ||$ the operator norm, 
$ \Big{|}\Big{|}  e^{ -\alpha_q \sqrt{\frac{\lambda}{N^{D-1}}} (\sigma^{i(q)_c} - \sigma^{i(q)_c\dagger}) }   \Big{|}\Big{|}  \le 1  $
and 
\bea 
&& \Big{|} \Tr \Big[ \prod^{\rightarrow}_{ q \in q(f^c)} e^{ -\alpha_q \sqrt{\frac{\lambda}{N^{D-1}}} (\sigma^{i(q)_c} - \sigma^{i(q)_c\dagger}) } \Big] \Big{|} \le
N \Big{|}\Big{|}\prod^{\rightarrow}_{ q \in q(f^c)} e^{ -\alpha_q \sqrt{\frac{\lambda}{N^{D-1}}} (\sigma^{i(q)_c} - \sigma^{i(q)_c\dagger}) }   \Big{|}\Big{|}  \crcr
&& \le N  \prod^{\rightarrow}_{ q \in q(f^c)}
\Big{|}\Big{|}  e^{ -\alpha_q \sqrt{\frac{\lambda}{N^{D-1}}} (\sigma^{i(q)_c} - \sigma^{i(q)_c\dagger}) }   \Big{|}\Big{|}  \le N  \; .
\eea 
The integrals over $\alpha_q$ and $u^{ij}$ are bounded by $1$ as well as the Gaussian integral (as $ \mu_{w^{ij}( T_n,u)  1^{ \otimes D}  } (\sigma ) $ is normalized and positive).
The Weingarten function is bounded, from eq. \eqref{eq:weingbound}, by 
$ {\rm Wg}(N,\tau_c \sigma_c^{-1}(\cB) ) \le \frac{   2^{2k} }{  N^{2k-C( \tau_c \sigma_c^{-1}(\cB) )  }  } $
and we get a bound
\bea
 |{\mathfrak K} (  \cB  ,\mu^{(4)}_N )| && \le  \sum_{n\ge k} \frac{1}{n!} |\lambda|^{n-1} 
 \frac{1}{  N^{(k+n-1)(D-1)} }  \sum_{ \stackrel{ i_{ 1} , i_{ 2} \dots , i_{ k} =1 }{i_d\neq i_{d'}} }^n
\sum_{ \cT^{\circlearrowright}_{n,\iota}  }  \sum_{ \vec \tau } \crcr
&& \quad \times k! 2^{2Dk} N^{-2Dk + \sum_c C( \tau_c \sigma_c^{-1}(\cB) ) + \big{|}  \cF ( \cT_{n,\iota,\vec \tau}^{\circlearrowright} )  \big{ | }    }
\eea 
The total scaling with $N$ is therefore, using eq. \eqref{eq:facestreeext},
\bea
 && - (k+n-1)(D-1) -2Dk + \sum_c C( \tau_c \sigma_c^{-1}(\cB) ) \crcr
 && +  D + (n-1)(D-1) -\sum_c C(\xi_c) +\sum_c C( \tau_c\xi_c)\crcr
 && = D -2Dk -k(D-1) + \sum_c C( \tau_c \sigma_c^{-1}(\cB) ) -\sum_c C(\xi_c) +\sum_c C( \tau_c\xi_c) \; ,
\eea 
which is bounded from Lemma \ref{lem:faces} by 
\bea
 D -2Dk -k(D-1) + (D+1)k - C(\cB_{\vec \sigma}) = D - 2(D-1)k - C(\cB_{\vec \sigma}) \; .
\eea 
Using the bounds in eq. \eqref{eq:counttrees} and eq. \eqref{eq:n} we get 
\bea
 |{\mathfrak K} (  \cB  ,\mu^{(4)}_N )|  && \le  N^{ D - 2(D-1)k - C(\cB_{\vec \sigma}) }
 k!^{D+1} 2^{2Dk} \crcr
&&  \sum_{n\ge k} |\lambda|^{n-1}   (2D)^{n-1} \frac{ (2n+k-3)! }{(n-k)!(n+k-1)!} \; .
\eea 
A (not tight) bound on the combinatorial factor is
\bea
  \frac{ (2n+k-3)!  }{ (n-k)!  ( n+k-1)! } \le (2n+k-3)^{k-2} \frac{(2n-1)!}{ (n-k)!  ( n+k-1)! } \le (3n)^{k-2} 2^{2n} \; ,
\eea 
and we get 
\bea
  |{\mathfrak K} (  \cB  ,\mu^{(4)}_N )| \le   N^{D -2 k(D-1)-C(\cB)}  k!^{D+1} 2^{2Dk}  3^{k-2} \sum_{n\ge k}   n^{k-2}  |8D\lambda|^{n-1} 
   \; .
\eea 
The series is absolutely convergent for $|\lambda|<2^{-3} D^{-1}$ and the cumulants are bounded by 
\bea
 |{\mathfrak K} (  \cB  ,\mu^{(4)}_N )| \le   N^{D -2 k(D-1)-C(\cB)} |\lambda|^{k-1} K(\cB)\;,
\eea 
for some constant $K(\cB)$ independent of $N$ (and independent of $\lambda$ for $|\lambda|$ small enough).

\qed

\subsubsection{Proof of corollary \ref{col:continuation} }

Consider first the simpler problem of proving the function of a complex variable $z= re^{\imath \varphi}$ defined as
\bea
f(z)= \int_{0}^{\infty} e^{-\imath \frac{\varphi}{2}}d\alpha \; e^{-\alpha e^{-\imath \frac{\varphi}{2}}   - \imath \alpha r^{1/2}  x }
\eea 
is analytic for $\varphi\in (-\pi ,\pi )$. First it is easy to see that $f$ is bounded as $|f(z)|< \frac{1}{\cos \frac{\varphi}{2} } <\infty$. Second, 
$f$ respects the Cauchy-Riemann equations
\bea
&& r\frac{\partial}{\partial r} f(z) = \int_{0}^{\infty} d\alpha \; e^{-\imath \frac{\varphi}{2}} e^{-\alpha e^{-\imath \frac{\varphi}{2}} } \frac{1}{2} \alpha \frac{d}{d\alpha} e^{  - \imath \alpha r^{1/2}  x }
 \crcr 
&& =  \int_{0}^{\infty} d\alpha \; e^{-\imath \frac{\varphi}{2}} \frac{1}{2} \frac{d}{d\alpha} \Bigl( \alpha  e^{-\alpha e^{-\imath \frac{\varphi}{2}} }   \Bigr) e^{  - \imath \alpha r^{1/2}  x } \crcr
&& = \int_{0}^{\infty} d\alpha \; e^{-\imath \frac{\varphi}{2}} \frac{1}{2}  \Bigl(  e^{-\alpha e^{-\imath \frac{\varphi}{2}} } 
- \alpha e^{-\imath \frac{\varphi}{2}}   e^{-\alpha e^{-\imath \frac{\varphi}{2}} }  \Bigr) e^{  - \imath \alpha r^{1/2}  x }
= \imath \frac{\partial}{\partial \varphi} f \; ,
\eea 
hence it is analytic. For our case similar partial integration with respect to $\alpha$ shows that each function 
\bea
&& {\mathfrak T}^{\mathfrak{E}}( \cT^{\circlearrowright}_{n,\iota,\vec \tau} )  =  k!  \Bigl( \prod_{c=1}^D {\rm Wg}\bigl(N,\tau_c \sigma_c^{-1}(\cB) \bigr) \Bigr) 
\int_{0}^1 \Bigl( \prod_{ (i,j) \in T_n } du^{ij} \Bigr) 
 \int d\mu_{w^{ij}( T_n,u)  1^{ \otimes D}  } (\sigma ) \crcr
&& \qquad \qquad \times
 \int \Big( \prod_{q=1}^{2n-2+k} e^{-\imath \frac{\varphi}{2}}d\alpha_q\Big)
 e^{-\sum_{q=1}^{2n-2+k} e^{-\imath \frac{\varphi}{2}}  \alpha_q } \crcr
&& \qquad \qquad  \times \prod_{f^c \in \cF ( \cT_{n,\iota,\vec \tau}^{\circlearrowright} ) }
\Tr \Big[ \prod^{\rightarrow}_{ q \in q(f^c)} e^{ -\alpha_q \sqrt{\frac{|\lambda|}{N^{D-1}}} (\sigma^{i(q)_c} - \sigma^{i(q)_c\dagger}) } \Big] 
 \; , 
\eea 
respects the Cauchy-Riemann equations and, being bounded, is analytic in $\lambda$. 
It follows that the full cumulant 
\bea\label{eq:analyticcont}
 && {\mathfrak K} (  \cB  ,\mu^{(4)}_N ) =   \sum_{n\ge k} \frac{1}{n!} (-\lambda)^{n-1}
   \frac{1}{  N^{(k+n-1)(D-1)} }
\sum_{ \stackrel{ i_{ 1} , i_{ 2} \dots , i_{ k} =1 }{i_d\neq i_{d'}} }^n
\sum_{  \cT^{\circlearrowright}_{n,\iota} } \sum_{ \vec \tau }   {\mathfrak T}^{\mathfrak{E}}( \cT^{\circlearrowright}_{n,\iota,\vec \tau} ) \; ,
\eea
which is a sum of products of analytic functions, is analytic whenever the sum over $n$ converges. 
Reproducing step by step the proof of Theorem \ref{thm:absconv} but taking into account that the 
integrals over $\alpha$ are bounded by $\frac{1}{ \cos\frac{\varphi}{2} }$ instead of $1$, we obtain that the series in 
eq.\eqref{eq:analyticcont} is absolutely convergent for 
\bea
|\lambda| < \bigl(\cos\frac{\varphi}{2}\bigr)^2 2^{-3} D^{-1}  \; .
\eea 

\qed

\subsubsection{Proof of the mixed expansion theorem \ref{thm:mixedexp}}

Consider the contribution of a tree with external edges 
\bea
&&   T^{E}( \cT^{\circlearrowright}_{n,\iota,\vec\tau}  )  =   N^{ - D +2 k(D-1) + C(\cB) - (k+n-1)(D-1)  } \; k!
 \Bigl( \prod_{c=1}^D {\rm Wg}\bigl(N,\tau_c \sigma_c^{-1}(\cB) \bigr) \Bigr) \crcr
&& \qquad \qquad \times \int_{0}^1 \Bigl( \prod_{ (i,j) \in T_n } du^{ij} \Bigr)   \int d\mu_{w^{ij}( T_n,u)  1^{ \otimes D}  } (\sigma )  \crcr
&& \qquad \qquad  \times \int \Big( \prod_{q=1}^{2n-2+k} e^{-\imath \frac{\varphi}{2}}d\alpha_q\Big)
 \; e^{-\sum_{q=1}^{2n-2+k} e^{-\imath \frac{\varphi}{2}}  \alpha_q } \crcr
&& \qquad \qquad  \times \prod_{f^c \in \cF ( \cT_{n,\iota,\vec \tau}^{\circlearrowright} ) }
\Tr \Big[ \prod^{\rightarrow}_{ q \in q(f^c)} e^{ -\alpha_q \sqrt{\frac{|\lambda|}{N^{D-1}}} (\sigma^{i(q)_c} - \sigma^{i(q)_c\dagger}) } \Big] \; .
\eea 
We use a Taylor expansion 
\bea
&& f (  \sqrt{|\lambda|}e^{\imath \frac{\varphi}{2}} ) = 
  \sum_{q=0}^{s-1} \frac{1}{q!} \Bigl[ \frac{d^q}{dt^q} f ( \sqrt{t |\lambda|}e^{\imath \frac{\varphi}{2}} ) \Bigr]_{t=0} \crcr
&& \qquad    + \frac{1}{(s-1)!} \int_{0}^1 (1-t)^{s-1} \frac{d^{s} }{dt^s} \Bigl(  f ( \sqrt{ t |\lambda|}e^{\imath \frac{\varphi}{2}} )    \Bigr) dt \; .
\eea 
We must evaluate the derivative with respect to $t$ acting on 
$ T^{E}( \cT^{\circlearrowright}_{n,\iota,\vec\tau}  ) $. In each operator $\sqrt{|\lambda|}$ multiplies a difference $\sigma - \sigma^{\dagger}$
and the derivative with respect to $t$ acting on an exponential computes to 
\bea
&& \frac{d}{dt} \prod_{f^c \in \cF ( \cT_{n,\iota,\vec \tau}^{\circlearrowright} ) }
 \Tr \Big[ \prod^{\rightarrow}_{ q \in q(f^c)} e^{ -\alpha_q \sqrt{t} \sqrt{\frac{|\lambda|}{N^{D-1}}} (\sigma^{i(q)_c} - \sigma^{i(q)_c\dagger}) } \Big] \crcr
&&  =\sum_{ f^c \in \cF ( \cT_{n,\iota,\vec \tau}^{\circlearrowright} )  }
\prod_{f^{c'} \in \cF ( \cT_{n,\iota,\vec \tau}^{\circlearrowright} ) \;, \;  f^{c'}\neq f^c }
\Tr \Big[ \prod^{\rightarrow}_{ q \in q(f^c)} e^{ -\alpha_q \sqrt{t} \sqrt{\frac{|\lambda|}{N^{D-1}}} (\sigma^{i(q)_c} - \sigma^{i(q)_c\dagger}) } \Big] \crcr
&& \qquad \times 
  \sum_{q \in q(f^c) }  \frac{1}{2t} \sum_{ab}\Bigl( \sigma^{i(q)_c}_{ab} \frac{\partial}{\partial  \sigma^{i(q)_c}_{ab}} + \sigma^{i(q)_c\dagger}_{ba}
   \frac{\partial}{\partial  \sigma^{i(q)_c\dagger}_{ba} }  \Bigr) \crcr
&& \qquad \qquad \times   \Tr \Big[ \prod^{\rightarrow}_{ q \in q(f^c)} e^{ -\alpha_q \sqrt{t} \sqrt{\frac{|\lambda|}{N^{D-1}}} (\sigma^{i(q)_c} - \sigma^{i(q)_c\dagger}) } \Big] \crcr
  && = \sum_{i=1}^n\sum_{c=1}^D  \frac{1}{2t} \sum_{ab} \Bigl( \sigma^{(i)_c}_{ab} \frac{\partial}{\partial  \sigma^{(i)_c}_{ab}} + \sigma^{(i)_c\dagger}_{ba}
   \frac{\partial}{\partial  \sigma^{(i)_c\dagger}_{ba} } \Bigr) \crcr
   && \qquad \times \prod_{f^c \in \cF ( \cT_{n,\iota,\vec \tau}^{\circlearrowright} ) }
\Tr \Big[ \prod^{\rightarrow}_{ q \in q(f^c)} e^{ -\alpha_q \sqrt{t} \sqrt{\frac{|\lambda|}{N^{D-1}}} (\sigma^{i(q)_c} - \sigma^{i(q)_c\dagger}) } \Big] \; ,
\eea 
Integrating by parts the Gaussian integral we get
\bea
&&  \frac{d}{d t} 
 \int  d\mu_{w^{ij}( T_n,u)  1^{ \otimes D}  } (\sigma )   
   \prod_{f^c \in \cF ( \cT_{n,\iota,\vec \tau}^{\circlearrowright} ) }
\Tr \Big[ \prod^{\rightarrow}_{ q \in q(f^c)} e^{ -\alpha_q t \sqrt{\frac{|\lambda|}{N^{D-1}}} (\sigma^{i(q)_c} - \sigma^{i(q)_c\dagger}) } \Big] \crcr
&& =  
 \int  d\mu_{w^{ij}( T_n,u)  1^{ \otimes D}  } (\sigma )  \crcr
 && \qquad \times \bigg[ \sum_{i,j=1}^n w^{ij}(T_n,u) \sum_{c=1}^D \frac{1}{2t} \sum_{ab}
 \bigg(  \frac{\partial}{\partial \sigma^{(j)_c\dagger}_{ba}} 
\frac{\partial}{\partial \sigma^{(i)_c}_{ab}} +  \frac{\partial}{\partial \sigma^{(j)_c }_{ab}}
   \frac{\partial}{\partial \sigma^{(i)_c\dagger}_{ba}}
 \bigg) \bigg]  \crcr
&& \qquad \qquad  \times   \prod_{f^c \in \cF ( \cT_{n,\iota,\vec \tau}^{\circlearrowright} ) }
\Tr \Big[ \prod^{\rightarrow}_{ q \in q(f^c)} e^{ -\alpha_q t \sqrt{\frac{|\lambda|}{N^{D-1}}} (\sigma^{i(q)_c} - \sigma^{i(q)_c\dagger}) } \Big] \; .
\eea 
The action of the derivatives on a product of traces has been evaluated in lemma \ref{lem:contrib}.
The sums over $i,j$ and $c$ yields a sum over all the possible ways to add a colored loop edge to 
the tree $  \cT^{\circlearrowright}_{n,\iota,\vec \tau} $.
The two new  auxiliary parameters $\beta_1$ and $\gamma_1$ (which we relabel as two supplementary $\alpha$ parameters)
don't have any $e^{-\imath \frac{\varphi}{2}}$ in the measure, hence we need to explicitly add them. Thus we get 
\bea
&& \sum_{j_1j_1',c}  \frac{1}{2t} \Bigl( - \frac{t |\lambda| e^{\imath \varphi}}{ N^{D-1} } \Bigr) 2 w^{j_1j_1'}(T_n,u) 
 \int d\mu_{w^{ij}( T_n,u)  1^{ \otimes D}  } (\sigma )  \crcr
 && \qquad \times \int \Big( \prod_{q=1}^{2n-2+k+2} e^{-\imath \frac{\varphi}{2}}d\alpha_q\Big)
 \; e^{-\sum_{q=1}^{2n-2+k+2} e^{-\imath \frac{\varphi}{2}}  \alpha_q } \crcr
&& \qquad \qquad \times \prod_{f^c \in \cF ( \cT_{n,\iota,\vec \tau,\{j_1j_1'\}}^{\circlearrowright} ) }
\Tr \Big[ \prod^{\rightarrow}_{ q \in q(f^c)} e^{ -\alpha_q \sqrt{\frac{|\lambda|}{N^{D-1}}} (\sigma^{i(q)_c} - \sigma^{i(q)_c\dagger}) } \Big] \;.
\eea 
Note that both derivative terms lead to the same contribution (the loop edge does not have any orientation) hence 
\bea
&& \frac{d}{dt} T^{E}( \cT^{\circlearrowright}_{n,\iota,\vec\tau}  ) 
 =   N^{ - D +2 k(D-1) + C(\cB) - (k+n-1)(D-1)  } \; k!
  \Bigl( \prod_{c=1}^D {\rm Wg}\bigl(N,\tau_c \sigma_c^{-1}(\cB) \bigr) \Bigr) \crcr
&& \times \int_{0}^1 \Bigl( \prod_{ (i,j) \in T_n } du^{ij} \Bigr)   \int d\mu_{w^{ij}( T_n,u)  1^{ \otimes D}  } (\sigma ) 
\sum_{ {\cal L}  = \{(j_1j,_1')\};c_1}  \Bigl( - \frac{ \lambda }{ N^{D-1} } \Bigr)    w^{j_1j_1'}(T_n,u) 
\crcr && \times 
\int \Big( \prod_{q=1}^{2n-2+k+2} e^{-\imath \frac{\varphi}{2}}d\alpha_q\Big)
 \; e^{-\sum_{q=1}^{2n-2+k+2} e^{-\imath \frac{\varphi}{2}}  \alpha_q } \crcr
&& \times \prod_{f^c \in \cF ( \cT_{n,\iota,\vec \tau,{\cal L}}^{\circlearrowright} ) }
\Tr \Big[ \prod^{\rightarrow}_{ q \in q(f^c)} e^{ -\alpha_q \sqrt{t}\sqrt{\frac{|\lambda|}{N^{D-1}}} (\sigma^{i(q)_c} - \sigma^{i(q)_c\dagger}) } \Big] \; ,
\eea 
and the derivative of order $s$ is
\bea
 && \frac{d^s}{dt^s}  T^{E}( \cT^{\circlearrowright}_{n,\iota,\vec\tau}  ) =   N^{ - D +2 k(D-1) + C(\cB) - (k+n-1)(D-1)  } \; k!
   \Bigl( \prod_{c=1}^D {\rm Wg}\bigl(N,\tau_c \sigma_c^{-1}(\cB) \bigr) \Bigr) \crcr
&& \times \int_{0}^1 \Bigl( \prod_{ (i,j) \in T_n } du^{ij} \Bigr)   \int d\mu_{w^{ij}( T_n,u)  1^{ \otimes D}  } (\sigma ) 
\sum_{ {\cal L};c_1\dots c_s }  \Bigl( - \frac{ \lambda }{ N^{D-1} } \Bigr)^s    \prod_{p=1}^s w^{j_pj_p'}(T_n,u) 
\crcr && \times
\int_0^{\infty} \Big( \prod_{q=1}^{2n-2+k+2s} e^{-\imath \frac{\varphi}{2}}d\alpha_q\Big)
 \; e^{-\sum_{q=1}^{2n-2+k+2s} e^{-\imath \frac{\varphi}{2}}  \alpha_q } \crcr
&& \times \prod_{f^c \in \cF ( \cT_{n,\iota,\vec \tau,{\cal L}}^{\circlearrowright} ) }
\Tr \Big[ \prod^{\rightarrow}_{ q \in q(f^c)} e^{ -\alpha_q \sqrt{t}\sqrt{\frac{|\lambda|}{N^{D-1}}} (\sigma^{i(q)_c} - \sigma^{i(q)_c\dagger}) } \Big] \; ,
\eea 
where ${\cal L} \equiv \{(j_1,j_1'),\dots (j_s, j_s')\}$ runs over all the possible ways to decorate $ \cT_{n,\iota,\vec \tau}^{\circlearrowright}   $
with unoriented loop edges ${\cal L}$ and $c_1, \dots c_s$ run over the possible colorings of the $2s$ loop edges.
Taking into account that the Gaussian measures are normalized we evaluate 
\bea
&&  \Bigl[ \frac{d^q}{dt^q} T^{E}( \cT^{\circlearrowright}_{n,\iota,\vec\tau}  ) \Bigr]_{t=0}
  =  N^{ - D +2 k(D-1) + C(\cB) - (k+n-1)(D-1)  } \; k!
   \Bigl( \prod_{c=1}^D {\rm Wg}\bigl(N,\tau_c \sigma_c^{-1}(\cB) \bigr) \Bigr) \crcr
&& \times \int_{0}^1 \Bigl( \prod_{ (i,j) \in T_n } du^{ij} \Bigr)  \Bigl( - \frac{ \lambda }{ N^{D-1} } \Bigr)^q   
\sum_{ {\cal L} ;c_1\dots c_q} \prod_{p=1}^s w^{j_pj_p'}(T_n,u) N^{ | \cF ( \cT_{n,\iota,\vec \tau,{\cal L}}^{\circlearrowright} ) |  }  \crcr
&& =  \Bigl( - \frac{ \lambda }{ N^{D-1} } \Bigr)^q   k!
  \Bigl( \prod_{c=1}^D {\rm Wg}\bigl(N,\tau_c \sigma_c^{-1}(\cB) \bigr) \Bigr) \sum_{ {\cal L},c_1\dots c_q }  \int_{0}^1 \Bigl( \prod_{ (i,j) \in T_n } du^{ij} \Bigr) \prod_{p=1}^s w^{j_pj_p'}(T_n,u)
 \crcr
 && N^{ - D +2 k(D-1) + C(\cB) - (k+n-1)(D-1) +   | \cF ( \cT_{n,\iota,\vec \tau,{\cal L}}^{\circlearrowright} ) |    } \; ,
\eea 
which yields the terms $T^{(q)}( \cT_{n,\iota,\vec \tau }^{\circlearrowright} )$ and 
$T^{EL,(q)}(  \cT_{n,\iota,\vec \tau,{\cal L}}^{\circlearrowright} )  $ in Theorem \ref{thm:mixedexp}. 
The rest term is 
\bea
&& R^{(s)}( \cT_{n,\iota,\vec \tau }^{\circlearrowright}  ) =  \frac{1}{(s-1)!} \int_{0}^1 dt (1-t)^{s-1} \Bigg[ 
  N^{ - D +2 k(D-1) + C(\cB) - (k+n-1)(D-1)  } \crcr
&& \qquad \qquad \times k! \Bigl( \prod_{c=1}^D {\rm Wg}\bigl(N,\tau_c \sigma_c^{-1}(\cB) \bigr) \Bigr) 
\sum_{ {\cal L};c_1\dots c_s }    \Bigl( - \frac{ \lambda }{ N^{D-1} } \Bigr)^s 
\crcr
&& \qquad\qquad\times 
   \int_{0}^1 \Bigl( \prod_{ (i,j) \in T_n } du^{ij} \Bigr)  \prod_{p=1}^s w^{j_pj_p'}(T_n,u)  
    \int d\mu_{w^{ij}( T_n,u)  1^{ \otimes D}  } (\sigma )
   \crcr
&& \qquad \qquad \times \int_0^{\infty} \Big( \prod_{q=1}^{2n-2+k+2s} e^{-\imath \frac{\varphi}{2}}d\alpha_q\Big)  \; e^{-\sum_{q=1}^{2n-2+k+2s} e^{-\imath \frac{\varphi}{2}}  \alpha_q }  
\crcr && 
\qquad\qquad \times 
 \prod_{f^c \in \cF ( \cT_{n,\iota,\vec \tau,{\cal L}}^{\circlearrowright} ) }
\Tr \Big[ \prod^{\rightarrow}_{ q \in q(f^c)} e^{ -\alpha_q \sqrt{t}\sqrt{\frac{|\lambda|}{N^{D-1}}} (\sigma^{i(q)_c} - \sigma^{i(q)_c\dagger}) } \Big] 
 \Bigg] \; ,
\eea 
which establishes the mixed expansion. Concerning the bounds, we first bound the term $T^{(q)}(  \cT_{n,\iota,\vec \tau }^{\circlearrowright} )$
\bea
 && T^{(q)}(  \cT_{n,\iota,\vec \tau }^{\circlearrowright}   ) =  
 \sum_{ {\cal L},c_1\dots c_q }  \frac{1}{q!} \Bigg[ \Bigl( - \frac{ \lambda }{ N^{D-1} } \Bigr)^q  k!
  \Bigl( \prod_{c=1}^D {\rm Wg}\bigl(N,\tau_c \sigma_c^{-1}(\cB) \bigr) \Bigr) \\
  &&  \times  \int_{0}^1 \Bigl( \prod_{ (i,j) \in T_n } du^{ij} \Bigr) \prod_{p=1}^s w^{j_pj_p'}(T_n,u)
  N^{ - D +2 k(D-1) + C(\cB) - (k+n-1)(D-1) +   | \cF ( \cT_{n,\iota,\vec \tau,{\cal L}}^{\circlearrowright} ) |    }  \Bigg] \; . \nonumber
\eea
We bound the product of Weingarten functions by $2^{2Dk} N^{ - 2Dk+ C(\tau_c \sigma_c^{-1}(\cB)   )} $. The integrals 
over $ u^{ij}  $ are bounded by 1, hence
\bea 
 \big{|} T^{(q)}( \cT_{n,\iota,\vec \tau }^{\circlearrowright}) \big{|} \le && \frac{1}{q!} \frac{|\lambda|^q}{N^{(D-1)q}}  k! 2^{2Dk}
    \sum_{ {\cal L},c_1\dots c_q } \\
   &&  \times N^{ - D +2 k(D-1) + C(\cB) -2Dk + C(\tau_c \sigma_c^{-1}(\cB)   )- (k+n-1)(D-1) +   | \cF ( \cT_{n,\iota,\vec \tau,{\cal L}}^{\circlearrowright} ) |    } 
 \; . \nonumber
   \eea 
By eq.\eqref{eq:facestreeextloop} we find that the scaling with $N$ is bounded by
\bea
&& - D +2 k(D-1) + C(\cB) -2Dk + C(\tau_c \sigma_c^{-1}(\cB)   )- (k+n-1)(D-1)  \\
&& +   D+ (n-1)(D-1) -\sum_c C(\xi_c) + \sum_c C( \tau_c\xi_c) +q -q(D-1)  \le -q(D-2) \; , \nonumber
\eea 
where we used lemma \ref{lem:faces}. Thus 
\bea
 \big{|} T^{(q)}( \cT_{n,\iota,\vec \tau }^{\circlearrowright}) \big{|} \le \frac{|\lambda|^q}{N^{q(D-2)}}  k! 2^{2Dk} D^q 
 \frac{ (2n+2q+k-3)! }{q! (2n+k-3)!} \; .
\eea 
For the rest term, $R^{(s)} ( \cT_{n,\iota,\vec \tau }^{\circlearrowright}  )  $ we use similar bounds 
and taking into account that the integrals over $\alpha$ are bounded by $\frac{1}{\cos \frac{\varphi}{2} }  $ and the integral
over $t$ is bounded by $1$ we find 
\bea
 |R^{(s)} ( \cT_{n,\iota,\vec \tau }^{\circlearrowright}  ) |\le \frac{1}{\bigl( \cos \frac{\varphi}{2} \bigr)^{2n+2s+k-2}  }
 \frac{|\lambda|^s}{N^{s(D-2)}}  k! 2^{2Dk} D^s 
 \frac{ (2n+2s+k-3)! }{(s-1)! (2n+k-3)!} \; .
\eea 

\qed

\subsubsection{Proof of the Borel summability theorem \ref{thm:borelsumability}}

We now show that the rescaled cumulants
\bea
&& K(\cB, N) =  N^{ - D +2 k(D-1) + C(\cB)} {\mathfrak K} (  \cB  ,\mu^{(4)}_N ) \crcr
&& =   \sum_{n\ge k} \frac{1}{n!} (-\lambda)^{n-1} \sum_{ \stackrel{ i_{ 1} , i_{ 2} \dots , i_{ k} =1 }{i_d\neq i_{d'}} }^n
\sum_{  \cT^{\circlearrowright}_{n,\iota} } \sum_{\vec \tau}  T^{E}( \cT^{\circlearrowright}_{n,\iota,\vec\tau}  ) \; ,
\eea 
are Borel summable in $\lambda$ uniformly in $N$. First, the corollary \ref{col:continuation} ensures that the series 
\eqref{eq:cumulant1} is absolutely convergent 
for $|\lambda|< \bigl( \cos\frac{\varphi}{2}\bigr)^2 2^{-3} D^{-1}$, hence it certainly is absolutely convergent in a Borel disk of fixed radius.

Second, using the mixed expansion theorem we write
\bea
K(\cB, N)  = \sum_{n\ge k} \frac{1}{n!} (-\lambda)^{n-1} \sum_{ \stackrel{ i_{ 1} , i_{ 2} \dots , i_{ k} =1 }{i_d\neq i_{d'}} }^n
\sum_{  \cT^{\circlearrowright}_{n,\iota} } \sum_{\vec \tau} 
\Bigl[ \sum_{q=1}^{s-1} T^{(q)}(  \cT_{n,\iota,\vec \tau }^{\circlearrowright}  )  + 
R^{(s)} ( \cT_{n,\iota,\vec \tau }^{\circlearrowright}   )\Bigr] \; .
\eea 
We perform a Taylor expansion of $ K(\cB,N) $ in $\lambda$ up to order $r>k$. All the terms corresponding to trees with $n\ge r+1$ are in the reminder, hence
for them we use the mixed expansion for $s=0$.
For the terms corresponding to trees with $n<r+1$ we use the mixed expansion up to order $s=r-(n-1)$. The explicit terms 
$T^{(q)}  ( \cT_{n,\iota,\vec \tau }^{\circlearrowright}   ) $ yield 
a series in $\lambda$ which is nothing but the Taylor expansion of the rescaled cumulant. 
The reminder term of the Taylor expansion of $ K(\cB,N) $ writes as
\bea
 R_{N,r }(\cB, \lambda) && =   \sum_{n=k}^{r} \frac{1}{n!} (-\lambda)^{n-1} \sum_{ \stackrel{ i_{ 1} , i_{ 2} \dots , i_{ k} =1 }{i_d\neq i_{d'}} }^n
\sum_{  \cT^{\circlearrowright}_{n,\iota} }\sum_{\vec \tau } R^{(r+1-n)} (  \cT^{\circlearrowright}_{n,\iota,\vec \tau } ) 
\crcr
&& + \sum_{n\ge r+1} \frac{1}{n!} (-\lambda)^{n-1} \sum_{ \stackrel{ i_{ 1} , i_{ 2} \dots , i_{ k} =1 }{i_d\neq i_{d'}} }^n
\sum_{  \cT^{\circlearrowright}_{n,\iota} } \sum_{\vec \tau } R^{(0)} ( \cT^{\circlearrowright}_{n,\iota,\vec \tau} ) \; .
\eea 
The terms with $n\ge r+1$ then admit a bound
\bea
&&  \Bigl{|} \sum_{n\ge r+1} \frac{1}{n!} (-\lambda)^{n-1} \sum_{ \stackrel{ i_{ 1} , i_{ 2} \dots , i_{ k} =1 }{i_d\neq i_{d'}} }^n
\sum_{  \cT^{\circlearrowright}_{n,\iota} } \sum_{\vec \tau } R^{(0)} ( \cT^{\circlearrowright}_{n,\iota,\vec \tau } )  \Bigl{|} \crcr
&& \le
\sum_{n\ge r+1}  |\lambda|^{n-1} (2D)^{n-1} \frac{(2n+k-3)!}{ (n-k)! ( n+k-1)! } \frac{1}{ \bigl(\cos \frac{\varphi}{2}\bigr)^{2n-2+k} }  
k!^{D+1} 2^{2Dk}\crcr
&& \le  k!^{D+1} 2^{2Dk} \sum_{n\ge r+1}  |\lambda|^{n-1} (2D)^{n-1} \frac{1}{ \bigl(\cos \frac{\varphi}{2}\bigr)^{2n-2+k} }
(3n)^{k-2} 2^{2n} \crcr
&& \le \frac{K}{ \bigl( \cos\frac{\varphi}{2}\bigr)^k  } \Big( \frac{|\lambda|}{ \bigl( \cos\frac{\varphi}{2}\bigr)^2 }\Big)^r \; ,
\eea 
for some constant $K$ and $|\lambda|< 2^{-3}D^{-1}\bigl( \cos\frac{\varphi}{2}\bigr)^2$.
The terms with $ r \le n $ admit a bound
\bea
&& \Bigl{|} \sum_{n=k}^{r} \frac{1}{n!} (-\lambda)^{n-1} \sum_{ \stackrel{ i_{ 1} , i_{ 2} \dots , i_{ k} =1 }{i_d\neq i_{d'}} }^n
\sum_{  \cT^{\circlearrowright}_{n,\iota} } R^{(r+1-n)} (  \cT^{\circlearrowright}_{n,\iota,\vec \tau} ) \Bigl{|} \crcr
&& \le \sum_{n = k}^r  |\lambda|^{n-1} (2D)^{n-1} \frac{(2n+k-3)!}{ (n-k)! ( n+k-1)! } \crcr
&& \frac{1}{\bigl(\cos \frac{\varphi}{2}\bigr)^{2n-2+k+2(r+1-n)}}  |\lambda|^{r+1-n} N^{-(r+1-n)(D-2)}    k!^{D+1} 
 2^{2Dk}\crcr
&& \qquad \times D^{r+1-n} \frac{ [2n+k+2(r+1-n)-3]!  }{ (r -n)! [2n+k-3]!   } 
\eea 
Taking into account that 
\bea
&& \frac{ (2n+k-3 )! }{ (n-k)! (n+k-1)! }  <  3^{2n+k-1} k! < 3^{3r} k! \crcr
&&  \frac{ (2r+k-1)!}{ (r-n)!  [  2n+k -3  ]!}   <  3^{ 2r+k-1 } (r-n+2)! < 3^{3r} (r+1)! \
\eea 
these terms are bounded by
\bea 
 \frac{|\lambda|^r}{\bigl(\cos\frac{\varphi}{2}\bigr)^{2r+k}}   k!^{D+2} 2^{2Dk}   D^r 3^{6r} (r+1)! 2^r
\eea 
Overall we thus derive a bound
\bea
  R_{N,r }(\cB, \lambda) \le K \frac{|\lambda|^r}{\bigl(\cos\frac{\varphi}{2}\bigr)^{2r+k}} \sigma^r r! \; ,
\eea 
for some constants $K$ and $\sigma$ which proves the theorem \ref{thm:borelsumability}.

\qed

\subsubsection{Proof of the 1/N expansion theorem \ref{thm:1/N}}
Using the mixed expansion up to order $s$ for every $\cT^{\circlearrowright}_{n,\iota,\vec \tau}$ 
leads to the rest term
\bea
&&  R^{(s)}_{N}(\cB,\lambda) =  \sum_{n\ge k} \frac{1}{n!} (-\lambda)^{n-1} \sum_{ \stackrel{ i_{ 1} , i_{ 2} \dots , i_{ k} =1 }{i_d\neq i_{d'}} }^n
\sum_{  \cT^{\circlearrowright}_{n,\iota} } \sum_{\vec \tau} R^{(s)}( \cT^{\circlearrowright}_{n,\iota,\vec \tau}) \;,
\eea 
hence using the bound on $R^{(s)}( \cT^{\circlearrowright}_{n,\iota,\vec \tau})  $ we get 
\bea
&& |  R^{(s)}_{N}(\cB,\lambda) | \le \sum_{n\ge k} \frac{1}{n!} |\lambda|^{n-1} \sum_{ \stackrel{ i_{ 1} , i_{ 2} \dots , i_{ k} =1 }{i_d\neq i_{d'}} }^n
\sum_{  \cT^{\circlearrowright}_{n,\iota} } \sum_{\vec \tau} \crcr
&& \qquad \times \frac{1}{\bigl( \cos \frac{\varphi}{2} \bigr)^{2n+2s+k-2}  }
 \frac{|\lambda|^s}{N^{s(D-2)}} ( k! 2^{2Dk} D^s) 
 \frac{ (2n+2s+k-3)! }{(s-1)! (2n+k-3)!} \; .
 \eea 
By the combinatorial countings eq. \eqref{eq:counttrees} we get 
\bea
&& |  R^{(s)}_{N}(\cB,\lambda) | \le \frac{1}{N^{s(D-2)}}   ( k!^{D+1} 2^{2Dk} D^s)  \frac{1}{\bigl( \cos \frac{\varphi}{2} \bigr)^k  }\sum_{n\ge k} 
 \Bigl( \frac{|\lambda|}{ \bigl( \cos \frac{\varphi}{2} \bigr)^2   } \Bigr)^{s+n-1} \crcr
 && \qquad \qquad  \qquad \qquad \times  (2D)^{n-1} \frac{(2n+k-3)!}{ (n-k)! ( n+k-1)! }   \frac{ (2n+2s+k-3)! }{(s-1)! (2n+k-3)!} \; ,
\eea 
that is, denoting $K$ and $\sigma$ two constants depending only on $k$,
\bea
 && |  R^{(s)}_{N}(\cB,\lambda) | \le \frac{1}{N^{s(D-2)}}   K \sigma^s \frac{|\lambda|^{s+k-1}}{\bigl( \cos \frac{\varphi}{2} \bigr)^{2s+3k-1}  }\crcr
 &&  \qquad \times  \sum_{q\ge 0} 
 \Bigl( \frac{|2D\lambda|}{ \bigl( \cos \frac{\varphi}{2} \bigr)^2   } \Bigr)^{ q} 
   \frac{ (2q+2s+3k-3)! }{(s-1)! q ! ( q+2k-1)!  } \; .
\eea 
A non optimal bound on the combinatorial factor is
\bea
 \frac{ (2q+2s+3k-3)! }{(s-1)! q ! ( q+2k-1)!  } \le (s+1)! k! 5^{ 2q+2s+3k-3  } \; ,
\eea 
and, as the sum over $q$ converges absolutely for $|\lambda|< 5^{-2}2^{-1}D^{-1} \bigl(\cos \frac{\varphi}{2} \bigr)^2  $, we get 
\bea
  |  R^{(s)}_{N}(\cB,\lambda) | \le K  \sigma^s  s!  \frac{1}{N^{s(D-2)}}  \frac{|\lambda|^{s+k-1}}{\bigl( \cos \frac{\varphi}{2} \bigr)^{2s+3k-1}  }\; .
\eea 
for some constants $K$ and $\sigma$.

\qed


\begin{thebibliography}{99}


\bibitem{review}
  R.~Gurau and J.~P.~Ryan,
  ``Colored Tensor Models - a review,''
  SIGMA {\bf 8}, 020 (2012)
  [arXiv:1109.4812 [hep-th]].
  %%CITATION = ARXIV:1109.4812;%%
  %46 citations counted in INSPIRE as of 08 Apr 2013

\bibitem{Mehta}
   M.~L.~Mehta, ``Random Matrices'', Elsevir, 2004
   Pure and Applied Mathematics (Amsterdam), 142. Elsevier/Academic Press, (2004).
  
\bibitem{matrix}
  P.~Di Francesco, P.~H.~Ginsparg and J.~Zinn-Justin,
  ``2-D Gravity and random matrices,''
  Phys.\ Rept.\  {\bf 254}, 1 (1995)
  [hep-th/9306153].
  %%CITATION = HEP-TH/9306153;%%
  %454 citations counted in INSPIRE as of 08 Apr 2013
  
\bibitem{'tHooft:1973jz}
  G.~'t Hooft,
  ``A Planar Diagram Theory for Strong Interactions,''
  Nucl.\ Phys.\ B {\bf 72}, 461 (1974).
  %%CITATION = NUPHA,B72,461;%%
  %3410 citations counted in INSPIRE as of 08 Apr 2013
  
\bibitem{Brezin:1977sv}
  E.~Brezin, C.~Itzykson, G.~Parisi and J.~B.~Zuber,
  ``Planar Diagrams,''
  Commun.\ Math.\ Phys.\  {\bf 59}, 35 (1978).
  %%CITATION = CMPHA,59,35;%%
  %991 citations counted in INSPIRE as of 08 Apr 2013
 
\bibitem{Kazakov:1985ds}
  V.~A.~Kazakov,
  ``Bilocal Regularization of Models of Random Surfaces,''
  Phys.\ Lett.\ B {\bf 150}, 282 (1985).
  %%CITATION = PHLTA,B150,282;%%
  %300 citations counted in INSPIRE as of 08 Apr 2013

\bibitem{mm}
   F.~David,
  ``A Model of Random Surfaces with Nontrivial Critical Behavior,''
  Nucl.\ Phys.\ B {\bf 257}, 543 (1985).
  %%CITATION = NUPHA,B257,543;%%
  %309 citations counted in INSPIRE as of 08 Apr 2013
  
\bibitem{oldgft0.5} 
  D.~Oriti,
  ``The microscopic dynamics of quantum space as a group field theory,''
  arXiv:1110.5606 [hep-th].
  %%CITATION = ARXIV:1110.5606;%%
  %32 citations counted in INSPIRE as of 01 Apr 2013

\bibitem{oldgft1}
  N.~Sasakura,
  ``Tensor model for gravity and orientability of manifold,''
  Mod.\ Phys.\ Lett.\  A {\bf 6}, 2613 (1991).
  %%CITATION = MPLAE,A6,2613;%%

\bibitem{oldgft2}
  J.~Ambjorn, B.~Durhuus and T.~Jonsson,
  ``Three-Dimensional Simplicial Quantum Gravity And Generalized Matrix
  Models,''
  Mod.\ Phys.\ Lett.\  A {\bf 6}, 1133 (1991).
  %%CITATION = MPLAE,A6,1133;%%
  
 \bibitem{oldgft3}
 N.~Sasakura,
  ``Tensor models and 3-ary algebras,''
   J.\ Math.\ Phys.\  {\bf 52}, 103510 (2011)
  [arXiv:1104.1463 [hep-th]].
  %%CITATION = ARXIV:1104.1463;%%

\bibitem{oldgft4}
    N.~Sasakura,
   ``Tensor models and hierarchy of n-ary algebras,''
  Int.\ J.\ Mod.\ Phys.\ A {\bf 26}, 3249 (2011),
  arXiv:1104.5312 [hep-th].
  %%CITATION = ARXIV:1104.5312;%%

\bibitem{oldgft5}
D.~V.~Boulatov,
   A Model of three-dimensional lattice gravity,
  Mod.\ Phys.\ Lett.\ A {\bf 7} (1992) 1629
  [hep-th/9202074].
  %%CITATION = HEP-TH/9202074;%%

\bibitem{oldgft6}
H.~Ooguri,
   Topological lattice models in four-dimensions,
  Mod.\ Phys.\ Lett.\ A {\bf 7} (1992) 2799
  [hep-th/9205090].
  %%CITATION = HEP-TH/9205090;%%
   
\bibitem{oldgft10}
A.~Baratin and D.~Oriti,
  {\it Group field theory with non-commutative metric variables,}
  Phys.\ Rev.\ Lett.\  {\bf 105} (2010) 221302
  [arXiv:1002.4723 [hep-th]].
  %%CITATION = ARXIV:1002.4723;%%
 
\bibitem{color}
   R.~Gurau,
  ``Colored Group Field Theory,''
  Commun.\ Math.\ Phys.\  {\bf 304}, 69 (2011)
  [arXiv:0907.2582 [hep-th]].
  %%CITATION = ARXIV:0907.2582;%%
  %79 citations counted in INSPIRE as of 08 Apr 2013

\bibitem{lost}
  R.~Gurau,
  ``Lost in Translation: Topological Singularities in Group Field Theory,''
  Class.\ Quant.\ Grav.\  {\bf 27}, 235023 (2010)
  [arXiv:1006.0714 [hep-th]].
  %%CITATION = ARXIV:1006.0714;%%
  %56 citations counted in INSPIRE as of 01 Apr 2013

\bibitem{uncoloring}
  V.~Bonzom, R.~Gurau and V.~Rivasseau,
  ``Random tensor models in the large N limit: Uncoloring the colored tensor models,''
  Phys.\ Rev.\ D {\bf 85}, 084037 (2012)
  [arXiv:1202.3637 [hep-th]].
  %%CITATION = ARXIV:1202.3637;%%
  %21 citations counted in INSPIRE as of 08 Apr 2013

\bibitem{expansion1}
  R.~Gurau,
  ``The 1/N expansion of colored tensor models,''
  Annales Henri Poincare {\bf 12}, 829 (2011)
  [arXiv:1011.2726 [gr-qc]].
  %%CITATION = ARXIV:1011.2726;%%
  %54 citations counted in INSPIRE as of 08 Apr 2013

\bibitem{expansion2} 
  R.~Gurau and V.~Rivasseau,
  ``The 1/N expansion of colored tensor models in arbitrary dimension,''
  Europhys.\ Lett.\  {\bf 95}, 50004 (2011)
  [arXiv:1101.4182 [gr-qc]].
  %%CITATION = ARXIV:1101.4182;%%
  %43 citations counted in INSPIRE as of 08 Apr 2013
  
\bibitem{expansion3}
  R.~Gurau,
  ``The complete 1/N expansion of colored tensor models in arbitrary dimension,''
  Annales Henri Poincare {\bf 13}, 399 (2012)
  [arXiv:1102.5759 [gr-qc]].
  %%CITATION = ARXIV:1102.5759;%%
  %53 citations counted in INSPIRE as of 08 Apr 2013

\bibitem{expansion4}
 V.~Bonzom,
   New 1/N expansions in random tensor models,
  arXiv:1211.1657 [hep-th].
  %%CITATION = ARXIV:1211.1657;%%

\bibitem{expansioin5} 
  S.~Dartois, V.~Rivasseau, A.~Tanasa and ,
  ``The 1/N expansion of multi-orientable random tensor models,''
  arXiv:1301.1535 [hep-th].
  %%CITATION = ARXIV:1301.1535;%%
  %1 citations counted in INSPIRE as of 01 Apr 2013   
  
\bibitem{critical}
  V.~Bonzom, R.~Gurau, A.~Riello and V.~Rivasseau,
  ``Critical behavior of colored tensor models in the large N limit,''
  Nucl.\ Phys.\ B {\bf 853}, 174 (2011)
  [arXiv:1105.3122 [hep-th]].
  %%CITATION = ARXIV:1105.3122;%%
  %50 citations counted in INSPIRE as of 08 Apr 2013

\bibitem{melbp} 
  R.~Gurau, J.~P.~Ryan and ,
  ``Melons are branched polymers,''
  arXiv:1302.4386 [math-ph].
  %%CITATION = ARXIV:1302.4386;%%
  %1 citations counted in INSPIRE as of 01 Apr 2013

\bibitem{sefu2}
  J.~B.~Geloun, J.~Magnen and V.~Rivasseau,
  ``Bosonic Colored Group Field Theory,''
  Eur.\ Phys.\ J.\  C {\bf 70}, 1119 (2010),
  arXiv:0911.1719 [hep-th].
  %%CITATION = EPHJA,C70,1119;%%

\bibitem{Ryan:2011qm}
  J.~P.~Ryan,
  ``Tensor models and embedded Riemann surfaces,''
  Phys.\ Rev.\ D {\bf 85}, 024010 (2012)
  [arXiv:1104.5471 [gr-qc]].
  %%CITATION = ARXIV:1104.5471;%%

\bibitem{Carrozza:2011jn} 
  S.~Carrozza and D.~Oriti,
  ``Bounding bubbles: the vertex representation of 3d Group Field Theory and the suppression of pseudo-manifolds,''
  Phys.\ Rev.\ D {\bf 85}, 044004 (2012)
  [arXiv:1104.5158 [hep-th]].
  %%CITATION = ARXIV:1104.5158;%%
 
\bibitem{Carrozza:2012kt} 
  S.~Carrozza and D.~Oriti,
  ``Bubbles and jackets: new scaling bounds in topological group field theories,''
  JHEP {\bf 1206}, 092 (2012)
  [arXiv:1203.5082 [hep-th]].
  %%CITATION = ARXIV:1203.5082;%%

\bibitem{IsingD}
  V.~Bonzom, R.~Gurau and V.~Rivasseau,
  ``The Ising Model on Random Lattices in Arbitrary Dimensions,''
  arXiv:1108.6269 [hep-th].
  %%CITATION = ARXIV:1108.6269;%%

\bibitem{EDT}
  D.~Benedetti and R.~Gurau,
  ``Phase Transition in Dually Weighted Colored Tensor Models,''
  Nucl.\ Phys.\ B {\bf 855}, 420 (2012)
  arXiv:1108.5389 [hep-th].
  %%CITATION = ARXIV:1108.5389;%%

\bibitem{doubletens} 
  R.~Gurau,
  ``The Double Scaling Limit in Arbitrary Dimensions: A Toy Model,''
   Phys.\ Rev.\ D {\bf 84}, 124051 (2011) arXiv:1110.2460 [hep-th].
  %%CITATION = ARXIV:1110.2460;%%

  \bibitem{sdequations}
  R.~Gurau,
  A generalization of the Virasoro algebra to arbitrary dimensions,
  Nucl.\ Phys.\ B {\bf 852} (2011) 592
  [arXiv:1105.6072 [hep-th]].
  %%CITATION = ARXIV:1105.6072;%%

  \bibitem{sdequations1}
   R.~Gurau,
  The Schwinger Dyson equations and the algebra of constraints of random tensor models at all orders,
  Nucl.\ Phys.\ B {\bf 865} (2012) 133
  [arXiv:1203.4965 [hep-th]].
  %%CITATION = ARXIV:1203.4965;%%

  \bibitem{sdequations2}
  T.~Krajewski,
   Schwinger-Dyson Equations in Group Field Theories of Quantum Gravity,
  arXiv:1211.1244 [math-ph].
  %%CITATION = ARXIV:1211.1244;%%

   \bibitem{sdequations3}
   V.~Bonzom,
  ``Revisiting random tensor models at large N via the Schwinger-Dyson equations,''
  arXiv:1208.6216 [hep-th].
  %%CITATION = ARXIV:1208.6216;%%
  
\bibitem{Bonzom:2012sz} 
  V.~Bonzom,
  ``Multicritical tensor models and hard dimers on spherical random lattices,''
  arXiv:1201.1931 [hep-th].
  %%CITATION = ARXIV:1201.1931;%%

\bibitem{Bonzom:2012qx} 
  V.~Bonzom and H.~Erbin,
   ``Coupling of hard dimers to dynamical lattices via random tensors,''
  arXiv:1204.3798 [cond-mat.stat-mech].
  %%CITATION = ARXIV:1204.3798;%%

\bibitem{BenGeloun:2011rc}
J.~Ben Geloun and V.~Rivasseau,
``A Renormalizable 4-Dimensional Tensor Field Theory,''
arXiv:1111.4997 [hep-th].
%%CITATION = ARXIV:1111.4997;%%
  
\bibitem{BenGeloun:2012pu}
  J.~Ben Geloun and D.~O.~Samary,
  ``3D Tensor Field Theory: Renormalization and One-loop $\beta$-functions,''
  arXiv:1201.0176 [hep-th].
  %%CITATION = ARXIV:1201.0176;%%

\bibitem{BenGeloun:2012yk} 
  J.~Ben Geloun,
  ``Two and four-loop $\beta$-functions of rank 4 renormalizable tensor field theories,''
  arXiv:1205.5513 [hep-th].
  %%CITATION = ARXIV:1205.5513;%%
  
\bibitem{jbgelon}
  J.~B.~Geloun,
   Asymptotic Freedom of Rank 4 Tensor Group Field Theory,
  arXiv:1210.5490 [hep-th].
  %%CITATION = ARXIV:1210.5490;%%

\bibitem{Samary:2013xla} 
  D.~O.~Samary,
  ``Beta functions of $U(1)^d$ gauge invariant just renormalizable tensor models,''
  arXiv:1303.7256 [hep-th].
  %%CITATION = ARXIV:1303.7256;%%
  
\bibitem{Geloun:2012bz} 
  J.~B.~Geloun and E.~R.~Livine,
  ``Some classes of renormalizable tensor models,''
  arXiv:1207.0416 [hep-th].
  %%CITATION = ARXIV:1207.0416;%%

\bibitem{Carrozza:2012uv} 
  S.~Carrozza, D.~Oriti and V.~Rivasseau,
  ``Renormalization of Tensorial Group Field Theories: Abelian U(1) Models in Four Dimensions,''
  arXiv:1207.6734 [hep-th].
  %%CITATION = ARXIV:1207.6734;%%

\bibitem{tt2} 
  S.~Carrozza, D.~Oriti, V.~Rivasseau and ,
  ``Renormalization of an SU(2) Tensorial Group Field Theory in Three Dimensions,''
  arXiv:1303.6772 [hep-th].
  %%CITATION = ARXIV:1303.6772;%%  
  
\bibitem{Rivasseau:2011hm}
  V.~Rivasseau,
  ``Quantum Gravity and Renormalization: The Tensor Track,''
  arXiv:1112.5104 [hep-th].
  %%CITATION = ARXIV:1112.5104;%%

\bibitem{Rivasseau:2012yp} 
  V.~Rivasseau,
  ``The Tensor Track: an Update,''
  arXiv:1209.5284 [hep-th].
  %%CITATION = ARXIV:1209.5284;%%
  %8 citations counted in INSPIRE as of 01 Apr 2013
  
\bibitem{universality}
  R.~Gurau,
  ``Universality for Random Tensors,''
  arXiv:1111.0519 [math.PR].
  %%CITATION = ARXIV:1111.0519;%%
  %23 citations counted in INSPIRE as of 08 Apr 2013

\bibitem{GlimmJaffe}
  J.~Glimm and A.~Jaffe,
  ``Quantum physics. A functional integral point of view,''
   Springer, 2nd edition (1987).

\bibitem{LVE1}
  V.~Rivasseau,
  ``Constructive Matrix Theory,''
  JHEP {\bf 0709}, 008 (2007)
  [arXiv:0706.1224 [hep-th]].
  %%CITATION = ARXIV:0706.1224;%%

\bibitem{LVE2}
  V.~Rivasseau and Z.~Wang,
  ``Loop Vertex Expansion for Phi**2K Theory in Zero Dimension,''
  J.\ Math.\ Phys.\  {\bf 51}, 092304 (2010)
  [arXiv:1003.1037 [math-ph]].
  %%CITATION = ARXIV:1003.1037;%%
  
\bibitem{Magnen:2009at} 
  J.~Magnen, K.~Noui, V.~Rivasseau and M.~Smerlak,
  ``Scaling behaviour of three-dimensional group field theory,''
  Class.\ Quant.\ Grav.\  {\bf 26}, 185012 (2009)
  [arXiv:0906.5477 [hep-th]].
  %%CITATION = ARXIV:0906.5477;%%
  %62 citations counted in INSPIRE as of 08 Apr 2013
  
\bibitem{collins}
   B.~Collins, ``Moments and cumulants of polynomial random variables on unitary groups, the Itzykson-Zuber integral,
   and free probability,'' Int. Math. Res. Not. {\bf 17}, (2003) 953 [arXiv:math-ph/0205010].

   
\bibitem{ColSni}
  B.~Collins and P.~Sniady ``Integration with respect to the Haar measure on unitary, orthogonal and symplectic group,''
  Commun.\ Math. \ Phys.\ {\bf 264}, 773 (2006) [arXiv:math-ph/0402073 ].
 
\bibitem{pezzana}
   M.~Pezzana,
    ``Sulla struttura topologica delle variet\`a compatte,''
   Atti Sem.\ Mat.\ Fis.\ Univ.\ Modena, {\bf 23} (1974), 269-277.

\bibitem{FG}
  M. ~Ferri and C.~Gagliardi
  ``Crystallisation moves,''
  Pacific\ Journal\ of\ Mathematics\ Vol. 100, No. 1, 1982.
 
\bibitem{NevSok}
   A.~D.~Sokal, ``An improvement of Watson's theorem on Borel summability,'' J.\ Math.\ Phys. {\bf 21}, 261 (1980).

\bibitem{BKAR}
   A.~Abdesselam and V.~Rivasseau, ``Trees, forests and jungles: a botanical
   garden for cluster expansions,'' in Constructive Physics, ed by V. Rivasseau,
   Lecture Notes in Physics 446, Springer Verlag, 1995.

  
\end{thebibliography}
\end{document}